\documentclass[useAMS,usenatbib]{mn2e}
\usepackage{natbib}
\usepackage{graphicx}
\usepackage{aecompl}

\newcommand{\mdot}{\mbox{M$_{\odot}$ yr$^{-1}$}}
\newcommand{\msun}{M$_{\odot}$}
\newcommand{\lsun}{L$_{\odot}$}
\newcommand{\rsun}{$R_{\odot}$}
\newcommand{\teff}{$T_{\mathrm{eff}}$}
\newcommand{\mbol}{$M_{\mathrm{bol}}$}
\newcommand{\tcont}{$T_{\mathrm{cont}}$}

\newcommand{\hii}{H\,{\sc ii}}
\newcommand{\ha}{H\,$\alpha$}

\newcommand{\vrad}{$v_{\rm rad}$}

\newcommand{\kband}{$K_{S}$}
\newcommand{\jhk}{\emph{JHK}$_{S}$}
\newcommand{\jmink}{$J-K_{S}$}

\newcommand{\iras}{\emph{IRAS}}
\newcommand{\iso}{\emph{ISO}}
\newcommand{\spitzer}{{\emph{Spitzer}}}
\newcommand{\herschel}{\emph{Herschel}}
\newcommand{\wise}{\emph{WISE}}
\newcommand{\akari}{\emph{AKARI}}

\newcommand{\sssmc}{S$^{3}$MC}

\newcommand{\kms}{km s$^{-1}$}
\newcommand{\um}{$\mu$m}

\newcommand{\hh}{\hbox{$^{h}$}}
\newcommand{\hm}{\hbox{$^{m}$}}
\newcommand{\hs}{\hbox{$^{s}$}}

\newcommand{\x}{$\times$\ }

\title[\spitzer-IRS point source classification in the SMC]{\spitzer\ Infrared Spectrograph point source classification in the Small Magellanic Cloud}

\author[Paul M. E. Ruffle et al.]
	{Paul~M.~E.~Ruffle,$^{1,2}$\thanks{Paul M. E. Ruffle passed away on 21 November 2013. His co-authors have finished the manuscript on his behalf, and would like to dedicate it to his memory. Paul was a very enthusiastic scientist, and a wonderful friend with a great sense of humour. We miss him tremendously. }
	F.~Kemper,$^2$\thanks{Email: ciska@asiaa.sinica.edu.tw}
	O.~C.~Jones,$^{1,3}$
	G.~C.~Sloan,$^4$
	K.~E.~Kraemer,$^5$
\newauthor        
        Paul~M.~Woods,$^{6}$         
	M.~L.~Boyer,$^{7}$
        S.~Srinivasan,$^{2}$
        V.~Antoniou,$^8$
	E.~Lagadec,$^{9}$       
\newauthor
        M.~Matsuura,$^{10,11}$
	I.~McDonald,$^1$
	J.~M.~Oliveira,$^{12}$
	B.~A.~Sargent,$^{13}$
        M.~{Sewi{\l}o},$^{14,15}$        
\newauthor
        R.~Szczerba,$^{16}$
	J.~Th.~van~Loon,$^{12}$
	K.~Volk$^{3}$	
    and
	A.~A.~Zijlstra$^1$ \\
	\\
	$^1$Jodrell Bank Centre for Astrophysics, The University of Manchester, Alan Turing Building, 
	    Oxford Road, Manchester M13 9PL \\
	$^2$Academia Sinica, Institute of Astronomy and Astrophysics, Taipei 10617, Taiwan \\
	$^{3}$Space Telescope Science Institute, 3700 San Martin Drive, Baltimore, MD 21218, USA \\
	$^4$Department of Astronomy, Cornell University, Ithaca, NY 14853, USA \\
	$^5$Institute for Scientific Research, Boston College, 140 Commonwealth Avenue, Chestnuthill, MA 02467, USA \\
    $^{6}$Astrophysics Research Centre, School of Mathematics \& Physics, 
        Queen's University Belfast, University Road, Belfast, BT7 1NN\\
    $^7$Observational Cosmology Lab, Code 665, NASA Goddard Space Flight Center, Greenbelt, MD 20771, USA \\
	$^8$Harvard-Smithsonian Center for Astrophysics, 60 Garden Street, Cambridge, MA 02138, USA\\
    $^{9}$Laboratoire Lagrange, Universit\'e de Nice - Sophia Antipolis, Observatoire de la C\^ote d'Azur, CNRS, 06304 Nice, France\\
    $^{10}$School of Physics and Astronomy,
Cardiff University, The Parade, Cardiff CF24 3AA\\
    $^{11}$Department of Physics and Astronomy, University College London, Gower Street, London WC1E 6BT\\
	$^{12}$School of Physical and Geographical Sciences, Lennard-Jones Laboratories, 
	    Keele University, Staffordshire ST5 5BG \\
	$^{13}$Laboratory for Multiwavelength Astrophysics, Rochester Institute of Technology, 
	    54 Lomb Memorial Drive, Rochester, NY 14623, USA \\ 
    $^{14}$Space Science Institute, 4750 Walnut Street, Suite 205, Boulder, CO 80301, USA\\
    $^{15}$Johns Hopkins University, Department of Physics and Astronomy, 366 Bloomberg Center,
        3400 N.~Charles Street, Baltimore, MD 21218, USA\\
    $^{16}$N.~Copernicus Astronomical Center, Rabianska 8, 87-100, Torun, Poland
	}

\begin{document}

\maketitle

\begin{abstract}
  The Magellanic clouds are uniquely placed to study the stellar
  contribution to dust emission. Individual stars can be resolved in
  these systems even in the mid-infrared, and they are close enough to
  allow detection of infrared excess caused by dust.  We have searched
  the \spitzer\ Space Telescope data archive for all Infrared
  Spectrograph (IRS) staring-mode observations of the Small Magellanic
  Cloud (SMC) and found that 209 Infrared Array Camera (IRAC) point
  sources within the footprint of the Surveying the Agents of Galaxy
  Evolution in the Small Magellanic Cloud (SAGE-SMC) \spitzer\ Legacy
  programme were targeted, within a total of 311 staring mode
  observations. We classify these point sources using a decision tree
  method of object classification, based on infrared spectral
  features, continuum and spectral energy distribution shape,
  bolometric luminosity, cluster membership and variability
  information. We find 58 asymptotic giant branch (AGB) stars, 51
  young stellar objects (YSOs), 4 post-AGB objects, 22 Red Supergiants
  (RSGs), 27 stars (of which 23 are dusty OB stars), 24 planetary
  nebulae (PNe), 10 Wolf-Rayet (WR) stars, 3 \hii\ regions, 3 R
  Coronae Borealis (R CrB) stars, 1 Blue Supergiant and 6 other
  objects, including 2 foreground AGB stars. We use these
  classifications to evaluate the success of photometric
  classification methods reported in the literature.
\end{abstract}


\section{Introduction}\label{intro}

The Mega-Surveying the Agents of Galaxy Evolution (Mega-SAGE) project
has obtained infrared photometric and spectroscopic inventories of the
Magellanic Clouds with the \spitzer\ Space Telescope (hereafter
\spitzer ), using \spitzer\ and \herschel\ Legacy Programmes.  The
initial SAGE survey \citep{Meixner_06_SAGE} detected and catalogued
$\sim$6.9 million point sources in the Large Magellanic Cloud
(LMC)\footnote{{http://irsa.ipac.caltech.edu/data/SPITZER/docs/ \\spitzermission/observingprograms/legacy/sage}},
while the SAGE-SMC survey \citep{Gordon_11_SAGE-SMC} detected and
catalogued $\sim$2.2 million point sources in the Small Magellanic
Cloud
(SMC)\footnote{{http://irsa.ipac.caltech.edu/data/SPITZER/docs/ \\spitzermission/observingprograms/legacy/sagesmc}}. Both
surveys used all bands of the Infrared Array Camera \citep[IRAC; 3.6,
4.5, 5.8, 8.0\,\um ;][]{Fazio_04_IRAC} and the Multi-Band Imaging
Photometer for \spitzer\ \citep[MIPS; 24, 70, 160\,\um
;][]{Rieke_04_MIPS} instruments on board \spitzer\
\citep{Werner_04_Spitzer}.  The resolution of the IRAC observations is
$\sim$2\arcsec, while for the MIPS bands, three different resolutions apply:
6\arcsec, 18\arcsec, and 40\arcsec\ for the 24, 70 and 160\,\um\ bands
respectively \citep{Gordon_11_SAGE-SMC}. To follow up on these
programmes, the SAGE-Spec project \citep{Kemper_10_SAGE-Spec} obtained
196 staring-mode pointings using \spitzer 's Infrared Spectrograph
\citep[IRS;][5.2--38\,\um ]{Houck_04_IRS} of positions selected from
the SAGE catalogue.  SAGE-Spec will relate SAGE photometry to the
spectral characteristics of different types of objects in both Magellanic Clouds,
and ultimately, allow us to classify photometric point sources
in both the LMC and SMC. This characterisation of the point sources
observed in the SAGE-Spec survey, and the IRS data archive,
builds an inventory of dusty sources and their interrelation in each
of the Magellanic Clouds.  In a first step towards this goal,
\citet{Woods_11_classification} classified the initial 196 LMC point
sources, using a decision tree method of object classification, based
on infrared (IR) spectral features, continuum and spectral energy
distribution (SED) shape, bolometric luminosity, cluster membership
and variability information. The initial classification of LMC objects
is being extended to $\sim$1,000 point sources, covering all archival
IRS observations within the SAGE footprint (Woods et al.~\emph{in
  prep.}).

To extend the LMC classifications to the SMC, we have searched the
\spitzer\ data archive for IRS staring-mode observations and found 311
spectra, yielding 209 unique and genuine point sources with IRS data
within the footprint of the SAGE-SMC \spitzer\ Legacy programme.  The
data used in the classification process are described in
Section~\ref{dataprep}. In Section~\ref{method} we discuss the
classification method, and in Section~\ref{class} we classify the 209
SMC point sources using the decision tree method.  Finally, in
Section~\ref{pops} we compare spectral versus colour classifications
by means of colour-magnitude diagrams (CMDs). We use our spectroscopic
classifications to test photometric classification methods, e.g.~those
by \citet{Boyer_11_SMC}; \citet{Sewilo_13_YSOs} and
\citet{Matsuura_13_SMC}.  The classification of each of these 209
sources is part of the data delivery of the SAGE-Spec Legacy project
to the \spitzer\ Science Center and the
community.\footnote{{http://irsa.ipac.caltech.edu/data/SPITZER/docs/ \\spitzermission/observingprograms/legacy/sagespec}}
These classifications will also be used to benchmark a
colour-classification scheme that will be applied to all point sources
in the SAGE and SAGE-SMC surveys (Marengo et al.~\emph{in prep}).

\section{Data preparation}\label{dataprep}

\subsection{\spitzer\ IRS staring mode observations}

The IRS on board \spitzer\ covers the wavelength range 5--38\,\um. For
the low-resolution mode, the spectrum splits in two bands,
short-low (SL: 5.2--14.5\,\um ) and long-low (LL: 14.0--38.0\,\um ),
with almost perpendicular slits. Each segment splits into a range covered at
second order (SL2, LL2), and one at first order (SL1, LL1). The
resolution varies between 60 and 130. The high resolution mode covers
the wavelength range from 10--19.6\,\um\ (short-high; SH) and from 
18.7--37.2\,\um\ (long-high; LH) with a spectral resolution of $R \sim 600$.

\begin{figure*}
\hbox{ 
\includegraphics[height=65mm]{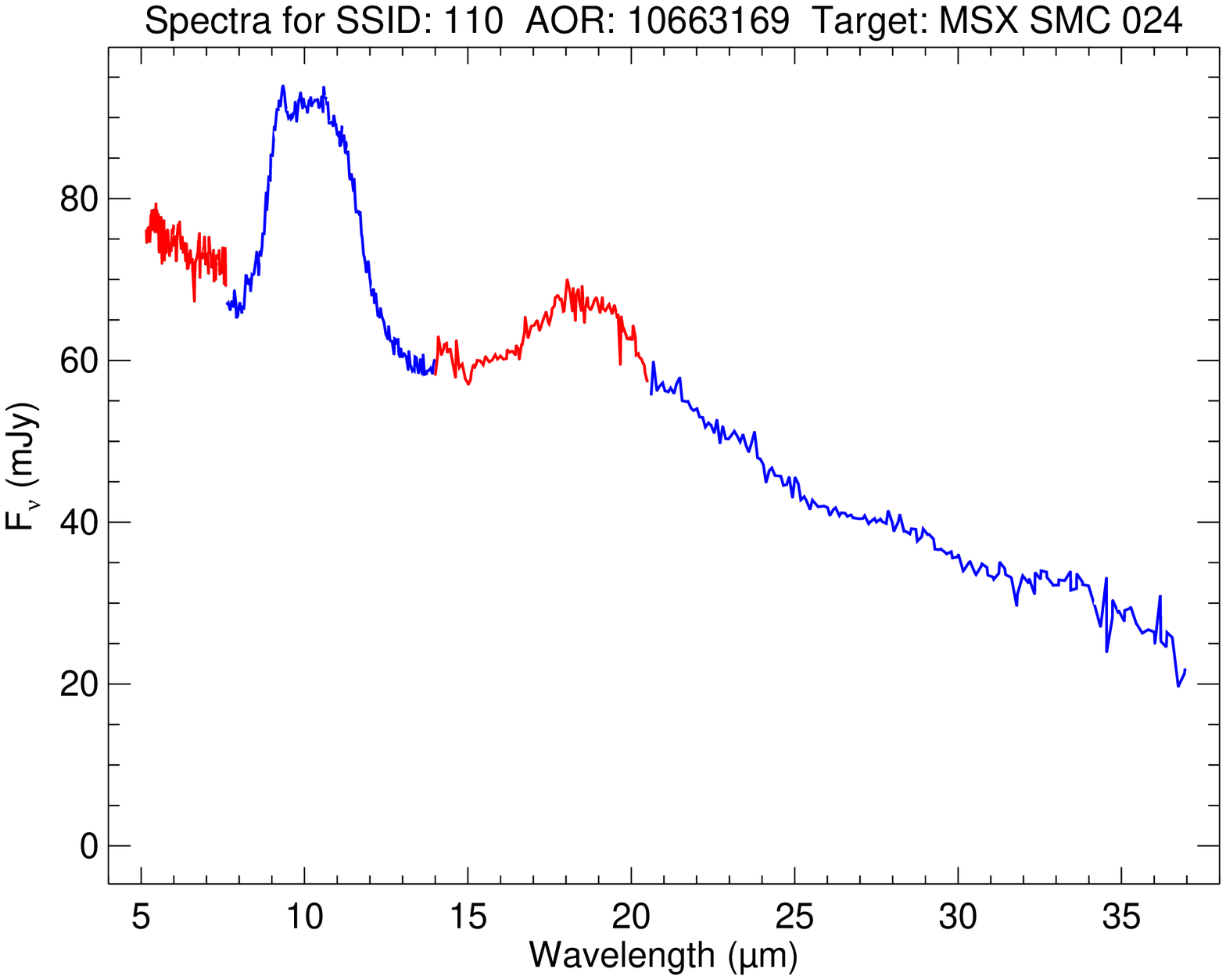}
\hspace{17mm}
\includegraphics[height=65mm]{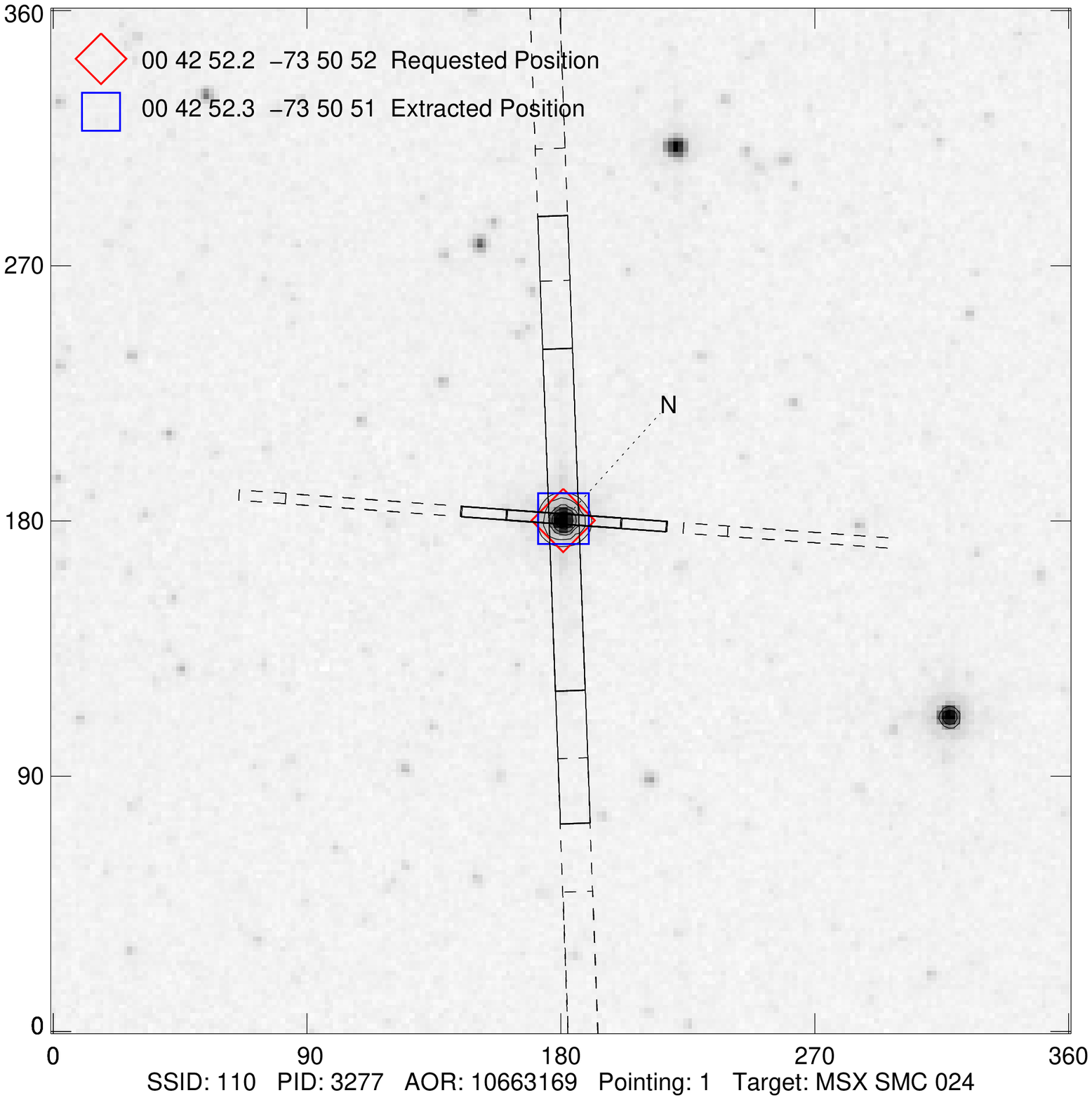}}
\vspace{4mm}
\hbox{ 
\includegraphics[height=65mm]{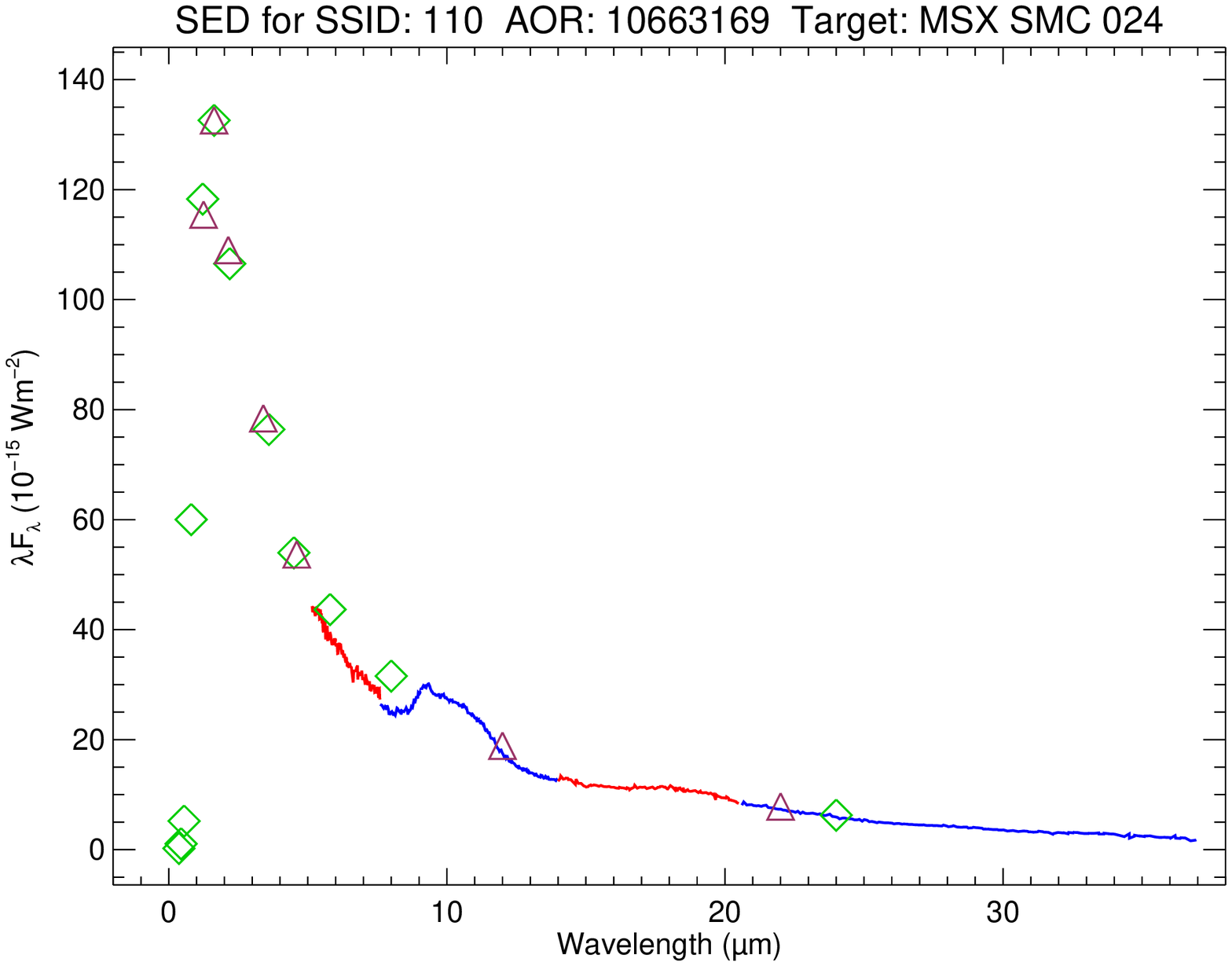}
\hspace{5mm}
\includegraphics[height=65mm]{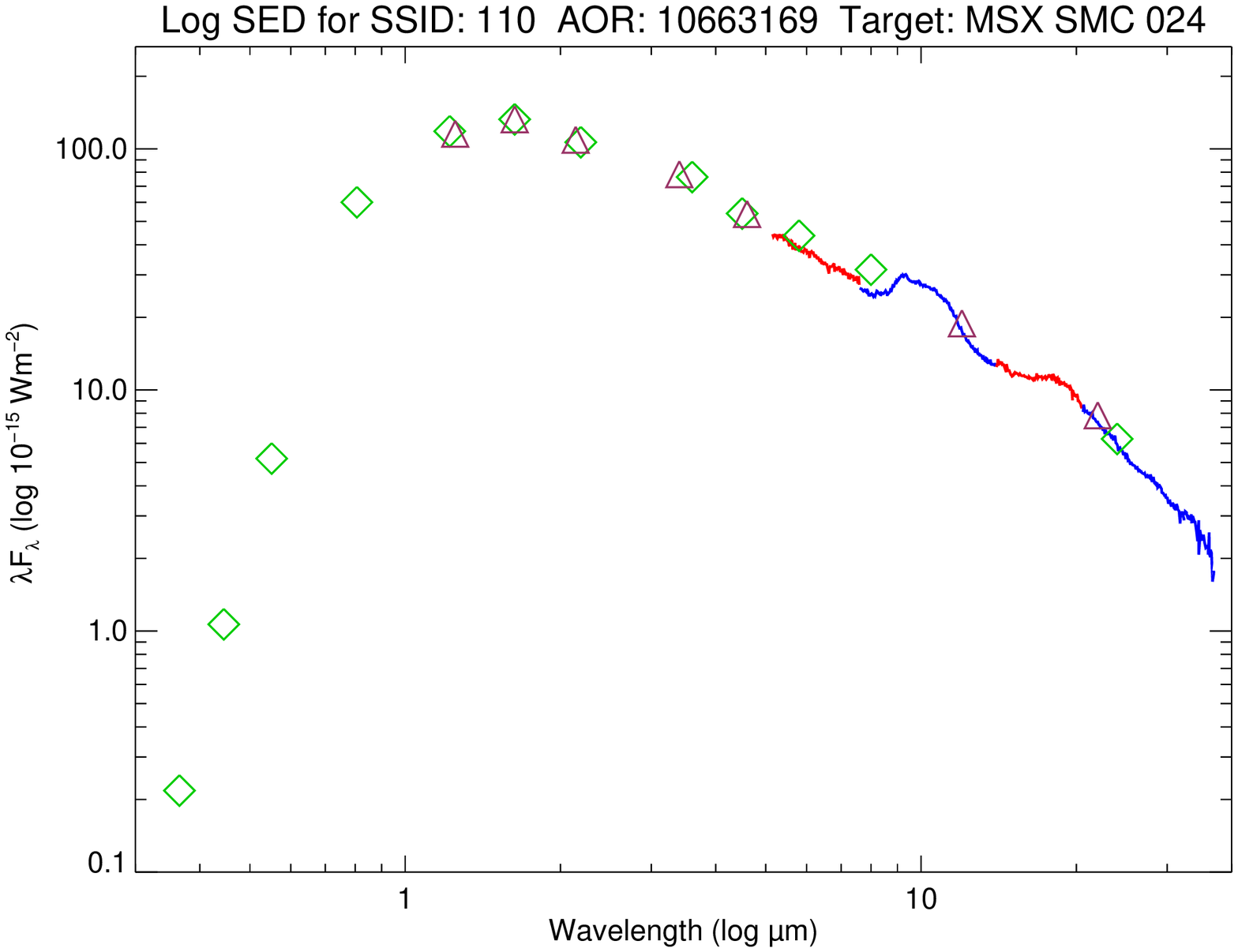}}
\caption{From upper left to lower right: Example spectrum, slit image,
  SED and log SED plots for point source SMC IRS 110.  The spectrum
  plots were generated using CASSIS-reduced IRS data from the
  following low resolution modules: 
     short-low 2nd order 5.2--7.6\,\um\ (red); 
     short-low 1st order 7.6--14.0\,\um\ (blue); 
     long-low 2nd order 14.0--20.5\,\um\ (red); 
     long-low 1st order 20.5--37.0\,\um (blue).  
  Slit images were created by over-plotting IRS short-low and
  long-low slit positions on a 360\arcsec\ \x 360\arcsec\ image
  extracted from the SAGE-SMC 8~\um\ image; the REQ and EXT or FOV
  positions were also over-plotted.  SED and log SED plots (lower left
  and lower right panels, respectively) were generated by combining
  the above IRS data with the following photometric data points:
    \emph{SAGE-SMC catalogue} (green diamonds): 
      $U$, $B$, $V$, $I$ (MCPS), 
      $J$, $H$, \kband\ (IRSF); 
      3.6, 4.5, 5.8, 8.0\,\um\ (IRAC), 24\,\um\ (MIPS); 
    \emph{WISE catalogue} (magenta triangles): 
      $J$, $H$, \kband\ (2MASS), 
      3.4, 4.6, 12, 22\,\um\ (\wise ).}
\label{example_plots}
\end{figure*}

We identified 311 \spitzer\ IRS low- and high-resolution staring mode
observations within the footprint of the SAGE-SMC survey
\citep{Gordon_11_SAGE-SMC}, not necessarily associated with a point
source. We numbered these SMC IRS 1--311, by ordering them by
observing program (Project ID; PID) first, and then the Astronomical
Observation Request (AOR) number (Table~\ref{tab:irsobservations}).
Where available (for SL and LL observations only), the reduced spectra
were downloaded from the Cornell Atlas of \spitzer\ IRS
Sources\footnote{http://cassis.astro.cornell.edu}
\citep[CASSIS;][]{Lebouteiller_11_CASSIS}, in a full resolution grid,
using the optimal extraction method. At the moment, the CASSIS
database only contains SL and LL data, but it turns out that there are
no point sources in the SMC targeted with only SH and LH, so we will
use the SL and LL data only.  The data were downloaded in the Infrared
Processing and Analysis Center (IPAC) Table
format\footnote{{http://irsa.ipac.caltech.edu/applications/DDGEN/Doc/ \\ipac\_{}tbl.html}};
the file names are also given in Table~\ref{tab:irsobservations}.  As
an example, the upper left panel of Fig.~\ref{example_plots} shows the
spectrum for point source SMC IRS 110, with the SL2 and LL2 data from
CASSIS in red and the SL1 and LL1 data in blue.

\begin{table*}
  \caption{IRS staring mode targets
    in the SMC. The first ten lines are presented to demonstrate the
    format of this table; the full table is available on-line.}
\label{tab:irsobservations}
\begin{tabular}{crrclll}
 SMC IRS  &       AOR  &    PID  &  PI            &  \multicolumn{2}{c}{spectrum position}  &  \emph{CASSIS} file name                       \\
& & & & RA (\hh\ \hm\ \hs) &  Dec (\degr\ \arcmin\ \arcsec) &\\
\hline
                    1  &   3824640  &     18  &  Houck         &  00 46 40.32           &  $-$73 06 10.80           &  NULL                                                   \\
                    2  &   3824896  &     18  &  Houck         &  01 09 16.80           &  $-$73 12 03.60           &  NULL                                                   \\
                    3  &   4384000  &     63  &  Houck         &  01 24 07.68           &  $-$73 09 03.60           &  NULL                                                   \\
                    4  &   4384000  &     63  &  Houck         &  01 24 07.68           &  $-$73 09 03.60           &  NULL                                                   \\
                    5  &   4384000  &     63  &  Houck         &  01 24 07.68           &  $-$73 09 03.60           &  NULL                                                   \\
                    6  &   4384768  &     63  &  Houck         &  00 59 09.84           &  $-$72 10 51.60           &  NULL                                                   \\
                    7  &   4385024  &     63  &  Houck         &  00 58 52.25           &  $-$72 09 25.92           &  cassis\_{}tbl\_{}spcf\_{}4385024\_{}1.tbl    \\
                    8  &   4385024  &     63  &  Houck         &  00 58 58.22           &  $-$72 09 50.76           &  cassis\_{}tbl\_{}spcf\_{}4385024\_{}2.tbl    \\
                    9  &   4385024  &     63  &  Houck         &  00 58 58.80           &  $-$72 10 25.32           &  cassis\_{}tbl\_{}spcf\_{}4385024\_{}3.tbl    \\
                   10  &   4385024  &     63  &  Houck         &  00 59 06.62           &  $-$72 10 25.68           &  cassis\_{}tbl\_{}spcf\_{}4385024\_{}4.tbl    \\
\hline
\end{tabular}
\end{table*}

The CASSIS-reduced IRS data header provides the original PI's
requested (REQ) position and the field of view (FOV) position,
i.e. where the telescope actually pointed (usually, but not always
coincident within 1\arcsec\ of the REQ position). In many cases,
however, neither of these two positions is the same as the position at
which the spectrum is actually extracted from the slit (EXT), using
the optimal extraction method.  In order to check the spectrum
position for each source, slit images were generated by over-plotting
the IRS SL and LL slit positions (recorded in each \spitzer\ AOR's BCD
FITS header) on a 360\arcsec\ \x 360\arcsec\ image extracted from the
SAGE-SMC 8~\um\ image.  The REQ and EXT positions were also
over-plotted on the image. In cases where the spectrum was extracted
at the FOV position (and thus an EXT position is lacking), the FOV
position was overplotted instead (see upper right panel of
Fig.~\ref{example_plots} for an example for SMC IRS 110). The slit
images are useful in determining the origin of the emission seen in
the IRS spectra. The coordinates in Table~\ref{tab:irsobservations}
represent, if available, the EXT position.  The next preference is the
FOV position, and if neither of these is available, the REQ position
is given.

\subsection{Photometric matching}

In order to find matching photometry for the IRS spectra, we searched
the SAGE-SMC {Single Frame +
Mosaic Photometry (SMP) Archive} v1.5 \citep{Gordon_11_SAGE-SMC} available on
\emph{Gator}\footnote{Gator is the general catalogue query engine
  provided by the NASA/IPAC Infrared Science Archive, which is
  operated by the Jet Propulsion Laboratory, California Institute of
  Technology, under contract with the National Aeronautics and Space
  Administration.}, using, in order of preference, the EXT, FOV or REQ
spectrum positions. We searched for IRAC point source matches within
3\arcsec\ of the spectrum positions, which corresponds with the
pointing accuracy of the IRS mode on \spitzer. In cases where the
SAGE-SMC point source catalogue did not provide a match, we also
searched the \spitzer\ Survey of the Small Magellanic Cloud (\sssmc )
catalogue\footnote{At the time of writing only the \sssmc\ Young
  Stellar Object Catalog is available in the public domain
  (A.~Bolatto, priv.~comm.).}  \citep{Bolatto_07_S3MC} for IRAC
matches within 3\arcsec. We found three sources in \sssmc\ without a
SAGE-SMC catalogue counterpart. Although the \sssmc\ data are included
in the SAGE-SMC result, both teams used different point source
extraction pipelines and the source catalogues therefore do not
provide a one-to-one match.

\begin{figure*}
\includegraphics[width=153mm]{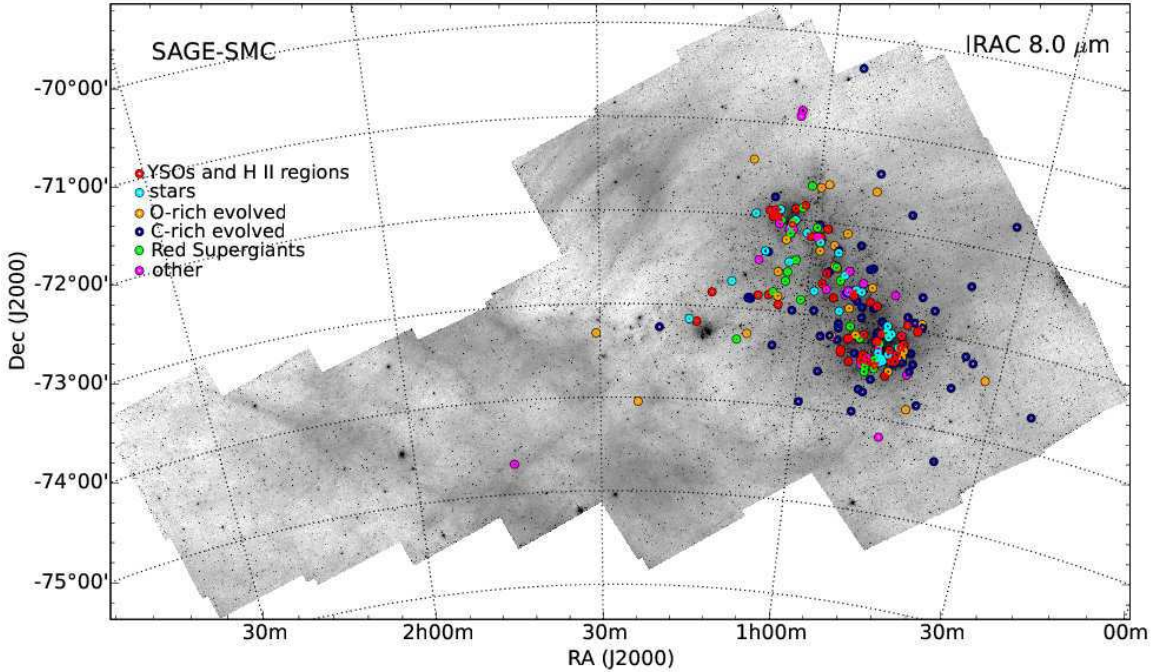}
\caption{The SMC IRS targets distributed on the sky, overlaid upon a
  SAGE-SMC IRAC 8 \um\ map. The colour of the points represent our
  classifications, according to the legend. All four YSO subcategories
  and the \hii\ regions are grouped together in red. The class
  ``stars'' contains objects classified as STAR and the sub-category
  of dusty OB stars. The ``O-rich evolved'' category contains O-EAGB,
  O-AGB, O-PAGB and O-PN objects, and likewise the ``C-rich evolved''
  category groups together C-AGB, C-PAGB and C-PN objects. The Red
  Supergiants are a group by themselves, and ``other'' contains all
  other categories (see Table~\ref{classgroups}).}
\label{smcskymap}
\end{figure*}

Of the original list of 311 IRS staring mode observations within the
SAGE-SMC footprint, we discarded all {44} spectra for which we could not
identify an IRAC point source within 3\arcsec\ in either the
SAGE-SMC or the \sssmc\ surveys.  In cases where multiple matches were
present within 3\arcsec, we manually compared the magnitudes with the
flux levels of the spectra, and used the slit images to establish
which source was responsible for the spectrum. We also consolidated
duplicate measurements of the spectrum of a given source, as evidenced
by their SAGE-SMC or \sssmc\ identification, into a single entry in
our analysis; this is sufficient for spectral identification purposes.
{This further reduced the number by
58 to} a list of 209 unique \spitzer -IRAC point sources,
with either SAGE-SMC or \sssmc\ identifications, for which IRS staring
mode observations are available.  We compiled all relevant information
in a table available online.  Table~\ref{tab:metatable} describes the
columns of the online table. Fig.~\ref{smcskymap} shows the
distribution of the 209 sources over the SMC.

\begin{table*}
  \caption{Numbering, names and description of the columns present in the
classification table which is available on-line only. }
\label{tab:metatable}
\begin{tabular}{lll}
  \hline
  Column & Name & Description \\
  \hline
  1 & smc\_{}irs & SMC IRS identification number of the target \\
  2 & name & Name of point source targeted \\
  3 & sage\_{}spec\_{}class & Source classification determined in this paper \\
  4--5 & ra\_{}spec, dec\_{}spec & Position of the extracted spectrum \\
  6 & aor & \spitzer\ Astronomical Observation Request \\ &&(AOR) number\\
  7 & pid & \spitzer\ observing program identification number \\
  8 & pi & Last name of the PI of the \spitzer\ PID\\
  9 & cassis\_{}file\_{}name & Name of the file containing the \emph{CASSIS}-reduced\\ && \spitzer -IRS spectrum \\
  10 & irac\_{}des & SAGE-SMC or \sssmc\ IRAC point source designation matching\\ && the extracted spectrum\\
  11--12 & ra\_{}ph, dec\_{}ph & RA and Dec in degrees of the IRAC point source\\
  13 & dpos\_{}ph & Distance in arcsec between the IRAC point source\\ && and the position of the extracted spectrum\\
  14--17 & irac1, irac2, irac3, irac4 & IRAC magnitudes in bands 1--4\\
  18--20 & tycho\_{}des, b\_{}tycho, v\_{}tycho & TYCHO counterpart and its $B$ and $V$ magnitudes\\
  21--25 & m2002\_{}des, u\_{}m2002, b\_{}m2002, v\_{}m2002, r\_{}m2002 & \citet{Massey_02_UBVR} counterpart and its $U$, $B$, $V$, $R$\\ && magnitudes \\
  26--29 & u\_{}mcps, b\_{}mcps, v\_{}mcps, i\_{}mcps & Matching MCPS $U$, $B$, $V$, $I$ magnitudes\\
  30--33 & denis\_{}des, i\_{}denis, j\_{}denis, k\_{}denis & DENIS counterpart and its $I$, $J$, \kband\ magnitudes\\
  34--37 & irsf\_{}des, j\_{}irsf, h\_{}irsf, k\_{}irsf& IRSF counterpart and its $J$, $H$, \kband\ magnitudes\\
  38--41 & tmass\_{}des, j\_{}tmass, h\_{}tmass, k\_{}tmass& 2MASS 6X counterpart and its $J$, $H$, \kband\ magnitudes\\
  42--46 & wise\_{}des, wise1, wise2, wise3, wise4& \wise\ counterpart and its magnitudes in the four\\ && \wise\ bands\\
  47--52 & akari\_{}n3, akari\_{}n4, akari\_{}s7, akari\_{}s11, akari\_{}l15, akari\_{}l22 & Matching \akari\ magnitudes in bands N3, N4, S7,\\ && S11, L15 and L22 from \citet{Ita_10_AKARI_SMC}\\
  53--54 & mips24\_{}des, mips24& SAGE-SMC MIPS-[24] designation and magnitude\\
  55--56 & mips70\_{}des, mips70& SAGE-SMC MIPS-[70] designation and magnitude\\
  57--58 & mips160\_{}des, mips160& SAGE-SMC MIPS-[160] designation and magnitude\\
  59 & mbol\_{}phot &\mbol\ calculated by interpolation of \jhk, IRAC and\\ && MIPS-[24] photometry, with a Wien and\\ && Rayleigh-Jeans tail\\
  60 & mbol\_{}phwi &as \#59, but with \wise\ photometry added\\
  61 & mbol\_{}phsp &as \#59, but with the IRS spectrum added\\
  62 & mbol\_{}pwsp &as \#59, but with the IRS spectrum and \wise\ \\ && photometry added\\
  63--65 &  mbol\_{}mcd, lum\_{}mcd, teff\_{}mcd  & \mbol\ calculated using the SED fitting code from\\ && \citet{McDonald_09_omegaCen,McDonald_12_Hipparcos}; only good fits are included\\
  66--67 & id\_{}groen, per\_{}groen & Source ID and variability period in days from\\ && \citet{Groenewegen_09_luminosities}\\
  68--72 & ogle3id, ogle3mean\_{}i, ogle3mean\_{}v, ogle3amp\_{}i, ogle3period & Source ID and variability information from the\\ && OGLE survey\\
  73 & boyer\_{}class & Colour classification from \citet{Boyer_11_SMC}\\
  74 & matsuura\_{}class & Colour classification from \citet{Matsuura_13_SMC}\\
  75 & sewilo\_{}class & Colour classification from \citet{Sewilo_13_YSOs}\\
\hline
\end{tabular}
\end{table*}

\begin{table}
\caption{Classification types used in the decision tree shown in Fig.~\ref{dectree}, and counts for a total of 209 SMC point sources. The last section of the table shows a breakdown of other known types (OTHER).}
\label{classgroups}
\setlength{\tabcolsep}{10pt} 
\begin{tabular}{lll}
\hline 
Code	&  Object type				&  Count	\\
\hline 
YSO-1	&  Embedded Young Stellar Objects	&  14		\\
YSO-2	&  Young Stellar Objects		&  5		\\
YSO-3	&  Evolved Young Stellar Objects	&  22		\\
YSO-4	&  HAeBe Young Stellar Objects		&  10		\\
HII &  \hii\ regions			&  3		\\
O-EAGB  &  Early-type O-rich AGB stars          &  8 \\
O-AGB	&  Oxygen-rich AGB stars		&  11		\\
RSG	&  Red Supergiants			&  22		\\
O-PAGB	&  Oxygen-rich post-AGB stars		&  1		\\
O-PN	&  Oxygen-rich planetary nebulae	&  4		\\
C-AGB	&  Carbon-rich AGB stars		&  39		\\
C-PAGB	&  Carbon-rich post-AGB stars		&  3		\\
C-PN	&  Carbon-rich planetary nebulae	&  20		\\
STAR	&  Stellar photospheres			&  4		\\
        &  Dusty OB stars  & 23 \\
{RCRB}	&  {R CrB stars}		&  3		\\
BSG	&  Blue Supergiant	&  1		\\
WR	&  Wolf-Rayet stars			&  10		\\
OTHER	&  B[e] stars			& 2  		\\
	&  Foreground stars & 2					\\
	&  S stars				&  1		\\
        &  symbiotic stars                      & 1 \\
\hline 
\end{tabular}
\end{table}

Preferring the IRAC coordinates over the spectrum coordinates, we then
matched the IRAC point sources to a number of other infrared and
optical photometric surveys. We obtained MIPS-[24], [70] and [160]
matches, within a search radius of 3\arcsec, 9\arcsec, and 20\arcsec,
respectively, from the SAGE-SMC survey \citep{Gordon_11_SAGE-SMC},
corresponding to half a resolution element in these bands. We also
searched the Wide-Field Infrared Survey Explorer (\wise ) All-Sky
Source Catalog for matches within 3\arcsec. We also searched for
\akari\ matches in the N3, N4, S7, S11, L15, and S22 bands within
3\arcsec\ using the catalogue provided by \citet{Ita_10_AKARI_SMC}.
In the near-infrared, the Two Micron All Sky Survey (2MASS) Long
Exposure (6X) survey was searched for matches within 2\arcsec\ of the
IRAC positions (SAGE-SMC matches with 2MASS), and we also used this
search radius with the InfraRed Survey Facility (IRSF) catalogue
\citep{Kato_07_IRSF}.  The Deep Near Infrared Survey (DENIS) of the
Southern Sky catalogue \citep[3rd release;][]{Epchtein_99_DENIS} was
also searched with a 2\arcsec\ search radius.  In the optical, many of
our sources have matches in the Magellanic Clouds Photometric Survey
\citep[MCPS;][]{Zaritsky_02_MCPS-SMC} and the catalogue published by
\citet{Massey_02_UBVR}. In both catalogues we looked for matches
within 1.5\arcsec\ of the IRAC position. Some of the objects in our
sample are actually too bright for those two optical surveys, and a
search of the TYCHO catalogue with a radius of 3\arcsec\ filled in
some of these gaps.  All tabulated photometry is available from the
online database (see Table~\ref{tab:metatable}). We only provide the
magnitudes for the purpose of evaluating the shape of the SED. Further
information, including the photometric uncertainties, can be found in
the respective source tables using the designations provided, {as well
as Appendix~\ref{appendixa}.} 

\subsection{Bolometric magnitudes, variability and colour classifications}

For each source bolometric magnitudes were calculated via a simple
trapezoidal integration of the SED, to which a Wien tail was fitted to
the short-wavelength data, and a Rayleigh-Jeans tail was fitted to the
long-wavelength data. The following SED combinations were calculated:
\begin{enumerate}
\item MCPS or \citet{Massey_02_UBVR} optical photometry; \jhk\ photometry; and IRAC and MIPS-[24] photometry, all as available (\emph{mbol\_{}phot} in
  Table~\ref{tab:metatable});
\item like (i) but combined with the \wise\ photometry
  (\emph{mbol\_{}phwi});
\item like (i) but  combined with the IRS
  spectrum (\emph{mbol\_{}phsp});
\item like (i) but combined with both the \wise\ photometry and the
  IRS spectrum (\emph{mbol\_{}pwsp});
\end{enumerate}

For sources where there is little reprocessing of the optical
emission, i.e.~little infrared excess, bolometric magnitudes were
calculated using an SED fitting code
\citep{McDonald_09_omegaCen,McDonald_12_Hipparcos}. This code performs
a $\chi^2$-minimisation between the observed SED (corrected for
interstellar reddening) and a grid of {\sc bt-settl} stellar
atmosphere models \citep{Allard_11_browndwarfs}, which are scaled in
flux to derive a bolometric luminosity. This SED fitter only works
effectively where a Rayleigh-Jeans tail is a good description of the 3
to 8\,\um\ region, and provides a better fit to the optical and
near-IR photometry than a Planck function. For the most enshrouded
stars, fitting the SED with `naked' stellar photosphere models leads
to an underestimation of the temperature and luminosity, due to
circumstellar reddening, and hence the integration method (above) for
calculating \mbol\ is preferred for very dusty sources. Experience
shows that good fits can be separated from bad fits in the NIR: if the
model and observations differ at $I$, $J$, $H$ or \kband\ by more than
a magnitude in any band, the fit is considered bad. This retains the
cases where the difference between model and observations in the MIR
or FIR is large, but often in these cases the excess emission is
unrelated to the point sources. In cases where it is related to the
point source, making the source very red, the values calculated by
trapezoidal integration provide a better estimate of \mbol.  \teff,
\mbol\ and $L$ for the good fits are included in the online table as
\emph{teff\_{}mcd}, \emph{mbol\_{}mcd} and \emph{lum\_{}mcd},
respectively (see Table~\ref{tab:metatable}).

The sample was then matched to the Optical Gravitational Lensing
Experiment (OGLE-III) catalogue of long-period variables in the SMC
\citep{Soszynski_11_OGLE-III} and \citet{Groenewegen_09_luminosities}
to obtain variability periods, and the variability information is
included in the on-line table (see Table~\ref{tab:metatable}).

Finally, we included a number of colour classification schemes for
comparison.  First, \citet{Boyer_11_SMC} have extended the
classification scheme developed by \citet{Cioni_06_AGBselection} to
classify dusty mass-losing evolved stars into subcategories, using
IRAC, MIPS and NIR colours. We checked our source list against their
catalogue for matches. Their classifications (O-AGB, C-AGB, x-AGB,
aO-AGB, RSG, RGB and FIR) are included in the on-line table as
\emph{boyer\_{}class} (see Table~\ref{tab:metatable}). Definitions of
these classes can be found in \citet{Boyer_11_SMC}. Furthermore, we
also applied the colour classification scheme proposed by
\citet{Matsuura_13_SMC} to the sources in our list. This
classification scheme is also designed to distinguish between various
kind of very red objects, to estimate the dust production rate. We
applied the cuts described by \citet[Fig.~4, 5]{Matsuura_13_SMC} on
our sample and list the classifications that follow from these cuts
(O-AGB, C-AGB, RSG) in our online table, as \emph{matsuura\_{}class}
(see Table~\ref{tab:metatable}).  The last colour classification
scheme we apply is the one proposed by \citet{Sewilo_13_YSOs} for
YSOs, who applied classification cuts in the five different infrared
CMDs, followed by visual inspection of images and SED fitting to
select YSO candidates from the SAGE-SMC survey. We checked our source
list against their catalogue and identified `high-reliability' and
`probable' YSO candidates accordingly (\emph{sewilo\_{}class};
Table~\ref{tab:metatable}).

\section{The Classification Method}\label{method}

\begin{figure*}
\center
\includegraphics[width=150mm]{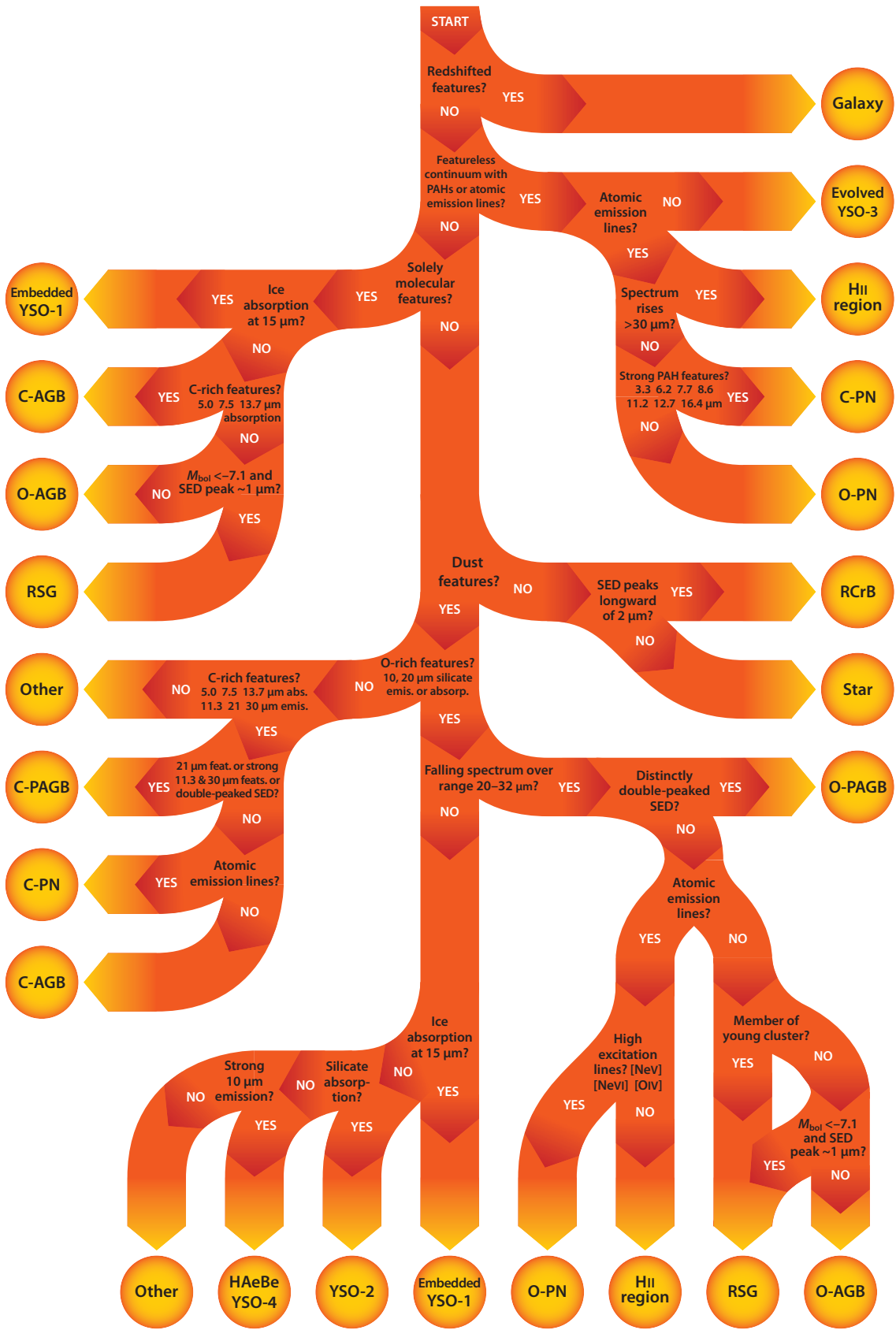} 
\caption{The logical steps of the classification decision tree, where
  \spitzer\ IRS spectra ($\lambda$ = 5.2--38\,\um ), associated
  optical, NIR, \wise, IRAC and MIPS photometry, luminosity,
  variability, age and other information are used to classify SMC
  infrared point sources. See Table~\ref{classgroups} for key to
  classification group codes. Figure restyled in appearance from the
  one published by \citet{Woods_11_classification}.}
\label{dectree}
\end{figure*}

To classify our sample of 209 SMC point sources for which IRS staring
mode data exist, we follow the method described by
\citet{Woods_11_classification}. Fig.~\ref{dectree} shows a restyled
version of the classification decision tree. We made enhancements to
the tree, which will be discussed in this section.

A literature search was performed for each object to retrieve other
information useful in the process of classification, including (but
not limited to) determination of stellar type, luminosity, age of
nascent cluster of stars (if the object was found to be a member of a
cluster), \ha\ detections, etc. This information was used in addition
to the spectroscopic data, the photometric matches and derived
bolometric luminosity, and the variability data, described in
Sec.~\ref{dataprep}, to classify the sources. Any existing
classification from the literature was used as a starting point before
our spectral classification.  Appendix \ref{appendixa} provides a
brief summary of the literature survey for each object.

As in \citet{Woods_11_classification}, we adopt the following
categories for our point source classification. Low- and
intermediate-mass ($M < 8$ \msun ) post-main-sequence stars are
classified by chemistry (O- or C-rich) and by evolutionary stage
(asymptotic giant branch, post-asymptotic giant branch and planetary
nebula), hence our groupings O-AGB, O-PAGB, OPN, C-AGB, C-PAGB,
C-PN. We propose an enhancement of the classification tree by
\citet{Woods_11_classification} to include early-type O-rich AGB
stars, namely O-EAGB. These stars do not show any evidence for
dust {features in their} infrared spectra, but {they do} show long period variability {in OGLE and MACHO and some evidence for continuum infrared excess.} {Although these stars are in the early stages of AGB evolution,
they are most likely more evolved than genuine early-AGB (E-AGB) stars, 
which have not yet started helium shell burning. E-AGB stars do not normally show long period variability.}
We assume that {O-EAGB stars} are {thermally-pulsing}
AGB-type objects prior to the onset of, {or just beginning}, significant dust formation.
More massive and luminous red supergiants have a class of their own,
RSG. Young stellar objects can be classified phenomenologically into
four groups, YSO-1, YSO-2, YSO-3, YSO-4 (see
\citealt{Woods_11_classification} for the definition of these
classes). Stars showing a stellar photosphere, but no additional dust
or gas features, or long-period variability, are classified as
STAR. One observational program focussed on stars showing a
far-infrared excess in their MIPS photometry
\citep{Sheets_13_dustyOB,Adams_13_dustyOB}, due to the illumination
and heating of interstellar dust (the Pleiades effect); these objects
are classified as a subcategory of STAR: dusty OB stars. The tree also
distinguishes galaxies (GAL; even though none are actually found in
the present work) and \hii\ regions (HII), and we furthermore have a
class for {R CrB stars (labeled RCRB in the classification table)}. Finally the classification OTHER exists for
objects of \emph{known type} which do not belong in the categories
above and do not follow from the classification tree (e.g.~B[e]
stars). The nature of these objects are usually identified by other
means, and as such reported in the astronomical literature.

\subsection{Classification Process}\label{classtool}

Each source was classified independently by at least
three of the co-authors. In cases where the classification was
unanimous, it was simply adopted as the final classification, whereas
in those cases where some discrepancies occurred, we asked additional
co-authors to classify the sources, to settle the issue. In some
cases, a discussion on the nature of the source ensued. We aimed to
reach consensus among the co-authors on the nature of a source in
cases where differences in classification occurred.

The lead author of this work (Paul Ruffle) developed an internal web
browser-based classification tool, to facilitate the classification
process. The decision tree was built into the tool as a series of
questions, and the collected data were available in tabulated
form. For each source the classification tool provided a slit image and
plots of its spectrum, SED, log SED and bolometric magnitudes (see
Fig.~\ref{example_plots} for examples). Each author was free to use either
the decision tree logic or their own method of classification.  
There was also room for the co-authors to add additional
comments to the table.\footnote{Although we were able to use this tool to its
conclusion and generate the final classification table prior to Paul
Ruffle's unexpected death, the tool was intended to be available online
indefintely, for application on other data collections. Unfortunately,
the internet service provider where this website was hosted recently stopped
providing this service and as of yet, the co-authors have not been 
able to reconstruct and resurrect the website with the classification tool.}

\section{Sage-Spec Point Source Classification}\label{class}

Table~\ref{classgroups} lists the classification types, used in the
decision tree shown in Fig.~\ref{dectree}, and counts for a total of
209 SMC point sources, for which \spitzer -IRS staring mode
observations are available. The classifications are also shown
overplotted on the map of the SMC (Fig.~\ref{smcskymap}).

\begin{figure}
\includegraphics[width=8cm]{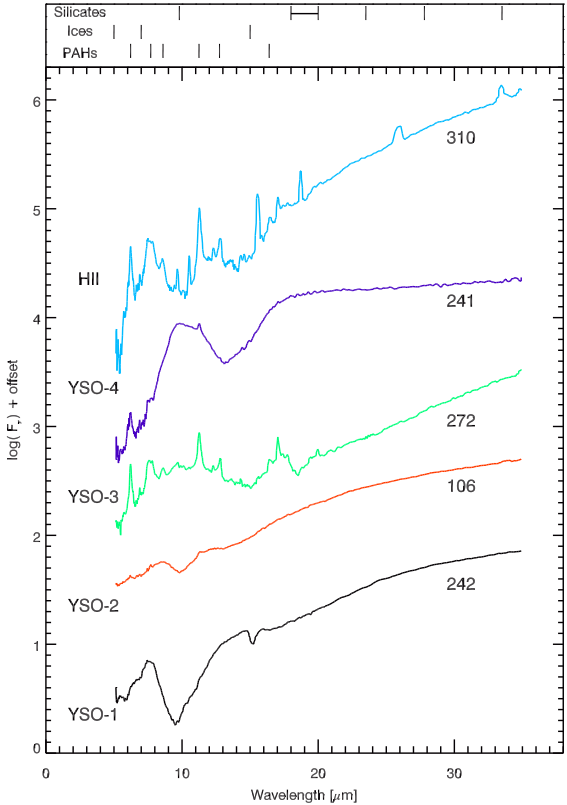}
\caption{Example spectra in the YSO and \hii\ categories. From bottom to top, 
we show examples of YSO-1 through 4, and a spectrum of an \hii\ region, in 
an evolutionary sequence from young to more evolved. The 
spectra are labelled with their SMC IRS number. Discernible spectral features
are indicated with tick marks and labels at the top of the diagram. The 
heavily embedded YSO-1 spectrum shows silicates and ices in absorption. The
YSO-2 spectrum shows silicate in self-absorption. Silicate emission and PAH 
features are visible in the spectrum of the YSO-3, YSO-4 and \hii\ objects,
albeit in different ratios. The \hii\ region has a rising SED slope. }
\label{fig:YSOspec}
\end{figure}

Fig.~\ref{fig:YSOspec} shows typical spectra of objects classified as
one of the four YSO type objects, as well as a typical IRS spectrum of
an \hii\ region. The numbering 1--4 for the YSO classes represent an
evolutionary sequence, with YSO-1 being the most embedded and YSO-4
the most evolved {type of YSO, namely Herbig Ae Be (HAeBe) stars.} This is evident from the spectral appearance of the
silicate feature, which appears in absorption towards YSO-1 objects,
then gradually in self-absorption (YSO-2), until it finally appears in
emission (YSO-3, YSO-4), for less embedded objects. The spectra of the
YSO-1 objects also show ice absorption features, for instance the
CO$_2$ ice feature at 15.2 \um, further evidence of their early
evolutionary phase. Polycyclic aromatic hydrocarbons (PAHs) are seen
in the spectra of the YSO-3, YSO-4 and \hii\ classes, indicative of a
UV radiation field. Atomic lines are also seen, particularly in YSO-3
and \hii\ objects, and the latter category shows a rising continuum
indicative of cold dust in the vicinity of the ionizing star.

\begin{figure}
\includegraphics[width=8cm]{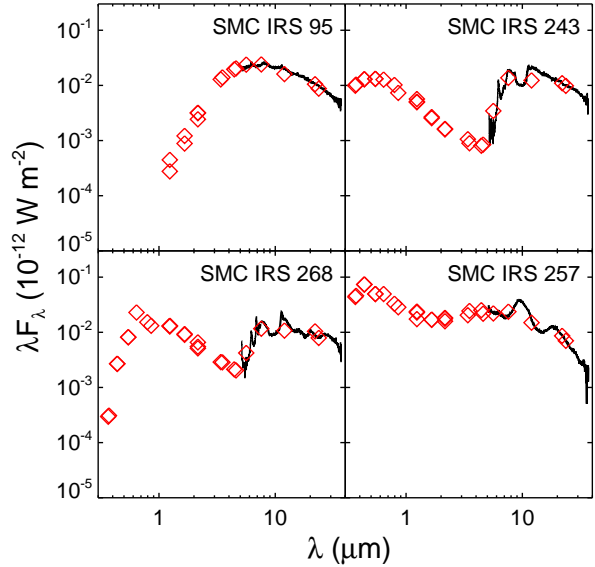}
\caption{{SEDs of the four post-AGB stars in the sample. The SMC IRS spectra
95, 243 and 268 correspond to C-PAGB objects, while SMC IRS 257 is the 
spectrum of a O-PAGB object. The black lines represent the IRS spectra, while
the red diamonds correspond to the collected photometric measurements for each
source. The double-peaked structure used as a distinguishing feature is visible in the SEDs of 243, 268 and 257. 2MASS
J00364631$-$7331351 (SMC IRS 95) does not show a double peaked structure, but
its C-PAGB nature is confirmed using near-infrared spectroscopy and the PAH
features in the IRS spectra (See Appendix~\ref{appendixa}).}}
\label{fig:pagb}
\end{figure}

\begin{figure}
\includegraphics[width=8cm]{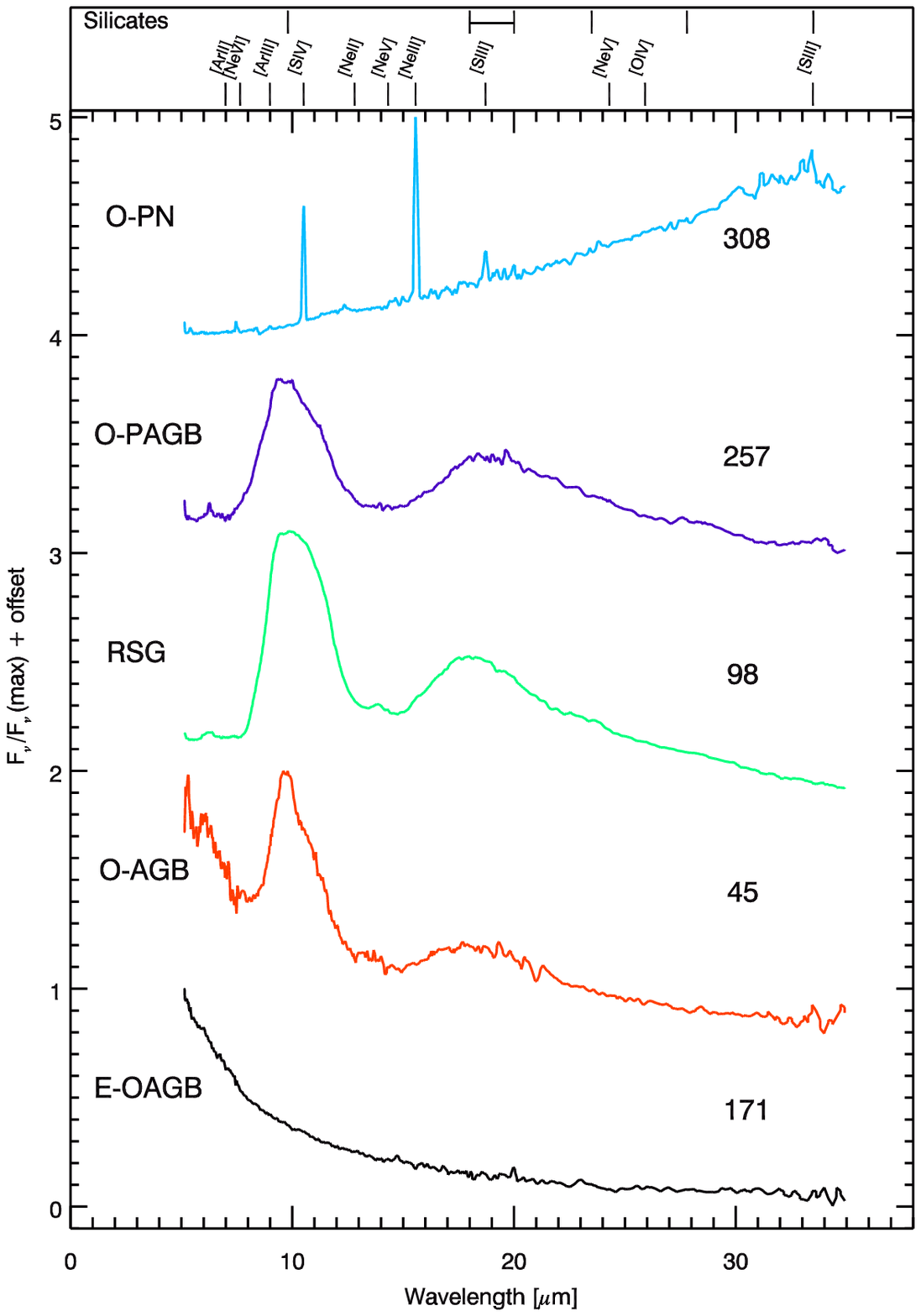}
\caption{Example spectra of different types of O-rich evolved
  stars. The spectra are labelled with their SMC IRS number. At the
  top the wavelengths of the silicate features and spectral lines are
  labelled. At the bottom of the plot the spectrum of the E-OAGB star
  closely resembles a stellar photosphere. The spectra of the O-AGB,
  RSG and O-PAGB star all show the silicate emission
  features. Distinguishing between these three types is not possible
  from IRS spectroscopy alone, and additional information on SED shape
  and bolometric luminosity is required.  The top spectrum is
  characeteristic of a O-PN. }
\label{fig:Ospec}
\end{figure}

Fig.~\ref{fig:Ospec} shows typical spectra in the group of oxygen-rich
evolved stars. The earliest type of O-rich AGB stars are shown at the
bottom of the plot (O-EAGB), and the spectra shown here do not show
any dust features, while the signature of oxygen-rich molecular
species may be present in the spectra. A slight change of slope due to
a small infrared excess caused by thermal dust emission, may be
visible in the SED, however.  Later type O-AGB stars, red supergiants
(RSG) and oxygen-rich post-AGB stars (O-PAGB), share spectroscopic
characteristics, such as the presence of silicate emission features,
although the detailed shape can be different. To distinguish between
the RSG and O-AGB category, a bolometric luminosity cut of \mbol
$=-7.1$ is used, while the distinction between O-AGB and O-PAGB is
based on the presence of a detached shell where a double-peaked SED is
used as a criterion for the latter category. {This is demonstrated 
in the lower right panel
of Fig.~\ref{fig:pagb} showing the SED of the sole O-PAGB object
in the sample, LHA 115-S 38 (SMC IRS 257).} O-rich PNe (O-PN) may
still show a distinguishable oxygen-rich chemistry in their dust
mineralogy (although not in the case shown), but they are
discriminated from C-PN using the presence of PAH lines in the spectra
of the carbon-rich objects.

\begin{figure}
\includegraphics[width=8cm]{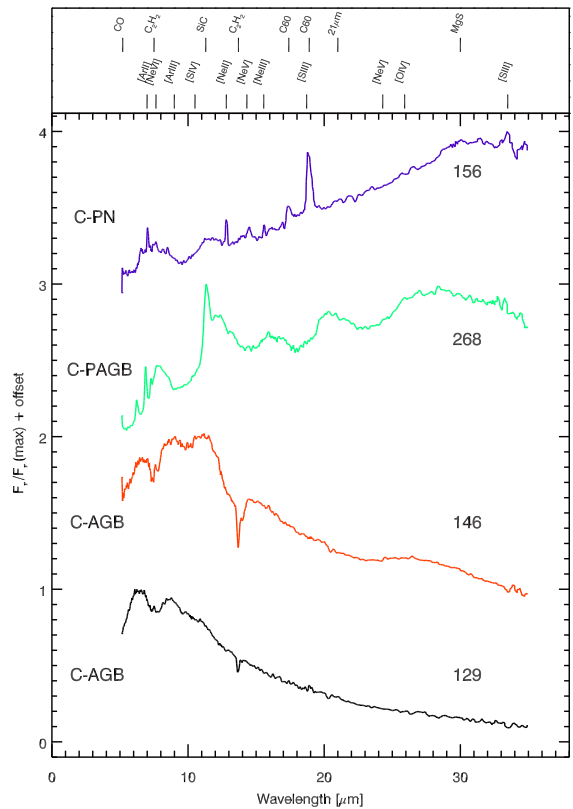}
\caption{Example spectra of C-rich evolved stars. From bottom to top
two C-AGB stars are shown, followed by a C-PAGB and a C-PN. All four spectra
are labelled with their SMC IRS number. The tick marks at the top show
the position of characteristic spectral features. In the C-AGB spectra
the molecular absorption bands of C$_2$H$_2$ are visible, as well as
emission due to SiC and possibly MgS (30\um, not in all sources). The C-PAGB
object no longer shows the molecular absoprtion bands, but the SiC and
the MgS (not always) are still visible. Other features include PAH bands
and the 21-\um\ feature. The infrared spectrum of C-PN objects shows
the spectral features due to PAHs and atomic lines.}
\label{fig:Cspec}
\end{figure}

Fig.~\ref{fig:Cspec} gives an overview of the 5--38\,\um\ spectral
appearance of carbon-rich evolved stars. Tracers of the carbon-rich
chemistry are the C$_2$H$_2$ molecular absorption bands at 5.0, 7.5
and 13.7\,\um, the SiC dust feature at 11.3\,\um, the 21-\um\ feature
(which remains unidentified and really peaks at 20.1\,\um), and the
so-called 30-\um\ feature, which has recently been suggested to be due
to the same carbonaceous compound that carries the continuum
\citep{Otsuka_14_C60}, while the MgS identification is under debate
\citep{Zhang_09_30mic,Lombaert_12_LLPeg}. Again, the distinction
between the C-PAGB and C-AGB categories is based on whether the SED is
double-peaked, which is evidence for a detached shell, indicating mass
loss has stopped. {Fig.~\ref{fig:pagb} shows two clear examples of this,
namely SMC IRS 243 and SMC IRS 268. SMC IRS 95 is confirmed to be a 
C-PAGB star \citep{Kraemer_06_post-AGB,vanLoon_08_dustproduction}, despite its absence of a double-peaked SED.} C-PAGB and C-PN also show the UV-excited PAH
features, and in case of the C-PN, the presence of atomic emission
lines.

\begin{figure}
\includegraphics[width=8cm]{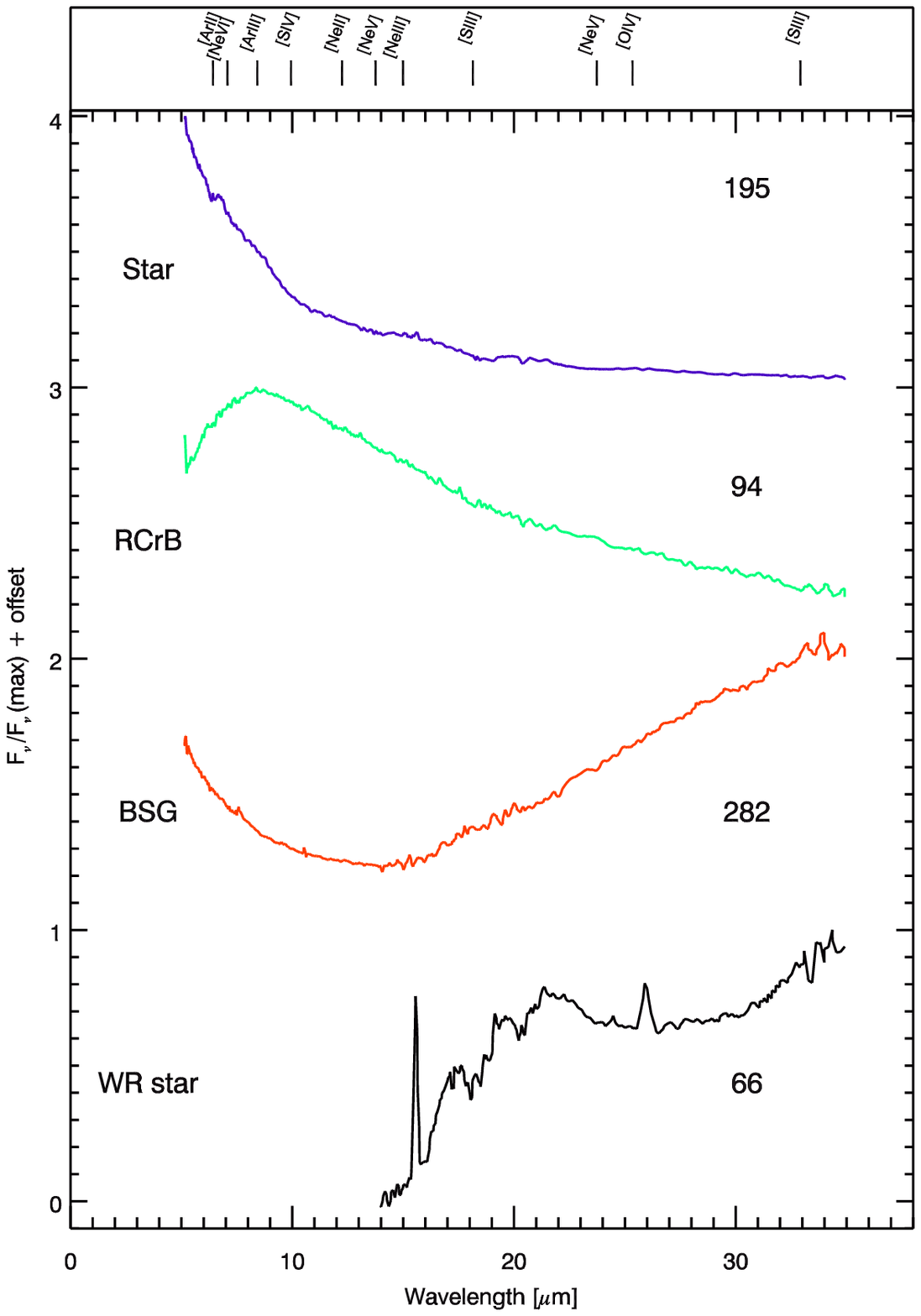}
\caption{Example spectra of objects with stellar photospheres or
  similar.  Shown here from bottom to top are the spectra of a WR
  star, a BSG, a R CrB star and a regular stellar photosphere, labelled
  with their SMC IRS number. The WR star shows some atomic lines in
  its spectrum (marked at the top of the diagram), but the other
  spectra are rather featureless. The infrared excess in the R CrB and
  BSG objects are caused by dust emission. }
\label{fig:starspec}
\end{figure}

\begin{figure}
\includegraphics[width=8cm]{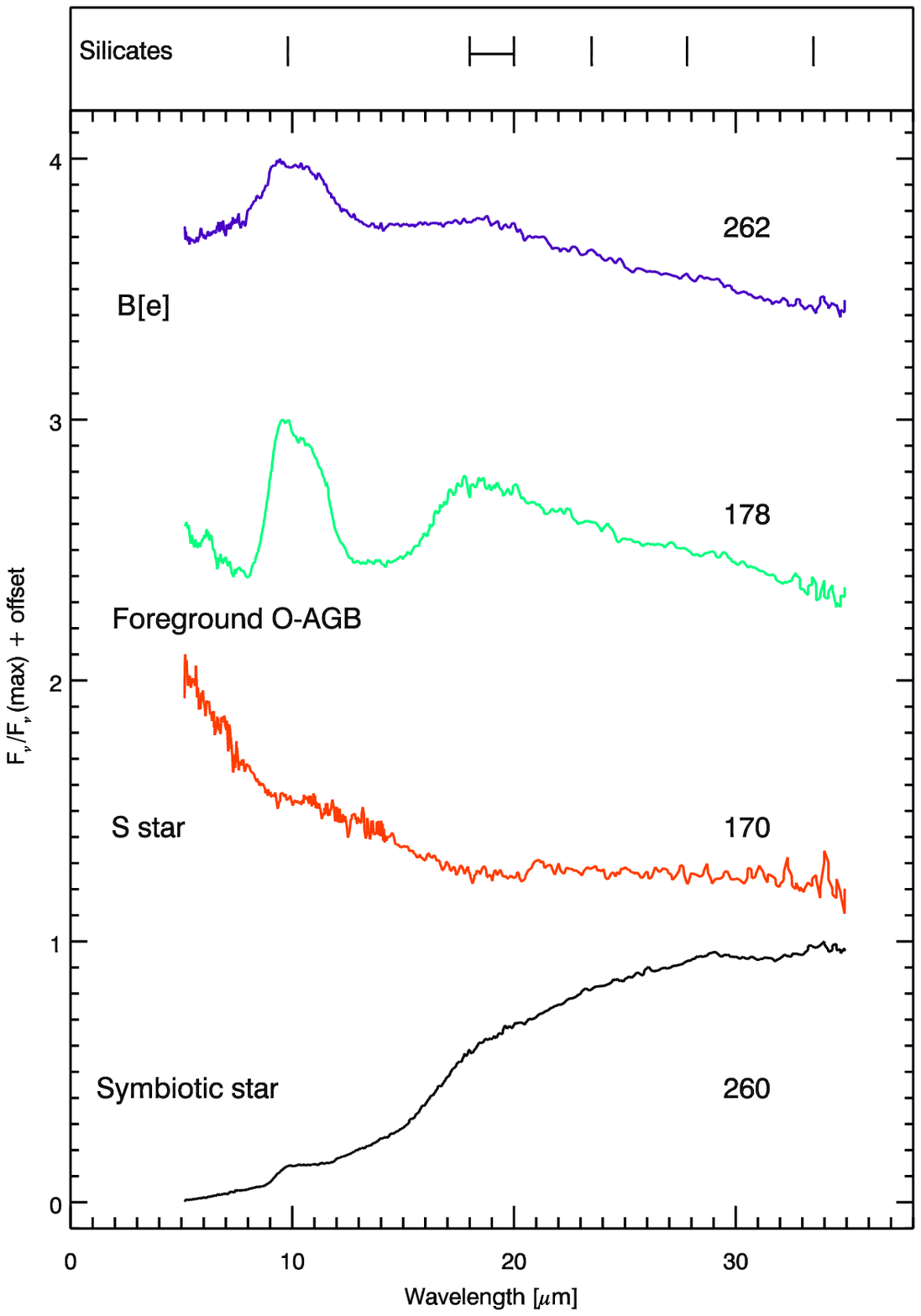}
\caption{Example IRS spectra of objects in the OTHER category. From
  bottom to top we show the spectra of a symbiotic star, an S star, a
  foreground (Galactic) O-AGB star and a B[e] star, all labelled with
  their SMC IRS number. Relevant spectral features due to silicates
  are labelled at the top of the figure.}
\label{fig:otherspec}
\end{figure}

Fig.~\ref{fig:starspec} shows typical spectra of a number of star-like
categories, namely (from top to bottom) stellar photospheres, with no
discernible dust features in the spectrum; R CrB stars, which only
appear to show a dust continuum with no spectral substructure; a Blue
Supergiant in our sample, which appears to have a strong far-infrared
excess on top of a stellar photosphere, and finally Wolf-Rayet stars,
which may form dust and show the corresponding infrared excess in the
spectrum. The Blue Supergiant and the Wolf-Rayet star classifications
are taken from the literature, and are not based on the infrared
spectroscopy.  Similarly, Fig.~\ref{fig:otherspec} shows the spectra
of the object types group under OTHER, of which the classifications
are taken from the literature (see Appendix~\ref{appendixa}). The
examples shown in Fig.~\ref{fig:otherspec} represent from top to
bottom a B[e] star, a foreground AGB star, an S star and a symbiotic
star. Their infrared spectra are not used to achieve this
classification, and the IRS data are plotted only for illustration.

\begin{figure*}
\includegraphics[width=170mm]{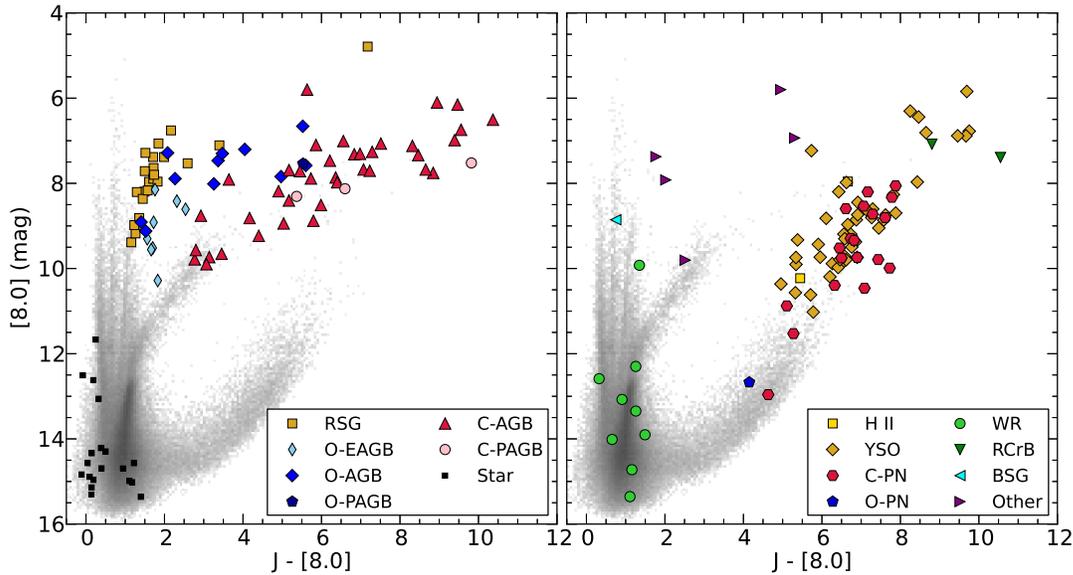}
\caption{[8.0] vs $J-[8.0]$ CMD. The gray scale in the background represents
a density plot of the SAGE-SMC point source catalogue, while the coloured
symbols are the sources classified in this work. For clarity, the plot is
duplicated with the coloured symbols spread out over the left- and right-hand
plot according to the legend.}
\label{fig:8vsJ-8}
\end{figure*}

\begin{figure*}
\includegraphics[width=170mm]{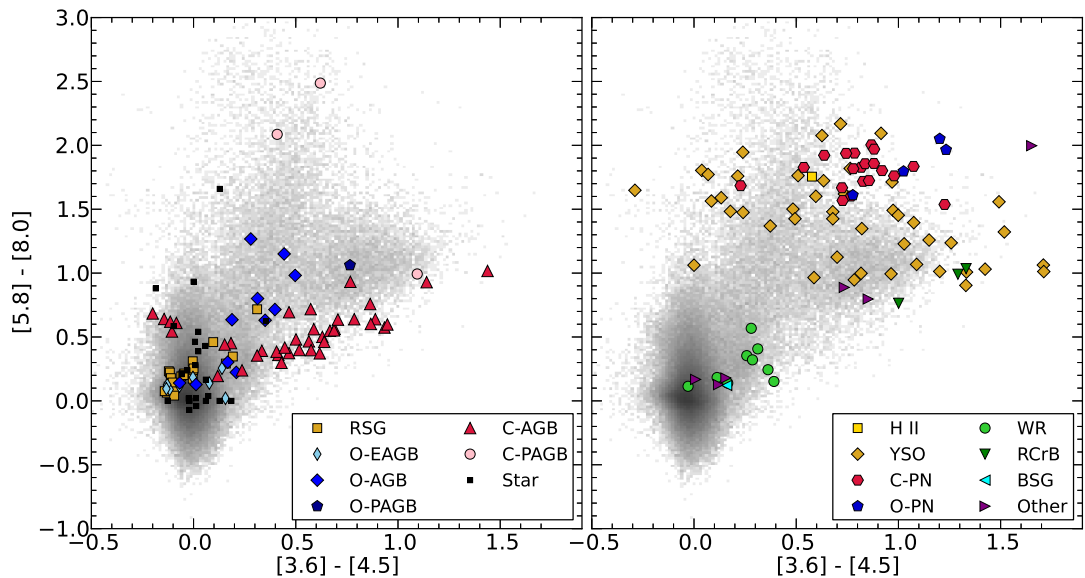}
\caption{[5.8]$-$[8.0] vs [3.6]$-$[4.5] all-IRAC CCD. Description
as in Fig.~\ref{fig:8vsJ-8}.}
\label{fig:iracccd}
\end{figure*}

\begin{figure*}
\includegraphics[width=170mm]{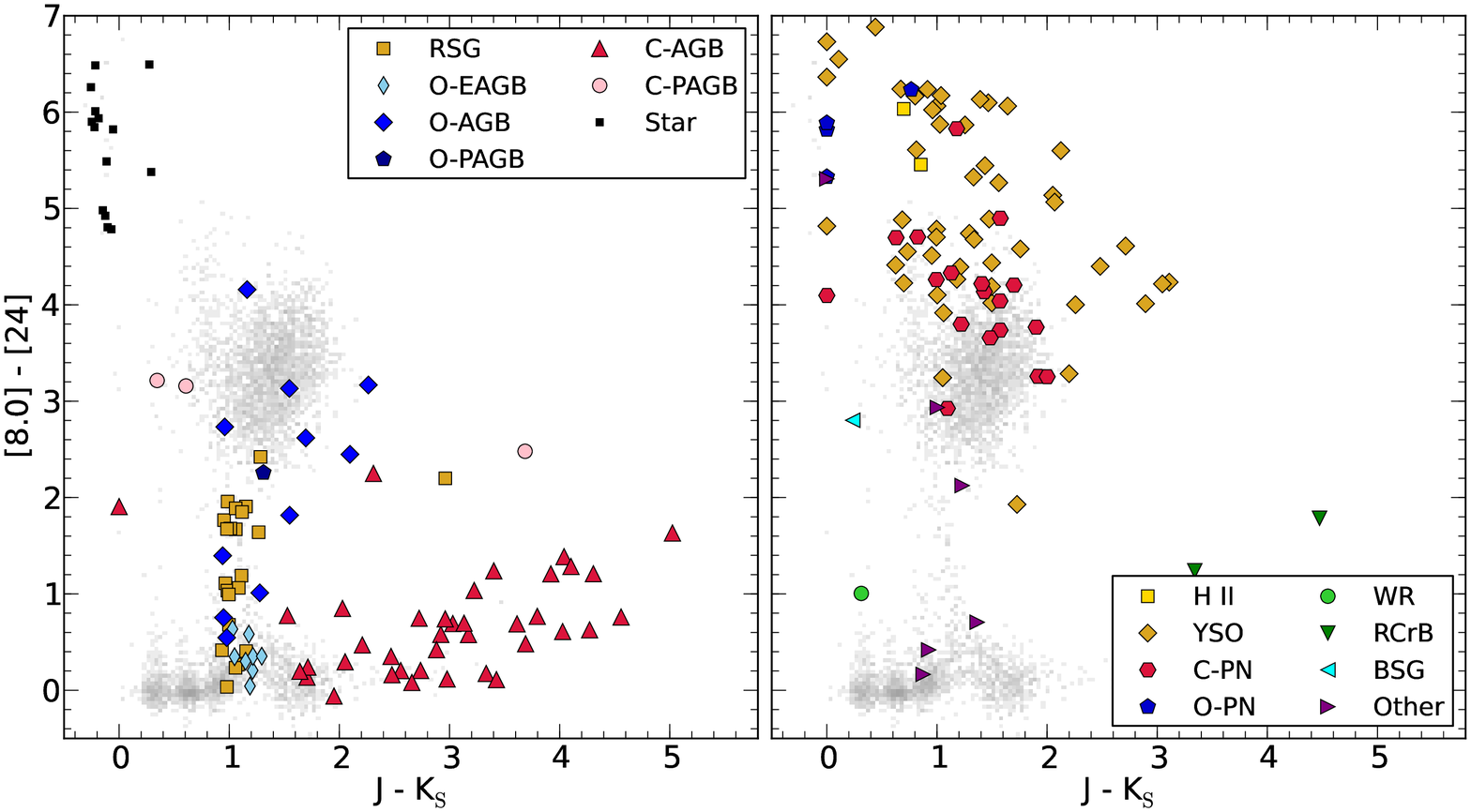}
\caption{[8.0]$-$[24] vs \jmink\ CCD. Description
as in Fig.~\ref{fig:8vsJ-8}.}
\label{fig:mipsccd}
\end{figure*}

Figs.~\ref{fig:8vsJ-8}--\ref{fig:mipsccd} show the 209 classified
point sources on the [8.0] versus $J-[8.0]$ CMD and two different
colour-colour diagrams (CCDs), overlayed on the SAGE-SMC point source
catalogue \citep{Gordon_11_SAGE-SMC} in gray scale.

The [8.0] versus $J-[8.0]$ CMD (Fig.~\ref{fig:8vsJ-8}) shows a large
spread for the population of 209 objects. The stellar atmospheres
(STAR) and Wolf-Rayet stars have colours more or less
indistinguishable from the bulk of the SAGE-SMC catalogue (with
$J-[8.0] \sim$ 0--1\,mag), and modest brightness at 8.0\,\um, even
though these stars are amongst the brightest stars in the optical in
the SMC.  All other categories displayed are bright in the IRAC [8.0]
band, and often show considerable redness in their $J-[8.0]$
colour. In this diagram RSG, O-AGB/O-EAGB, C-AGB and YSOs (all classes
combined), are reasonably well separated from each other, although
there are some interlopers. It appears to be difficult to separate PNe
and YSOs on the one hand, and the most extreme C-AGB stars and YSOs on
the other hand. Distinguishing between the four YSO classes is also
not possible in this diagram.

Fig.~\ref{fig:iracccd} is a CCD composed of the four IRAC bands,
namely the [3.6]$-$[4.5] versus [5.8]$-$[8.0] colours. The advantage
of using a CCD is that it is distance independent.  The coverage of
the sample of 209 objects fans out nicely over colour-colour space. In
this diagram, the stars and WR stars no longer separate well from the
rest of the sample, however it now appears easier to separate YSOs
from C-AGB stars on the one hand and PNe on the other hand. However,
C-PAGB and O-PAGB stars probably overlap with the colour-colour space
taken up by YSOs, as they transition from the AGB region to the PN
region in the diagram. But as this phase is short-lived, pollution of
the colour-selected YSO sample with post-AGB stars is limited.
Subdivision within the YSO category is still not possible, and also
the O-(E)AGB objects do not seem to occupy a unique part of
colour-colour space.

Finally, Fig.~\ref{fig:mipsccd} shows the \jmink\ versus [8.0]$-$[24]
CCD.  In this CCD, the C-AGB objects are easily separated from all
other types of objects, as their \jmink\ colour quickly increases,
with increasing [8.0]$-$[24].  For O-(E)AGB stars and RSGs this
increase is less steep, forming a separate branch in the middle of the
plot. O-PN and C-PN group together with YSOs towards the top of the
diagram, showing the highest values of [8.0]$-$[24], against modest
\jmink\ reddening.

\section{Comparison with existing colour classifications}\label{pops}

These 209 spectral classifications will allow us to verify existing
infrared photometric classification schemes that have come out of
recent studies of the Magellanic Clouds. We compare our results with
three distinct colour classification schemes: a) the \jhk\ colour
classification scheme for evolved stars by
\citet{Cioni_06_AGBselection}, {expanded} by
\citet{Boyer_11_SMC} {to include mid-IR wavelengths}; b) the IRAC classification scheme for AGB stars
and RSGs by \citet{Matsuura_13_SMC}, which is based on the previous
work by \citet{Matsuura_09_dustbudget} on the LMC; and c) the
\spitzer\ classification scheme to select YSO candidates by
\citet{Sewilo_13_YSOs}, based on earlier work by
\citet{Whitney_08_YSO}.

\subsection{\citet{Boyer_11_SMC}}\label{boyer_class}

\begin{figure*}
\includegraphics[width=170mm]{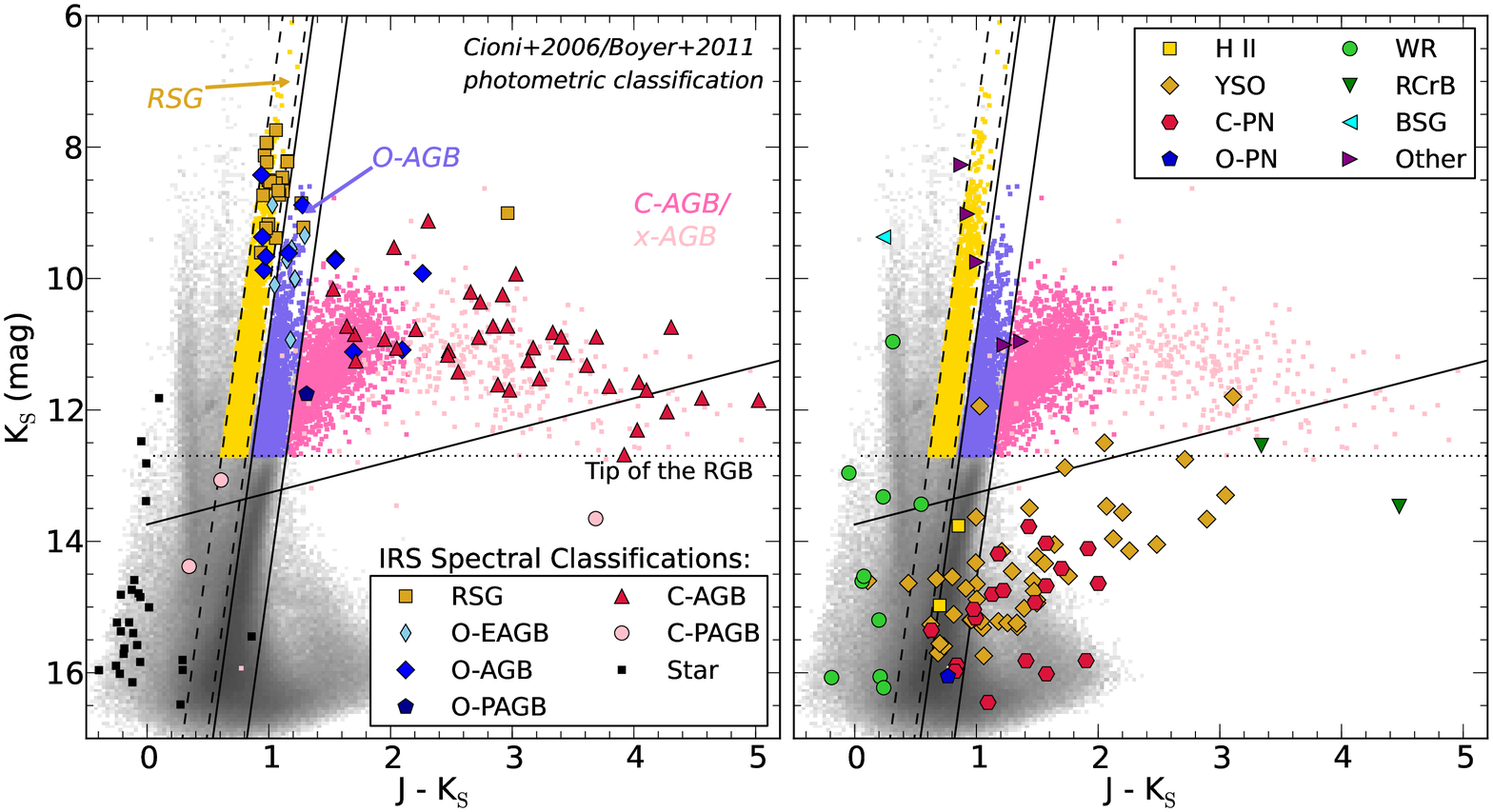}
\caption{The \citet{Boyer_11_SMC} photometric classification, based on
  the work by \citet{Cioni_06_AGBselection}, applied to the SMC. {\citet{Boyer_11_SMC} begin with the \jmink\ classification shown here, then use 3.6 and 8-\micron\ photometry to recover and classify the dustiest sources.} The
  slanted lines in this \kband\ vs \jmink\ CMD represent the
  boundaries of the C-AGB, O-AGB and RSG categories, and the
  horizontal dotted line shows the boundary between the tip of the RGB
  and the earlier type AGB stars. In gray scale the SAGE-SMC point
  source catalogue is shown in the form of a density plot. The lilac,
  yellow and dark pink dots represent the RSG, O-AGB and C-AGB
  objects, as they are classified in the \kband\ vs \jmink\ CMD. The
  light pink dots are the x-AGB stars, selected by
  \citet{Boyer_11_SMC}, replacing any \kband\ vs \jmink\
  classification. The x-AGB objects fall mostly within the C-AGB
  boundaries. The coloured symbols represent the objects spectrally
  classified in this work, following the legends and spread out over
  two panels for clarity.}
\label{cioniboyer}
\end{figure*}

In order to classify all evolved stars in the SAGE-SMC
\citep{Gordon_11_SAGE-SMC} data, \citet{Boyer_11_SMC} devised a
classification scheme, based on the 2MASS classification scheme
presented by \citet{Cioni_06_AGBselection}. The basis for this
classification scheme is the $K$ versus \jmink\ CMD
(Fig.~\ref{cioniboyer}), showing the cuts for the RSG, O-AGB and C-AGB
object classes, superposed on the SAGE-SMC point sources (grey
pixels). The dotted line refers to the tip of the Red Giant Branch
(RGB), and \citet{Boyer_11_SMC} use this line to separate RGB stars
from the AGB and RSG categories.  Photometrically classified objects
from \citet{Boyer_11_SMC} are shown as coloured pixels.  At the red
end of the diagram, some photometrically classified C-AGB stars appear
below the diagonal cut, or even below the dotted line corresponding to
the tip of the RGB. These objects are called \emph{extreme} AGB stars,
predominantly carbon-rich, which are defined as $J-[3.6] > 3.1$ mag
\citep[e.g.,][]{Blum_06_evolved}. Often these objects are so red, that
they are not detected in 2MASS \jhk, and alternative selection
criteria in the mid-infrared are required. The reader is referred to
\citet{Boyer_11_SMC} for a detailed description.  The larger symbols
in Fig.~\ref{cioniboyer} represent the sample of \spitzer -IRS
spectroscopically classified objects described in this work.

\subsubsection{C-AGB stars}

Objects showing the C-AGB spectral signature are all classified as
either C-AGB or x-AGB ({most of} which are expected to be C-rich), according to
the \citet{Boyer_11_SMC} classification, although three of these
objects are not included in the catalogue published by
\citet{Boyer_11_SMC}. These objects (OGLE SMC-LPV-7488 (SMC IRS 44;
SSTISAGEMA J004903.78$-$730520.1); 2MASS J01060330$-$7222322 (SMC IRS
109; SSTISAGEMA J010603.27$-$722232.1); and 2MASS J00561639$-$7216413
(SMC IRS 129; SSTISAGEMA J005616.36$-$721641.3)) are among a larger
set of objects that had not been properly matched between the IRAC Epochs
in the mosaicked catalogue (Srinivasan
et al.~\emph{in prep.}). These objects were thus
missing photometric measurements in the point source catalogue, in some bands,
 and were therefore not classified (correctly)
by \citet{Boyer_11_SMC}.  
The on-line table described by Table~\ref{tab:metatable}
shows the correct photometry for these three targets.

\subsubsection{Red Supergiants}

Most of the 21 spectroscopically classified RSGs indeed fall within
the RSG strip defined by
\citet{Boyer_11_SMC}. Only three of the objects classified as an RSG
by us were {classified differently} by \citet{Boyer_11_SMC} using only the
photometry: HV 11417 (SMC IRS 115) is designated as a FIR object by
\citet{Boyer_11_SMC}; Massey SMC 55188 (SMC IRS 232) as an O-AGB star;
and IRAS F00483$-$7347 (SMC IRS 98) as an x-AGB star.  HV 11417 is a
variable star with a period of 1092 days, and an amplitude in the I
band of 1.9 mag \citep{Soszynski_11_OGLE-III}. Due to this large
amplitude, the timing of the observations affects the IR colours of
the object significantly, which is why the photometry measurements
show a small positive slope between 8 and 24\,\um\ ($[8.0]-[24] >
2.39$\,mag), causing it to be classified as a FIR object by
\citet{Boyer_11_SMC}, while the slope of the IRS spectrum is
distinctly negative. If we disregard the FIR colour cut, which 
is meant to separate AGB stars from objects such as YSOs, PNe and
background galaxies, this object
would have been classified as an O-AGB star, which deviates from our
determination of RSG, and from past classifications
\citep{Elias_80_HV11417}. The bolometric luminosity of this object is
close to the RSG/O-AGB boundary, and could be affected by the
variability of the object too. The large amplitude favours a classification as luminous AGB star over a RSG. {RSGs were separated from O-AGB stars using the classical AGB bolometric magnitude limit (Fig.~\ref{dectree}), but this boundary is not absolute. AGB stars undergoing hot bottom burning can be brighter than this limit, while less-evolved RSGs can be fainter. This causes some disagreement in classifications of stars near the boundary.}

Alternatively, the luminosity boundary
between RSG and O-AGB may not be properly represented by the cuts in
the \kband\ vs \jmink\ diagram from \citet{Boyer_11_SMC}.
Similary, the misclassification of Massey SMC 55188 also suggests that
the bolometric cut we applied to distinguish between O-AGB stars and
RSGs does not correspond to the boundaries between these two
categories in the \kband\ vs \jmink\ CMD. IRAS F00483$-$7347 clearly shows an
oxygen-rich chemistry in its spectrum, with the presence of the
amorphous silicate bands at 9.7 and 18\,\um. The object is heavily
embedded, with a very red SED, and the 9.7-\um\ feature starts to
show signs of self-absorption. IRAS F00483$-$7347 demonstrates that
not all x-AGB objects are in fact carbon-rich AGB stars.

\subsubsection{O-AGB stars}
\label{sec:boyer-o-agb}

All objects classified as O-EAGB in this work are {similarly} classified
as O-AGB by \citet{Boyer_11_SMC}.  The spectrally-confirmed O-AGB
objects, however, show a much wider spread in colour-magnitude space
than the defined strip, and they encroach on the
photometrically-defined RSG, C-AGB and FIR categories. In particular,
only two spectroscopically classified O-AGB stars, HV 12149 (SMC IRS
45) and 2MASS J00444463$-$7314076 (SMC IRS 259) are
classified {as O-AGB based on} photometry. One object (HV 1375; SMC IRS 110)
is classified as a C-AGB based on its photometric colours, and a
further six objects (RAW 631, 2MASS J00463159$-$7328464, 2MASS
J00445256$-$7318258, BMB-B 75, IRAS F01066$-$7332 and HV 12956) are
apparently sufficiently red ($[8.0]-[24]>2.39$ mag) to be classified
as FIR, even though they are not background galaxies, YSOs or
PNe. These six objects are heavily embedded, and show the effect of a
high optical depth in the relative strength of the 18 \um\ silicate
band with respect to the 9.7-\um\ band. They also show evidence for
the presence of crystalline silicates in their spectrum, another sign
of high optical depth \citep{Kemper_01_xsilvsmdot,Jones_12_xsils}.
Finally, HV 2232 (SMC IRS 230) and HV 11464 (SMC IRS 309) are
photometrically classified as Red Supergiants \citep{Boyer_11_SMC}
(see the table described by
Table~\ref{tab:metatable}).

\subsubsection{Additional {sources}}

The further nine interlopers in the C-AGB/x-AGB section of the CMD
defined by \citet{Boyer_11_SMC} are a mixed bag of objects, all but
one falling in the x-AGB category. These objects reflect selection
biases and include rare categories of previously known types such as R CrB stars (2MASS J00461632$-$7411135, 2MASS J00571814$-$7242352 and
OGLE SMC-SC10 107856), an S star (BFM 1) and a B[e] star (Lin
250). These object types are not included in the \citet{Boyer_11_SMC}
classification scheme, and could therefore never have agreed with the spectral classification.
True misclassifications are the objects thought to be C-rich AGB
stars, only to be revealed to be something else based on their IRS
spectroscopy. These include O-PAGB star LHA 115-S 38, YSO-2 object
2MASS J01050732$-$7159427, and RSG IRAS F00483$-$7347. NGC 346:KWBBe
200 is a special case, as it was thought to be a B[e] supergiant
\citep{Wisniewski_07_discovery}, but its classification has recently
been revised to a YSO \citep{Whelan_13_N66}, in line with our
classification of YSO-3 (See Appendix~\ref{appendixa}).

The FIR category defined by \citet{Boyer_11_SMC} was introduced to
exclude YSOs, compact \hii\ regions, PNe and background galaxies from
the AGB/RSG sample, by applying the $[8]-[24]>2.39$ mag cut, corresponding
to a rising continuum. Indeed, 14 out of 23 FIR objects are either
C-PN (three objects) or YSOs (11 objects), according to their IRS
spectra.  The remaining nine FIR objects include the six O-AGB stars
discussed in Sec.~\ref{sec:boyer-o-agb}, but also a symbiotic star
(SSTISAGEMA J005419.21$-$722909.7; SMC IRS 260) and a C-PAGB (2MASS
J00364631$-$7331351; SMC IRS 95), both of which are classified based 
on their IRS spectra.

\subsection{\citet{Matsuura_13_SMC}}

\begin{figure*}
\includegraphics[width=170mm]{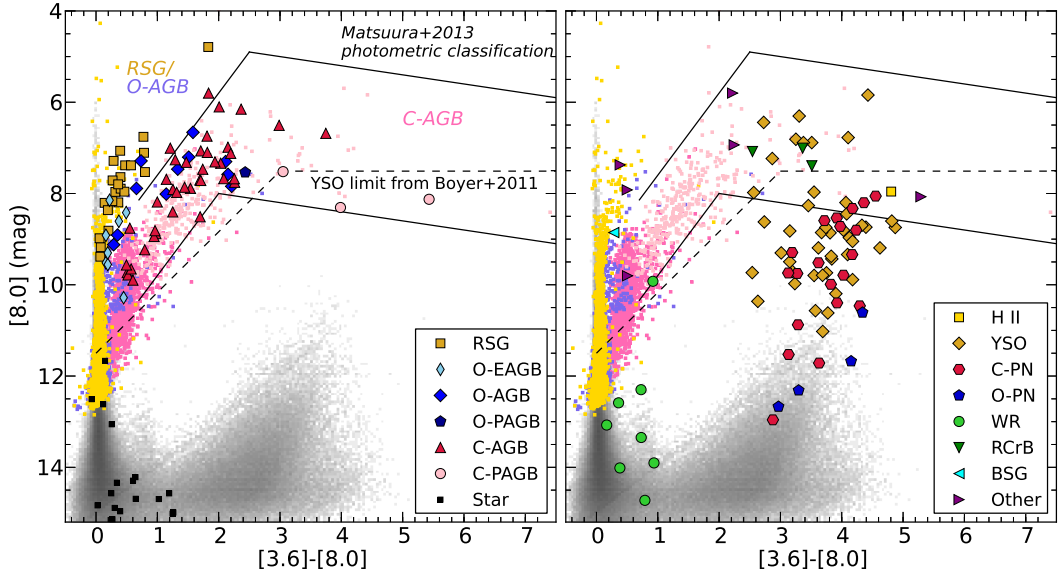}
\caption{First step of the \citet{Matsuura_13_SMC} photometric
  classification method, based on the work by
  \citet{Matsuura_09_dustbudget}, applied to the SMC.  The solid lines
  in this [8.0] vs [3.6]-[8.0] CMD represent the boundaries of the
  C-AGB category, and provide a lower boundary to the RSG/O-AGB
  category. The dashed line shows the position of the YSO limit used
  by \citet{Boyer_11_SMC}. All colours have the same meaning as in 
  Fig.~\ref{cioniboyer}.}
\label{matsuura1}
\end{figure*}

\begin{figure*}
\includegraphics[width=170mm]{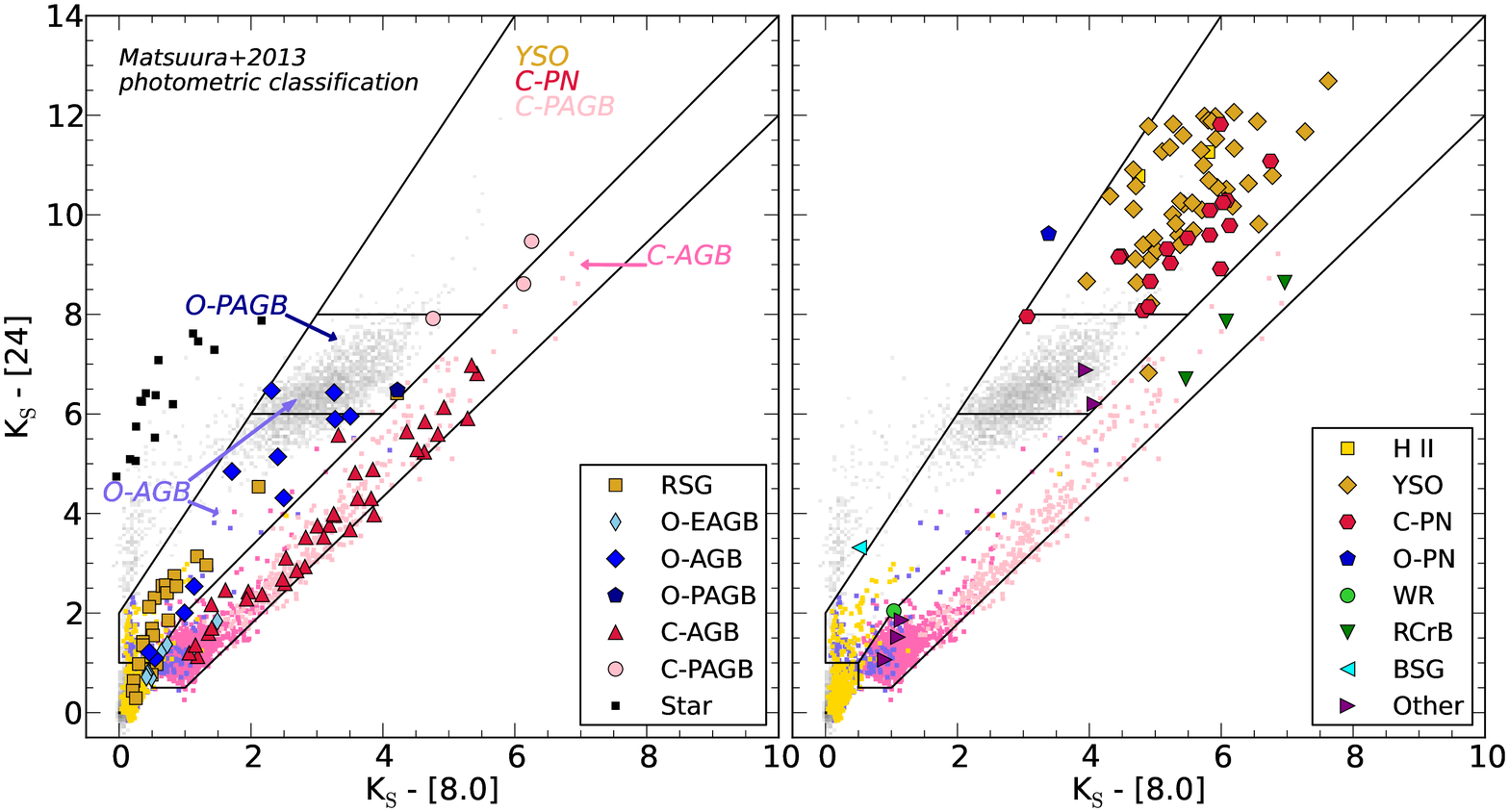}
\caption{Second step of the \citet{Matsuura_13_SMC} photometric
  classification method.  The solid lines in this \kband--[24] vs \kband--[8.0]
  CCD represent the boundaries of the various
  evolved star categories.  All colours have the same meaning as
in Fig.~\ref{cioniboyer}.}
\label{matsuura2}
\end{figure*}

The second classification scheme was proposed by
\citet{Matsuura_09_dustbudget} to identify mass-losing O-rich and
C-rich AGB stars, as well as Red Supergiants, in order to estimate the
dust production budget in the LMC, separated out by C-rich and O-rich
chemistries. The scheme is based on the IRAC bands, using the [8.0] vs
[3.6]$-$[8.0] CMD, and separates out the O-rich AGB
stars and RSGs on the one hand, and C-rich AGB stars on the other hand
(Fig.~\ref{matsuura1}). \citet{Matsuura_09_dustbudget} used only this diagram
to classify the IRAC point sources in the LMC, but for the SMC,
\citet{Matsuura_13_SMC} added an extra step to overcome pollution
between the categories for redder [3.6]$-$[8.0] sources. Indeed, from
Fig.~\ref{matsuura1} it is clear that the red part of the C-AGB
section is dominated by sources identified as YSOs (tan diamonds),
while a significant fraction of the O-AGB objects (blue diamonds) also
fall within the C-AGB section.  In the second step, MIPS and NIR data
are included, and a complex series of cuts is made in the \kband--[24] vs
\kband--[8.0] CCD  \citep[see Fig.~5 in][; the
equations are not provided]{Matsuura_13_SMC}.  In cases where the
classification using the \kband--[24] vs \kband--[8.0] diagram
(Fig.~\ref{matsuura2}) deviates from the [8.0] vs [3.6]$-$[8.0]
classification, the \kband--[24] vs \kband--[8.0] classification takes
preference. Most {spectral classifications agree with the photometric classification} in the
second step (see Fig.~\ref{matsuura2}). In case where sources were
only classified in one of the two steps, we used that classification.
We have executed this classification method for our 209 sources, and
included the results in the table described by Table~\ref{tab:metatable}.

\subsubsection{Carbon-rich AGB stars}
\label{sec:matsuuraCAGB}

The two-step classification described by \citet{Matsuura_13_SMC}
correctly identifies most of the spectroscopically classified C-AGB
stars in our sample as such. In only three cases was an object
photometrically classified as being oxygen-rich (RSG/O-AGB), while the
spectroscopy shows the carbon-rich nature of the source. These sources
are 2MASS J00515018$-$7250496 (SMC IRS 103); 2MASS J00524017$-$7247276
(SMC IRS 127) and IRAS 00350$-$7436 (SMC IRS 238). IRAS 00350$-$7436
is among the brightest mid-infrared objects in the SMC, and falls above the
C-AGB cut in Fig.~4 of \citet{Matsuura_13_SMC}, due to its exceptional 
brightness.

Furthermore, the classification {of several C-AGB stars} by \citet{Matsuura_13_SMC}
{disagrees with the spectral classification.}
These include: a C-rich post-AGB star (2MASS
J00364631$-$7331351; SMC IRS 95), four O-rich early-type AGB stars (HV
1366, HV 11303, HV 838 and HV 12122; SMC IRS 36, 38, 39 and 116,
respectively), a WR star (HD 5980; SMC IRS 281), a RSG (HV 11262; SMC
IRS 111), a YSO-3 object (NGC 346:KWBBe 200; SMC IRS 19), the three
R CrB stars (2MASS J00461632$-$7411135, 2MASS J00571814$-$7242352 and
OGLE SMC-SC10 107856; SMC IRS 94, 114 and 245, respectively), the two
foreground O-rich AGB stars NGC 362 SAW V16 and HV 206, the S star BFM
1 and the symbiotic star SSTISAGEMA J005419.21$-$722909.7. The last
four objects are all in the OTHER category, which contains subclasses
the work by \citet{Matsuura_13_SMC} does not seek to classify. The four
O-rich early-type objects have actually long been recognized as such,
in some cases their nature was already known in the 1980s, and {they fall only slightly outside}
the boundaries in Fig.~5 of
\citet{Matsuura_13_SMC}. The
RSG HV 11262 has very similar \kband--[8.0] and \kband--[24] colours to the four
O-EAGB objects.  2MASS J00364631$-$7331351 is not properly filtered
out by Fig.~5, as it falls just below the \kband--[24] = 8 mag cutoff that is
meant to exclude YSO, C-PN, and C-PAGB objects from the C-AGB
category.  The R CrB stars represent a rare
class which, not surprisingly, overlaps in infrared colours with the
C-AGB stars, as it is believed that the dust in these stars is
carbonaceous \citep{Feast_86_RCrB}.  HD 5980 is the only WR star that
is classified by \citet{Matsuura_13_SMC}.  Although other WR stars
do appear in Fig.~\ref{matsuura1}, they do not receive a
classification as they are too faint in the [8.0] band. HD 5980 is
considerably brighter in the [8.0], and is the only WR star with a
MIPS-[24] detection in the SMC \citep{Bonanos_10_SMC}.  Closer
inspection of the IRS spectrum seems to suggest that a chance
superposition with a compact \hii\ region gives rise to this 24\,\um\
detection, and is actually not a detection of the WR star itself.  Thus,
the position of HD 5980 in Fig.~\ref{matsuura2} and the classification
following from it, should be disregarded.  Finally, NGC 346:KWBBe 200
is a curious object, which, from its IRS spectrum appears to be a YSO,
but has characteristics in common with B[e] stars (See
Appendix~\ref{appendixa}). It falls well outside the box defined for
YSO, C-PN and C-PAGB, perhaps due to its unusual nature.

\subsubsection{Oxygen-rich AGB stars and RSGs}

Since the main purpose of the classification scheme by
\citet{Matsuura_13_SMC} is to determine the dust production by evolved
stars distinguished by carbon-rich and oxygen-rich chemistry, the
subdivision in types of evolved stars within the oxygen-rich class is
less important. In fact, in the first step, all types of oxygen-rich
evolved stars (AGB stars and RSGs) are lumped together as RSG/O-AGB
(Fig.~\ref{matsuura1}), while in the second step a subdivision is made
between O-AGB (which also includes RSG) and O-AGB/O-PAGB
(Fig.~\ref{matsuura2}).  The latter category contains the more evolved
O-AGB stars. Thus, a total of three partially overlapping categories
exist. The RSG/O-AGB category contains objects that are classified as
such in the first step, but remained unclassified in the second
step. This class contains only three objects (2MASS
J00515018$-$7250496, 2MASS J00524017$-$7247276 and IRAS 00350$-$7436),
which are all C-AGB according to their IRS spectroscopy, and are
already discussed in Sec.~\ref{sec:matsuuraCAGB}.

The 24 objects classified as O-AGB in the second step are indeed all
O-AGB or RSG objects according to their IRS spectroscopy. However, the
category O-PAGB/O-AGB contains a more diverse range of objects. Of the
eight SMC IRS objects classified by the method of
\citet{Matsuura_13_SMC} to be in this category only four are genuine
O-rich evolved stars: IRAS F00483$-$7347 (SMC IRS 98) is a RSG
according to our classification, LHA 115-S 38 (SMC IRS 257) is found
to be a O-PAGB object, and 2MASS J00463159$-$7328464 (SMC IRS 121) and
HV 12956 (SMC IRS 277) are O-AGB stars.  The other objects in this
category include two carbon-rich evolved stars, 2MASS
J00444111$-$7321361 (SMC IRS 268), a C-PAGB object, and Lin 343 (SMC
IRS 161), a C-PN. Both of these objects should have fallen in the
YSO/C-PN/C-PAGB, but with \kband--[24] just below 8\,mag they both just miss the
cutoff. Since Fig.~\ref{matsuura2} does not show any O-rich objects in
the vicinity of the K--[24] = 8\,mag cutoff between C-rich and O-rich post-AGB
objects, a case can be made to lower the cutoff slightly.  Finally,
the remaining two misclassified objects in the O-PAGB/O-AGB category
by \citet{Matsuura_13_SMC} are spectrally classified as B[e] stars in
the OTHER category. They are RMC 50 (SMC IRS 193) and Lin 250 (SMC IRS
262).

When checking the reverse direction, we find that all objects
classified by us as O-AGB are indeed identified as either
O-AGB or as O-PAGB/O-AGB according to the classification scheme by
\citet{Matsuura_13_SMC}. However, this classification scheme
misclassifies all objects identified as O-EAGB by us. According to the
scheme by \citet{Matsuura_13_SMC}, they are either C-AGB (See
Sect.~\ref{sec:matsuuraCAGB}; four out of eight objects), or remain
unclassified, as they have $[3.6]-[8.0] < 0.7$\,mag, rendering them
unclassifiable in Fig.~\ref{matsuura1}, and have \kband $-[8.0] <0.5$\,mag and
\kband $-[24]<0.5$\,mag, which causes them to fall outside all boundaries in
Fig.~\ref{matsuura2}.  This also happens to six of the 22 objects
classified by us to be RSGs; one is misclassified as a C-AGB (See
Sect.~\ref{sec:matsuuraCAGB}), and the other five do not receive a
classification because their colours are too blue (i.e.~their mass-loss
rates are too low).

Our sample of 209 targets with IRS spectra contains only one oxygen-rich
post-AGB star, which agrees with the O-PAGB/O-AGB classification from \citet{Matsuura_13_SMC}.

\subsection{\citet{Sewilo_13_YSOs}}
\label{sec:YSOclass}

\begin{figure*}
\includegraphics[width=170mm]{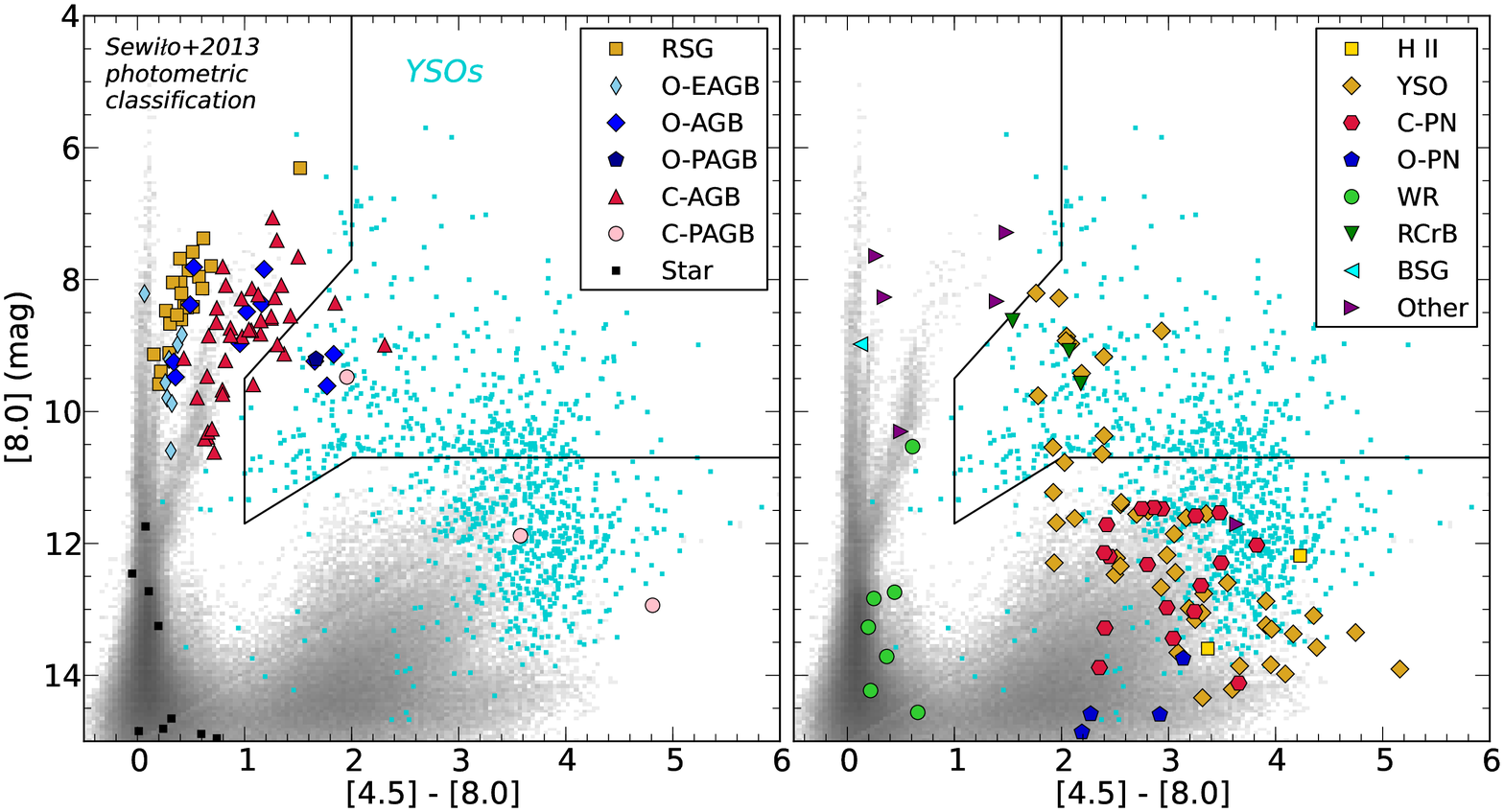}
\caption{One of the CMD diagrams used in the \citet{Sewilo_13_YSOs}
  selection method for candidate YSOs, based on the work by
  \citet{Whitney_08_YSO}.  This [8.0] vs [4.5]-[8.0] CMD diagram shows
  the cut used in solid black lines.  In gray scale the SAGE-SMC point
  source catalogue is shown in the form of a density plot. The light
  blue dots represent the YSO candidates finally selected by
  \citet{Sewilo_13_YSOs}. The coloured symbols show the location of
  the objects spectrally classified in this work, spread out over two
  panels for clarity.  \citep{Sewilo_13_YSOs}, based on
  \citep{Whitney_08_YSO}}
\label{sewilo1}
\end{figure*}

\begin{figure*}
\includegraphics[width=170mm]{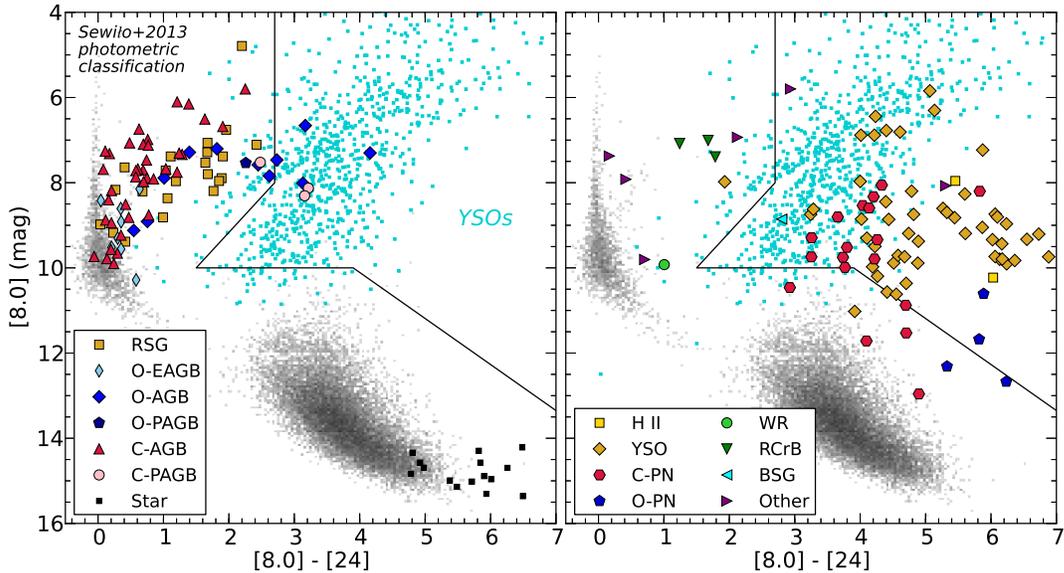}
\caption{As Fig.~\ref{sewilo1}, but now for [8.0] vs [8.0]-[24].}
\label{sewilo2}
\end{figure*}

The third classification scheme that we compare with is the method
developed by \citet{Whitney_08_YSO} and refined by
\citet{Sewilo_13_YSOs} to select YSO candidates. From the set of CMDs
used by \citet{Whitney_08_YSO}, \citet{Sewilo_13_YSOs} selected a combination of
five different CMDs to select YSO candidates in
the SMC. Two of these diagrams are reproduced in this work, with the
209 IRS point sources overplotted (Figs.~\ref{sewilo1} and
\ref{sewilo2}). After the initial colour selection of YSO candidates,
\citet{Sewilo_13_YSOs} performed additional tests, including a visual
inspection of the imaging, a check against the SIMBAD and other
catalogues for known non-YSO sources, and fitting against the YSO SED
grid calculated by \citet{Robitaille_06_grid}. \citet{Sewilo_13_YSOs}
arrive at a list of approximately 1000 `high-reliability' and
`probable' YSOs in the SMC.

Although Figs.~\ref{sewilo1} and \ref{sewilo2} appear to show
considerable amount of pollution from non-YSO IRS staring mode
targets, as well as many confirmed YSOs based on the IRS data straying
out of the defined boxes, it is the combination of the five
CMDs, along with the additional non-colour based
checks that yields a highly reliable YSO candidate list.  Indeed,
the \citet{Sewilo_13_YSOs} classification favours reliability over completeness,
and CMD areas with significant pollution due to other sources (background
galaxies, PNe)  have been excluded.  
The list of
209 IRS staring mode point sources contains nine objects classified as
probable YSOs by \citet{Sewilo_13_YSOs}, and 45 {high-reliability} YSO candidates.  Of
the high-reliability YSOs for {which} IRS observations are available, we find that the vast
majority are indeed YSOs (covering all four classes). Only two objects
turn out to be something different: \hii\ region IRAS 00436$-$7321
(SMC IRS 310) and C-PAGB object 2MASS J01054645$-$7147053 (SMC IRS
243). Among the probable YSO candidates, the success rate is lower: four out
of nine are not YSOs, upon inspection of their IRS spectra. RMC 50
(SMC IRS 193) is a B[e] star, while SMP SMC 11 (SMC IRS 32), LHA 115-N
43 (SMC IRS 155) and Lin 49 (SMC IRS 292) are actually C-PN objects.

Furthermore, there are three YSO-3 type objects, as spectrally
classified, that are not identified as YSO candidates by
\citet{Sewilo_13_YSOs}.  These are NGC 346:KWBBe 200 (SMC IRS 19),
2MASS J00465185$-$7315248 (SMC IRS 269) and LHA 115-N 8 (SMC IRS 274).

\section{Summary}
\label{sum}

We have analysed all 311 \spitzer -IRS staring mode observations
within the SAGE-SMC  IRAC and MIPS coverage
of the SMC \citep{Gordon_11_SAGE-SMC}.  After removing IRS observations of extended emission,
blank sky and duplicate observations, we find that 209 unique IRAC
point sources were targeted. We applied the infrared spectroscopic
classification method devised by \citet{Woods_11_classification}, with
the addition of one more category, namely O-EAGB stars (early-type
oxygen-rich AGB stars). We find that the \spitzer -IRS staring
mode sample of point sources in the SMC contains 51 YSOs, subdivided
in 14 embedded YSOs (YSO-1), 5 less-embedded YSOs (YSO-2), 22 evolved YSOs
(YSO-3) and 10 HAeBe type objects (YSO-4). Furthermore, we find the
sample contains 46 oxygen-rich evolved stars: 8 O-EAGB stars, 11 O-AGB
stars, 22 RSGs, 1 O-PAGB object and 4 O-PN objects. 62 objects turn
out to be carbon-rich evolved stars, namely 39 C-AGB stars, 3 C-PAGB
objects and 20 C-PN objects. The sample also includes 3 R CrB stars, 1
BSG and 10 WR stars. 27 objects show stellar photospheres, 23 of which
were selected based on their MIPS-[24] excess, and labelled by us as
dusty OB stars. It turns out that the 24-\um\ emission is in general
not related to the host OB star
\citep{Adams_13_dustyOB,Sheets_13_dustyOB}.  Finally, the sample
includes a small number of other objects (OTHER), which do not follow from the
classification method. These include 2 B[e] stars, 2 foreground
oxygen-rich AGB stars, an S star, and a symbiotic star.

In this work, we have compared the resulting spectral
classifications with the outcome of photometric classification
schemes.  It should be noted that the spectroscopic observations are obtained in
14 different observing programs, with a diverse range of science
goals. Thus, there is no homogeneous coverage of colour-magnitude
space, and it will be impossible to quantify the goodness of any given
photometric classification method.  Furthermore, the observing
programs tend to target rare types of sources in a disproportionate
amount, and some of these rare type of sources (R CrB stars, WR stars, B[e]
stars, etc.) are not included in photometric classification schemes,
precisely because they are rare. Thus, these objects tend to pollute
the classification schemes discussed here, but statistically they are
rather insignificant.

We reviewed three different photometric classification schemes for
infrared sources in the SMC: the schemes by \citet{Boyer_11_SMC} and
\citet{Matsuura_13_SMC} for evolved stars, and the classification
scheme to select candidate YSOs by \citet{Sewilo_13_YSOs}. The latter
scheme is not a pure photometric classification, as it includes
additional steps, such as visual inspection of the direct environment
of the point source in imaging, checks against existing catalogues and
fitting against the grid of YSO SEDs calculated by \citet{Robitaille_06_grid}.
However, as discussed in Sec.~\ref{sec:YSOclass}, the 54 overlapping
sources from this work with the resulting YSO candidate list are
mostly correctly classified. Only a few sources are misclassified in
either direction, i.e.~three spectroscopically confirmed YSOs were not
on the candidate list by \citet{Sewilo_13_YSOs}, and six sources on
the candidate list were found to be something else upon inspection of
their IRS spectroscopy. All-in-all we conclude that the YSO candidate
list produced by \citet{Sewilo_13_YSOs} is reliable, with 48/54
sources indeed being YSOs and the {high-reliability} sources doing better than the {probable}
sources. Only three spectroscopically confirmed YSOs were missed due
to unusual infrared colors.

The two photometric classification methods for evolved stars can be
directly compared to each other. The method by \citet{Boyer_11_SMC}
has its focus on identifying the entire dusty evolved star population,
while the \citet{Matsuura_13_SMC} method is mainly driven by the motivation
to determine the carbon-rich and oxygen-rich dust production rates. Thus,
in the later method, correct identification of lower mass-loss rate stars
is not so important. Both methods can be used to estimate the integrated
dust production rate.
 Due to the low metallicity of the SMC, carbon-rich evolved
stars are more numerous \citep{Blanco_78_carbon,Blanco_80_latetype, Lagadec_07_carbonstars}, and the {carbon stars have, on average, the highest mass-loss rates.} 
Thus, identifying
carbon stars correctly is important as they dominate the dust budget
{\citep{Matsuura_09_dustbudget,Boyer_12_metallicity,Riebel_12_dustproduction}}. Apart from rare object classes, \citet{Boyer_11_SMC}
very efficiently separate C-AGB stars from the other classes, only
classifying {one} non-C-AGB star as C-AGB
star, {while classifying all genuine C-AGB objects as C-AGB}. \citet{Matsuura_13_SMC} do slightly worse, with three genuine
C-AGB objects being classified as something else, and a number of
objects, including some rare types, incorrectly classified as C-AGB
star.

On the oxygen-rich side, we find that \citet{Matsuura_13_SMC} perform
better for the high-mass loss rate objects (O-AGB stars), but that
they perform rather poorly on the low-mass rate objects (O-EAGB
stars), while the performance of the \citet{Boyer_11_SMC}
classification method is {reversed because of overlap between high-mass AGB stars and RSGs near the classical AGB limit.}

Finally, we note that of the two steps involved in the classification
method by \citet{Matsuura_13_SMC}, the second step
(Fig.~\ref{matsuura2}) matches very well with the actual spectroscopic
classification. It has been introduced by \citet{Matsuura_13_SMC} as a
correction on the first step from \citet{Matsuura_09_dustbudget}, but
almost all photometric classifications currently included in the online table
described by Table~\ref{tab:metatable} correspond to the second step, and
most of the time match the spectroscopic classification, rendering the
first step practically {unnecessary}. Thus, in case of the
\citet{Matsuura_13_SMC} classification scheme, applying only the
second step, as demonstrated in Fig.~\ref{matsuura2}, would suffice {for dusty sources.}

\section*{Acknowledgements}
The authors wish to thank Paul Ruffle's partner Rose Wheeler. Rose
provided access to Paul's notes and files, which allowed us to finish
this work.  PMER thanks Academia Sinica Institute of Astronomy and
Astrophysics (ASIAA) for their financial support and hospitality
during the preparation of this work. The authors thank David Whelan
for making available \spitzer\ spectra of point sources described in
his 2013 paper.  Astrophysics at JBCA is supported by STFC.
F.K.~acknowledges support from the former National Science Council and
the Ministry of Science and Technology in the form of grants
NSC100-2112-M-001-023-MY3 and MOST103-2112-M-001-033-.
B.A.S.~acknowledges support from NASA ADP NNX11AB06G.
R.Sz.~acknowledges support from the Polish NCN grant
2011/01/B/ST9/02031. {The research presented here was conducted within the scope of the HECOLS International Associated Laboratory, supported in part by the Polish NCN grant
DEC-2013/08/M/ST9/00664 (E.L.; R.Sz).} This work is based (in part) on observations made
with the \spitzer\ Space Telescope, obtained from the NASA/IPAC
Infrared Science Archive, both of which are operated by the Jet
Propulsion Laboratory, California Institute of Technology under a
contract with the National Aeronautics and Space Administration.  This
publication makes use of data products from the Wide-field Infrared
Survey Explorer, which is a joint project of the University of
California, Los Angeles, and the Jet Propulsion Laboratory/California
Institute of Technology, funded by the National Aeronautics and Space
Administration.  Some of the data presented in this paper were
obtained from the Mikulski Archive for Space Telescopes (MAST). STScI
is operated by the Association of Universities for Research in
Astronomy, Inc., under NASA contract NAS5-26555. Support for MAST for
non-HST data is provided by the NASA Office of Space Science via grant
NNX13AC07G and by other grants and contracts.  This research has also
made use of the SAGE CASJobs database, which is made possible by the
Sloan Digital Sky Survey Collaboration; SAOImage DS9, developed by
Smithsonian Astrophysical Observatory; the VizieR catalogue access
tool, CDS, Strasbourg, France; the SIMBAD database, operated at CDS,
Strasbourg, France; and NASA's Astrophysics Data System Bibliographic
Services.

\bibliographystyle{mn2e}
\bibliography{smcpsc}

\appendix

\section{Literature-Based Classification Support For Sage-Spec Objects}\label{appendixa}

\noindent \emph{NGC 346 MPG 320 (SMC IRS 16)} 
derives its name from the catalogue by
\citet{Massey_89_NGC346}. \citet{Simon_07_NGC346} list it as entry
no.~79, a probable YSO, consistent with our classification as a
YSO-2. However, \citet{Lebouteiller_08_HIIregions} suggest that this
point source is not stellar but rather a Photon-Dominated Region
(PDR).  \citet{Whelan_13_N66}, who list this as PS12, follow that
suggestion. Furthermore, 
\citet{Kamath_14_post-AGBcandidates} list it as a PN
candidate. This source is also known to be an infrared variable
\citep{Polsdofer_15_variables}.

\noindent \emph{NGC 346 MPG 454 (SMC IRS 18)} 
derives its name from the catalogue by \citet{Massey_89_NGC346}. It
is located at the position of N66, an \hii\ region
\citep{Henize_56_Halpha}. This source also occurs 
in the list of candidate PNe by \citet{Kamath_14_post-AGBcandidates}. 
\emph{Chandra} X-ray observations reveal that NGC 346
MPG 454 is one of the brightest blue stars in the NGC 346 cluster
\citep{Naze_02_HD5980}. Indeed, the infrared part of the spectrum
is dominated by a massive YSO \citep{Sabbi_07_NGC346,
Whelan_13_N66}, in the latter work catalogued as PS9. 

\noindent \emph{NGC 346:KWBBe 200 (SMC IRS 19)} 
is also known as NGC 346 MPG 466 \citep{Massey_89_NGC346}. It is
known to show \ha\ emission \citep{Meyssonnier_93_Halpha}. It occurs
in the list of PN candidates by \citet{Kamath_14_post-AGBcandidates}. The
source was recognized as a classical Be star by
\citet{Keller_99_Be}, and it also derives its name from this
catalogue.  \citet{Wisniewski_07_classicalBe} performed a study of classical
Be stars in NGC 346, and in a detailed follow-up paper on NGC
346:KWBBe 200 alone they concluded that this object is in fact a dusty
B[e] supergiant \citep{Wisniewski_07_discovery}.  However, recently
\citet{Whelan_13_N66} suggested that NGC 346:KWBBe 200 is not an
B[e] supergiant, but rather a Herbig AeBe star, based on the presence
of silicate emission features, PAHs and cold dust. In this paper, we
classify this object as an YSO-3.

\noindent \emph{NGC 346 MPG 534 (SMC IRS 20)} 
derives its name from the catalogue by
\citet{Massey_89_NGC346}. The location coincides with that of the
N66A \hii\ region \citep{Henize_56_Halpha}, but also with two class I
protostars, of which the heaviest is thought to be 16.6\,\msun
\citep{Simon_07_NGC346}, later updated to 17.8\,\msun
\citep{Whelan_13_N66}. \citet{Heydari-Malayeri_10_starburst} report
that an O8 star is responsible for ionizing the N66A \hii\ region. This source is a known infrared variable \citep[][Riebel et al.~\emph{in prep.}]{Polsdofer_15_variables}.

\noindent \emph{NGC 346 MPG 605 (SMC IRS 21)} 
was identified for the first time as an \ha\ emitter
\citep[N66B;][]{Henize_56_Halpha}, and then in the survey of NGC 346
by \citet{Massey_89_NGC346} (star number 605) and classified as a red
star.  \citet{Kamath_14_post-AGBcandidates} list this object as a PN
candidate. The star (also known as NGC 346:KWBBe 448) has been then
observed by \citet{Keller_99_Be} and classified as having Be spectral
type.  The spectral type has been established also by
\citet{Martayan_10_Halpha} as being B0:. The star (95) has also been
observed with F555W (V) and F814W (I) \emph{Hubble Space Telescope}
ACS filters by \citet{Gouliermis_06_NGC346} \citep[see
also][]{Hennekemper_08_NGC346}.  The star seems to be located close to
the border between upper main sequence (UMS) and red giant branch
(RGB).  Similar observations have been performed by
\citet{Sabbi_07_NGC346} (source 17) and the obtained magnitudes may
suggest that the star is a red giant branch (RGB) source.  In the
infrared this source has been detected by \iso-CAM \citep[source F
-][]{Contursi_00_N66} and then by \spitzer\
\citep[see][]{Bolatto_07_S3MC}. Using SED fits \citet{Simon_07_NGC346}
classify the source as Class I YSO. The SL \spitzer\ spectrum has been
analyzed by \citet{Whelan_13_N66} (source PS6), and shows silicate
emission features, which is rare among young star clusters.  The
reduced spectrum presented by \citet{Whelan_13_N66} does not show
H$_2$ S(3) emission at 9.67 \um, while it seems to be present in our
reduced spectrum.  \citet{Whelan_13_N66} assume that the optical star
related to this IR source has spectral type O5.5V \citep[star number
37 in][]{Hennekemper_08_NGC346}. However this star is located more
than 3 arcsec (about 1 pc) from the IR source, and probably unrelated.

\noindent \emph{NGC 346 MPG 641 (SMC IRS 22)}. 
The optical counterpart, located at about 0.8 arcsec from the
extracted position, was detected for the first time by
\citet{Massey_89_NGC346}.  Better photometry has been obtained by
\citet{Heydari-Malayeri_10_starburst} using the NTT (star N66A-2). The star
has been classified as an UMS star
\citep{Gouliermis_06_NGC346}. Near-infrared observations with the VLT
have been performed by \citet[ID number 1043]{Gouliermis_10_N66},
who conclude that this is a classical Be star with approximately B0.5
V spectral type.  In the infrared this source has been detected by
\iso-CAM \citep[source H -][]{Contursi_00_N66}, \spitzer\
\citep{Bolatto_07_S3MC}, and \herschel\ \citep{Meixner_13_HERITAGE}.
The SL \spitzer\ spectrum has been analyzed by
\citet{Whelan_13_N66} (source PS4).  It shows that at this
position the extended emission is significant and `optimal extraction'
is required. Using SED fits
\citet{Simon_07_NGC346} classify the source as a Class I YSO,
while we conclude that this is an \hii\ region.

\noindent \emph{LHA 115-N 1 (SMC IRS 26)}  
has been detected as an \ha\ emission-line object by Henize
\citep{Henize_56_Halpha}, and classified by \citet{Lindsay_61_catalogue}
as a probable planetary nebula. Its nature as a planetary nebula has
been confirmed by \citet[][SMP SMC
1]{Sanduleak_78_PN}. \citet{Bernard-Salas_09_unusualdust} already
presented the IRS spectrum, showing that this object has an extremely
strong 11.3-\um\ feature due to SiC emission. Further discussion of
this object is given by \citet{Otsuka_13_M1-11}.

\noindent \emph{LHA 115-N 4 (SMC IRS 27)} 
is also catalogued as Lin 16, and SMP SMC 3.  It is a well known SMC
planetary nebula first catalogued by \citet{Henize_56_Halpha} as an
emission line object.  It was subsequently identified as a planetary
nebula in \citet{Lindsay_61_catalogue}.  It was noted as a lower
excitation object in \citet{Morgan_84_PNe}.  Various optical
spectroscopic observations were carried out in the late 1980s and
early 1990s, but the first determination of the C/O ratio was carried
out by \citet{Vassiliadis_96_PNe} showing that the nebula is
carbon-rich.  There were no mid-infrared observations of this object
prior to the \spitzer\ spectroscopy which was presented in
\citet{Bernard-Salas_09_unusualdust}.  Their classification of the
spectrum as being carbon-rich agrees with ours.

\noindent \emph{LHA 115-N 6 (SMC IRS 28)} 
is another well-studied planetary nebula in the SMC: it is also known
as SMP SMC 6, Lin 33, IRAS 00395$-$7403, LI-SMC 12, GSC
09141$-$007041, and 2MASS J00412777$-$7347063.  The object was first
noted as an emission-line object by \citet{Henize_56_Halpha} and
classified as a planetary nebula by \citet{Lindsay_61_catalogue}.  The
object is bright enough to have been detected by the Infrared
Astronomical Satellite (\iras ), and aside from the original \iras\ PSC
there are pointed observations for this object
\citep{Schwering_89_SMC}.  The star was found to be of Wolf-Rayet type
by E.~J.~Wampler \citep[unpublished, 1978: reported
by][]{Monk_88_abundances}.  The nebula was initially found to be
oxygen-rich by \citet{Vassiliadis_96_PNe}; however this was revised to
carbon-rich by \citet{Stanghellini_09_carbon} wherein the carbon
abundance is determined from HST observations and combined with a
previous ground-based oxygen abundance.  The latter result is
consistent with our spectral classification of C-PN, which is the same
as the classification from \citet{Bernard-Salas_09_unusualdust} where
the \spitzer\ spectrum was first published.

\noindent \emph{LHA 115-N 67 (SMC IRS 29)} 
was discovered by \citet{Henize_56_Halpha} and identified as a planetary
nebula by \citet{Lindsay_61_catalogue}.  The object has many other
names including Lin 333, SMP SMC 22, and DENIS-P J005837.0$-$7131548.
It also appears to be a super-soft X-ray source detected by \emph{Einstein},
\emph{ROSAT}, and \emph{XMM} \citep[i.e.~][]{Pellegrini_94_Xray,Sturm_13_XMM}.  Our
classification of the object as an O-rich PN based on the infrared
spectrum is consistent with the abundance determinations of
\citet{Leisy_96_PNe} which suggests C/O of 0.275 from combined optical
and IUE spectroscopy.  Similar abundance results were obtained by
\citet{Vassiliadis_98_UV}.

\noindent \emph{Lin 536 (SMC IRS 30)} 
is a PN with cold dust compared to other Magellanic Cloud PNe
\citep[SED peaks around 30 \um][]{Bernard-Salas_09_unusualdust}. HST
ACS prism spectra show it is somewhat carbon-poor, with an $A(C)\sim
7$, while other PNe in the SMC have $A(C)>8$
\citep{Smith_95_lithium}.

\noindent \emph{LHA 115-N 70 (SMC IRS 31)}  
was discovered by \citet{Henize_56_Halpha},
and identified as a planetary nebula by \citet{Lindsay_61_catalogue}
under the name Lin 347.  It is also known as 2MASS J00591608$-$7201598
and SMP SMC 24.  No carbon abundance was available for this nebula
before the measurement of \citet{Stanghellini_09_carbon} which, along
with a previous oxygen abundance determination from
\citet{Leisy_96_PNe}, suggests that the object has C/O just larger
than unity.  The mid-infrared spectrum of the object was considered as
being unusual (but carbon-rich) in
\citet{Bernard-Salas_09_unusualdust}, and some of the emission
features were identified as due to fullerenes by
\citet{Garcia-Hernandez_11_fullerenes}.  These are all consistent with
our classification as a C-PN.

\noindent \emph{SMP SMC 11 (SMC IRS 32)}.  
is a well known PN.  Other names for this object include LHA 115-N 29,
Lin 115, and IRAS 00467$-$7314.  The object was first discovered by
\citet{Henize_56_Halpha} as an emission-line object, and it was
classified as a planetary nebulae by \citet{Lindsay_61_catalogue}.  The
carbon-rich nature of the nebula was confirmed from HST observations
well before there was any mid-infrared spectroscopic information
available \citep[i.e.~][]{Stanghellini_03_clusters}.  Due to the relatively
strong mid-infrared emission the object was detected by \iras\ in the
PSC and was the subject of pointed observations as reported by
\citet{Schwering_89_SMC}.

\noindent \emph{HV 1366 (SMC IRS 36)}. 
This O-EAGB object has long been known to be an oxygen-rich AGB star
\citep{Catchpole_81_RSG}, with a period of 293\,days
\citep{Wood_83_LPVs}; it is also known to vary in the infrared \citep[][Riebel et al.~\emph{in prep.}]{Polsdofer_15_variables}. Repeated searches do not show detectable
enhanced Lithium abundances due to hot bottom burning in 4-8\,\msun\ AGB
stars \citep{Smith_90_lithium,Smith_95_lithium}.
\citet{Sloan_08_zoo} report a revised period of 305\,days based
on MACHO data, and classify the \spitzer\ IRS spectrum as 1.N. The
corresponding blackbody temperature was found to be (2270 $\pm$ 560) K.
OGLE variability data also exist \citep{Soszynski_11_OGLE-III}.
\citet{Boyer_12_metallicity} estimate that HV 1366 has a dust mass loss
rate of $10^{-10.1}$ \mdot.

\noindent \emph{HV 11303 (SMC IRS 38)}.  
This object is recognized as a variable oxygen-rich AGB star with a
period of 534 days by \citet{Wood_83_LPVs}, in agreement with
our classification of O-EAGB. The star shows enhanced Lithium
abundances due to hot bottom convective envelope burning
\citep{Smith_90_lithium,Smith_95_lithium}.  The red spectra are
dominated by CN bands, showing signs of enhancements in $^{13}$CN
\citep{Brett_91_envelopeburning}, which is further evidence for hot bottom
burning processes.  \citet{Sloan_08_zoo} assign a spectral
classification of 1.N, and fit a blackbody of 2130 $\pm$ 240\,K to the
SED.

\noindent \emph{HV 838 (SMC IRS 39)}. 
This object was initially thought to be a Red Supergiant, showing
strong hydrogen emission
\citep{Lloyd-Evans_71_SMC,Humphreys_79_abundances}, but later recognized
to be an oxygen-rich AGB star \citep{Wood_83_LPVs}.
\citet{Yang_12_PLrelation} consider HV 838 as an RSG candidate, but
notice that it is indeed an outlier with respect to the RSG
population.  A period of 654\,days was reported by
\citet{Lloyd-Evans_85_largeamplitude}, which was later revised to 629 days based
on the MACHO data \citep{Cioni_03_ISO}, however
\citet{Sloan_08_zoo} report a period of 622\,days, based on the
MACHO data. Riebel et al.~(\emph{in prep.}) also find it to vary in the infrared. HV 838 shows Li\,{\sc i} 6707-\AA\ absorption, a consequence of
hot bottom convective envelope (HBCE) burning
\citep{Smith_95_lithium}. The IRS spectrum is classified as 1.N by
\citet{Sloan_08_zoo}, with a blackbody temperature of 2320
$\pm$ 320\,K.  \citet{Garcia-Hernandez_09_Rb} detect Zr and derive its
abundance, and report an upper limit for the Rb abundance.

\noindent \emph{HV 11366 (SMC IRS 41)}. 
This is an oxygen-rich AGB star with a period of 366\,days
\citep{Catchpole_81_RSG,Wood_83_LPVs}, in agreement with our
classification of O-EAGB.  Its spectrum shows strong Li\,{\sc i} absorption
features at 6707 \AA, indicative of hot bottom burning
\citep{Smith_89_s-process,Smith_90_lithium,Plez_93_lithium,Smith_95_lithium}.
TiO bands are also visible in the red spectra, but there is no clear
sign of CN, and the C/O ratio is estimated to be 0.4-0.8
\citep{Brett_91_envelopeburning}.  This member of NGC 292 has an effective
temperature of 3450\,K, with log $g = 0.00$\,cm s$^{-2}$ and a metallicity
of $[$Fe/H$]=-0.42$ \citep{Soubiran_10_PASTEL}.
\citet{Sloan_08_zoo} classify the \spitzer\ IRS spectrum as
1.N:O::, with a blackbody temperature of 2230 $\pm$ 300\,K.  OGLE
\citep{Soszynski_11_OGLE-III} and MACHO \citep{Yang_12_PLrelation}
variability data are available.

\noindent \emph{OGLE SMC-LPV-7488 (SMC IRS 44)} 
is a Mira variable with a period of 434\,days, and an amplitude of
1.54 magnitude, derived from its OGLE ligthcurve
\citep{Soszynski_11_OGLE-III}. It is also variable in the infrared
\citep[][Riebel et al.~\emph{in prep.}]{Polsdofer_15_variables}. It is
known to be an carbon-rich object from the optical spectrum
\citep{Lloyd-Evans_80_spectra}, consistent with our classification of
C-AGB.

\noindent \emph{HV 12149 (SMC IRS 45)}. 
In agreement with our result, this object was recognized as an O-rich
AGB star \citep{Wood_83_LPVs}, with a period of 742\,days
\citep{Payne-Gaposchkin_66_variable}. This star is also an infrared variable \citep[][Riebel et al.~\emph{in prep.}]{Polsdofer_15_variables}. Enhanced Li\,{\sc i}  absorption indicative of
hot bottom burning is not detected
\citep{Smith_90_lithium,Smith_95_lithium}, however, very strong
TiO bands are visible in the red spectrum \citep{Brett_91_envelopeburning},
pointing to a low C/O ratio. VO bands are also present
\citep{Brett_91_envelopeburning}.  A search for OH maser emission only
yielded an upper limit of 0.04\,Jy at 1612\,MHz
\citep{Wood_92_OHIR}. \citet{Sloan_08_zoo} classify the
\spitzer\ IRS spectrum as 2.SE8, and the dust mass loss rate has been
estimated to be 10$^{-9.0}$ \mdot \citep{Boyer_12_metallicity}.  OGLE
\citep{Soszynski_11_OGLE-III} and MACHO \citep{Yang_12_PLrelation}
variability data are available.

\noindent \emph{HV 11223 (SMC IRS 46)} 
is known to be a semi-regular variable from its OGLE lightcurve, and
is also known as OGLE SMC-LPV-1141 \citep{Soszynski_11_OGLE-III}.
\citet{Smith_95_lithium} show that it is in fact an O-rich AGB star
with Li\,{\sc i} 6707-\AA\ detection, with log$_\epsilon$(Li) = 3.5.

\noindent \emph{Dachs SMC 2-37 (SMC IRS 47)}. 
This object, also known as Massey SMC 59803, is known to be an M
supergiant \citep{Humphreys_79_abundances}, in agreement with our
classification. Its spectral type is reported to be M0-1 I
\citep{Massey_03_RSGs}, although \citet{Levesque_06_metallicity}
believe it to be K2-3 I. Modelling of optical spectra results in
determinations of \teff\ = 4100\,K, $A_V=0.93$, log $g =0.0$
(model) or $-0.2$ (actual) and $R$ = 900\,\rsun\ 
\citep{Levesque_06_metallicity}. The \spitzer\ IRS spectrum is classified as
2.SE4u \citep{Sloan_08_zoo}, looking similar to the \iso-SWS
spectra of RSGs in the h and $\chi$ Perseus supergiant cluster
\citep{Sylvester_98_Msupergiants}, and a C class PAH spectrum
\citep{Peeters_02_6-9um}.  The period is derived from MACHO data
and found to be 547 $\pm$ 17\,days, with a long secondary period of
2543 $\pm$ 38\,days \citep{Yang_12_PLrelation}. This source is also variable in the infrared \citep{Polsdofer_15_variables}.

\noindent \emph{SMC-WR9 (SMC IRS 57)} 
is a WN3ha Wolf-Rayet star with a possible
spectroscopic binary orbit of 37.6\,days
\citep{Morgan_91_WRstar,Massey_01_WR,Foellmi_03_WR}. The star is
modelled to be 94,000--168,000\,\lsun\ and to have a wind velocity of
1616\,\kms and a mass-loss rate $\dot{M}$ of 1.4--3.5 \x\ 10$^{-6}$\,\mdot \citep{Nugis_07_massloss}.

\noindent \emph{SMC-WR10 (SMC IRS 58)}, 
also known as 2MASS J00452890$-$7304458, is a single WN3ha Wolf-Rayet
star, lying in nebulous knot e12 within NGC 249
\citep{Massey_01_WR,Foellmi_03_WR,Crowther_06_WR}. The star is
modelled to be 120,000--240,000\,\lsun\ and to have a wind
velocity of 1731\,\kms and $\dot{M} = 2-5 \times
10^{-6}$\,\mdot \citep{Nugis_07_massloss}. Its spectrum
is noted for having strong nebulosity \citep{Massey_01_WR}.
 
\noindent \emph{SMC-WR12 (SMC IRS 60)}, 
also known as Massey SMC 54730
\citep{Massey_02_UBVR}, was identified as a WN3--4.5 Wolf-Rayet star
by \citet{Massey_03_WRstar}.

\noindent \emph{GSC 09141-05631(SMC IRS
  61)}, 
also known as SMC-WR1, is a Wolf-Rayet star of spectral type WN3ha
\citep{Bonanos_10_SMC}.

\noindent \emph{SMC-WR2 (SMC IRS 64)} is 
a single WN5ha Wolf--Rayet star with surrounding nebulosity
\citep{Azzopardi_80_WR,Massey_01_WR,Foellmi_03_WR}. The star has an
estimated luminosity of 320,000\,\lsun\ and $\dot{M} = 7-8
\times 10^{-6}$\,\mdot \citep{Martins_09_WNh}.

\noindent \emph{SMC-WR3 (SMC IRS 66)} 
is a binary (WN3h+O9:) Wolf--Rayet star with
a spectroscopic period of 10\,days
\citep{Azzopardi_80_WR,Massey_01_WR,Foellmi_03_WR}. It is thought to
be a lower-mass, evolved, H-rich object evolved from a 40--50-\msun\ star \citep{Marchenko_04_hydrogen}. This star appears to be
interacting with the surrounding ISM, producing a kidney-shaped
nebulosity \citep{Gvaramadze_11_runaway}.

\noindent \emph{SMC-WR4 (SMC IRS 68)} is 
a binary WN6h+ Wolf--Rayet star with a
photometric period of 6.55\,days
\citep{Azzopardi_80_WR,Massey_01_WR,Foellmi_03_WR}. The system has an
estimated luminosity of 800,000\,\lsun\ and a combined effective
temperature of $\sim$42\,500\,K \citep{Martins_09_WNh}.

\noindent \emph{RMC 31 (SMC IRS 70)}, 
also known as SMC-WR6, is a 6.5-day binary (WN4:+O6.5I:)
Wolf--Rayet star showing X-ray emission and long-period variability
\citep{Azzopardi_80_WR,Massey_01_WR,Foellmi_03_WR}.

\noindent \emph{SMC-WR11 (SMC IRS 85)}, 
also known as 2MASS J00520738$-$7235385, is a single WN4h:a
Wolf--Rayet star with appreciable reddening
\citep{Massey_01_WR,Foellmi_03_WR}. The star is modelled to be
210,000--390,000\,\lsun\ and have a wind velocity of 1616\,\kms and 
$\dot{M} = 2-5 \times 10^{-6}$\,\mdot
\citep{Nugis_07_massloss}.

\noindent \emph{2MASS J00461632$-$7411135  (SMC IRS 94)}, 
also known as MSX SMC 014, was classified as an R CrB
candidate by \citet{Kraemer_05_RCrB}, based on the IRS spectrum and near-IR
variability. This classification is supported by its
3--4=-um\ spectrum from \citet{vanLoon_08_dustproduction}.

\noindent \emph{2MASS J00364631$-$7331351  (SMC IRS 95)}  
is also known as MSX SMC 029. PAH features in the IRS spectrum of this source
\citep{Kraemer_06_post-AGB} and the spectral appearance of the 3--4-\um\
wavelength range \citep{vanLoon_08_dustproduction} point to a C-rich post-AGB
star.

\noindent \emph{2MASS J00455394$-$7323411 (SMC IRS 96)} 
is also known as MSX SMC 036.
Spectra from \citet{Groenewegen_09_luminosities},
\citet{vanLoon_08_dustproduction}, and \citet{Sloan_06_SMC} all
reveal the presence of C$_2$H$_2$, demonstrating that this a C-rich
AGB star.  \citet{Sloan_06_SMC} also note
SiC and MgS dust features in the IRS spectrum.  OGLE-III
\citep{Soszynski_11_OGLE-III} measure a period of 553\,days, and an
$I$-band amplitude of 2.01\,mag. It is also variable in the infrared \citep{Polsdofer_15_variables}. \citet{Groenewegen_09_luminosities}
find $\dot{M} = 3.6\times10^{-6}$\,\mdot.

\noindent \emph{2MASS J00430590$-$7321406 (SMC IRS 97)}, 
also known as MSX SMC 054,
 is a C-rich AGB star, as demonstrated by spectra from
\citet{Groenewegen_09_luminosities},
\citet{vanLoon_08_dustproduction}, and \citet{Sloan_06_SMC}, which all
confirm the presence of C$_2$H$_2$.  \citet{Sloan_06_SMC} also note
SiC and MgS dust features in the IRS spectrum.
\citet{Groenewegen_09_luminosities} find $\dot{M}=3.9\times10^{-6}$\,\mdot. Its period is not known, although it is an infrared variable \citep{Polsdofer_15_variables}.

\noindent \emph{IRAS F00483$-$7347 (SMC IRS 98)} 
is also known as S9 and MSX SMC 055. IR spectra from
\citet{Groenewegen_09_luminosities} and
\citet{vanLoon_08_dustproduction} indicate an M star, with spectral
type M8. Optical spectra show ZrO$+$H$_2$O and TiO absorption
\citep{Groenewegen_98_IRASsources}. Variability information
\citep[$P=1749$\,d, $Amp_I=0.87$\,mag;][]{Groenewegen_09_luminosities},
a Li-overabundance \citep{Castilho_98_Li}, and Rb enhancement
\citep{Garcia-Hernandez_09_Rb} also point to an AGB star.
\citet{Groenewegen_09_luminosities} classify the star as a Mira, and a
candidate super-AGB star ($\dot{M} = 5.3\times10^{-6}$\,\mdot).  No maser emission has been detected
\citep{vanLoon_01_masers,Marshall_04_superwind}. We classify it as an
RSG by virtue of its \mbol $= -7.3$\,mag. It is also in the Riebel et al.~(\emph{in prep.}) list of infrared variables.

\noindent \emph{IRAS F00448$-$7332 (SMC IRS 99)}, 
  also known as S6 or MSX SMC 060 is known to be a carbon-rich
AGB star.  Spectra from
  \citet{Groenewegen_09_luminosities},
  \citet{vanLoon_08_dustproduction}, and \citet{Sloan_06_SMC} all
  confirm the presence of C$_2$H$_2$.  \citet{Sloan_06_SMC} also note
  SiC and MgS dust features in the IRS
  spectrum. \citet{Groenewegen_09_luminosities} measure a period of
  429 days, a $K$-band amplitude of 0.55\,mag, and 
  $\dot{M}=8.9\times10^{-6}$\,\mdot.  It is also variable in the infrared \citep[][Riebel et al.~\emph{in prep.}]{Polsdofer_15_variables}. \citet{Sloan_06_SMC} note that
  its pulsation properties and photometry make it look similar to
  R CrB stars, but its spectrum does not support that
  classification. We classify it as an C-AGB star, consistent with the
  literature.

  \noindent \emph{2MASS J00485250$-$7308568  (SMC IRS 100)}, 
also known as MSX SMC 066,
  was suggested to be a carbon star by \citet{Raimondo_05_Cstars}
  based on its near-IR colours, consistent with our classification and
  that of \citet{Sloan_06_SMC} as a carbon-rich AGB star. Although
  listed as a Mira by \citet{Soszynski_11_OGLE-III}, there is some
  question regarding its primary period, given as 519.5\,days therein,
  but as 267\,days by \citet{Groenewegen_04_LPV} with a secondary
  period of 512\,days. It is also in the Riebel et al.~(\emph{in prep.}) list of infrared variables.

\noindent \emph{2MASS J00592336$-$7356010 (SMC IRS 101)}, 
also known as MSX SMC 093,
 was identified as a carbon-rich
object by \citet{Sloan_06_SMC}, who noted it has a particularly weak SiC 
emission feature. \citet{Soszynski_11_OGLE-III} used the OGLE light curve and 
near-infrared photometry to classify it as a carbon-rich semi-regular 
variable with a primary period of 457\,days. We classify it here as C-AGB.

\noindent \emph{2MASS J00450214$-$7252243 (SMC IRS 102)}, 
also known as MSX SMC 105,
 was identified as a
carbon star candidate by \citet{Tsalmantza_06_virtual} by its 2MASS
near-IR colours.  Shortly thereafter, \citet{Sloan_06_SMC} found that
its IRS spectrum showed a moderately thick carbon-rich dust
shell. OGLE data show that it is a Mira with a period of $\sim$670\,days \citep{Groenewegen_04_LPV,Soszynski_11_OGLE-III}. It is also a known infrared variable \citep{Polsdofer_15_variables}. We classify it
here as a C-AGB, consistent with the previous works.

\noindent \emph{2MASS J00515018$-$7250496 (SMC IRS 103)}, 
also known as MSX SMC 125, was identified as a carbon star by
\citet{Raimondo_05_Cstars} based on its near-IR photometry. MACHO and
OGLE both find a primary period of $\sim$460\,days and
\citet{Soszynski_11_OGLE-III} call it a carbon-rich Mira. This source is also known to vary in the infrared \citep[][Riebel et al.~\emph{in prep.}]{Polsdofer_15_variables}. It was not
in the \citet{Sloan_06_SMC} sample of SMC carbon star observations
with the IRS due to contamination by Massey SMC 21202 in the IRS
slit. The IRS spectrum of 2MASS J00515018$-$725049 was recovered in
the present re-processing and is classified as a C-AGB, consistent
with the optical and near-IR data.
             
\noindent \emph{2MASS J00542228$-$7243296 (SMC IRS 104)}, 
also known as MSX SMC 159, is classified as a carbon-rich AGB star by
\citet{Sloan_06_SMC}, based on its IRS spectrum, its period
(560\,days) and its luminosity (7470\,\lsun). It is also known to vary
in the infrared (Riebel et al.~\emph{in
  prep.}). \citet{Kamath_14_post-AGBcandidates} list this object as a
candidate PN, but that is inconsistent with its firm C-AGB
classification.

\noindent \emph{2MASS J00510074$-$7225185 (SMC IRS 105)}, 
also known as MSX SMC 163,
is classified as a carbon-rich AGB star by
\citet{Sloan_06_SMC}, based on its IRS spectrum, its period
(660\,days) and its luminosity (13,000\,\lsun). This source was also found to be infrared-variable \citep[][Riebel et al.~\emph{in prep.}]{Polsdofer_15_variables}.
             
\noindent \emph{IRAS 01039$-$7305 (SMC IRS 106)}, 
also known as 2MASS J01053027$-$7249536 and MSX SMC 180, was first
detected with \iras\ as an infrared source towards the SMC
\citep{Schwering_89_SMC}. It was identified as being part of an OB
association by \citet{Battinelli_91_OB}, and classified as a YSO
candidate based its near-IR spectroscopy
\citep{vanLoon_08_dustproduction}. This classification was confirmed
using \spitzer\ spectroscopy \citep{Oliveira_13_SMC}. Here we classify
it as YSO-2.
             
\noindent \emph{2MASS J00571098$-$7230599 (SMC IRS 107)}, 
also known as MSX SMC 198, 
is classified as a carbon-rich AGB star by
\citet{Sloan_06_SMC}, based on its IRS spectrum, its period
(500\,days) and it luminosity (7810\,\lsun). This source is also known to vary in the infrared \citep[][Riebel et al.~\emph{in prep.}]{Polsdofer_15_variables}.

\noindent \emph{2MASS J00465078$-$7147393 (SMC IRS 108)}, 
also known as MSX SMC 200,
is classified as a carbon-rich AGB star by \citet{Sloan_06_SMC}, based
on its IRS spectrum and its luminosity (9130\,\lsun).
            
\noindent \emph{2MASS J01060330$-$7222322 (SMC IRS 109)}, 
also known as MSX SMC 232,
 is classified as a carbon-rich AGB star by
\citet{Sloan_06_SMC}, based on its IRS spectrum, its period
(460\,days) and its luminosity (7280\,\lsun). It is an infrared variable \citep[][Riebel et al.~\emph{in prep.}]{Polsdofer_15_variables}.

\noindent \emph{HV 1375 (SMC IRS 110)} 
is also known as MSX SMC 024.  IR spectra \citep[this
work,][]{Groenewegen_09_luminosities,vanLoon_08_dustproduction}
identify it as an M type AGB star.  \citet{Groenewegen_09_luminosities} find a
spectral type of M5, a period of 418\,days, an $I$-band amplitude of
0.11\,mag, and $\dot{M}=6.3\times10^{-7}$\,\mdot.
\citet{Smith_95_lithium} detect a strong Li overabundance.

\noindent \emph{HV 11262 (SMC IRS 111)}, 
which is also known as MSX SMC 067, is a red supergiant with an optical
spectral type of late K or early M
\citep{Massey_02_UBVR,Massey_09_OStars}, and was first noted as a
possible variable by \citet{Payne-Gaposchkin_66_variable}. This source is also known to vary in the infrared \citep[][Riebel et al.~\emph{in prep.}]{Polsdofer_15_variables}. Its IRS spectrum is
dominated by photospheric emission, and the fine structure lines
longward of 15\,\um\ are probably contamination from a nearby
source or the extended emission seen in the neighborhood.

\noindent \emph{IRAS 00469$-$7341 (SMC IRS 112)}, 
also known as MSX SMC 079, was identified by \citet{Bolatto_07_S3MC} as
a candidate YSO based on its IRAC and MIPS
photometry. Van Loon et al.~(\citeyear{vanLoon_08_dustproduction}) detected H$_2$O
absorption as well as Pf$\gamma$ and Br$\alpha$ emission in a 3--4-\um\ spectrum, and noted it is located near the young cluster Bruck
48. \citet{Oliveira_13_SMC} analyzed its IRS spectrum as part of an
IRS study of SMC YSOs and classified it with a group of early,
embedded YSOs with strong ice and silicate absorption, weak PAHs, and
no fine-structure lines. It is known to be an infrared variable \citep{Polsdofer_15_variables}.

\noindent \emph{{RAW 631} (SMC IRS 113)}, 
{also known as MSX SMC 134,}
was classified as a carbon star from its optical spectrum by
\citet{Rebeirot_93_carbonstars}. It also shows absorption features
from C$_2$H$_2$ and HCN in the 3--4-\um\ region
\citep{vanLoon_08_dustproduction}, although its spectrum is somewhat
atypical of the other carbon stars in that
sample. \citet{Kamath_14_post-AGBcandidates} also list it as a carbon
star.  In contrast, the IRS spectrum shows strong features from
crystalline silicates at $\sim$19, 23, 27, and 33\,\um, which are more
typical of oxygen-rich objects.  \citet{Jones_12_xsils} analyse {this
dual chemistry source} in
their sample of O-rich evolved stars, and suggest the strong
crystalline silicate features are indicative of dust processing in a
circumstellar disk, possibly around a binary system
\citep{Barnbaum_91_envelopes,Lloyd-Evans_91_Jstars}.  In this work, we
classify it as O-AGB, {because of the dominance of oxygen-rich features in 
the Spitzer spectrum.}

\noindent \emph{2MASS J00571814$-$7242352 (SMC IRS 114)}, 
also known as MSX SMC 155, 
 was classified as a R CrB candidate by
\citet{Tisserand_04_RCrB}, based on its EROS 2 optical
lightcurve. From \spitzer\ spectroscopy, \citet{Kraemer_05_RCrB}
confirmed that it is carbon-rich, with an almost featureless
mid-infrared spectrum, characteristic of R CrB stars. It is known to vary in the infrared \citep{Polsdofer_15_variables}.

\noindent \emph{HV 11417 (SMC IRS 115)} 
was classified as an SMC M supergiant by \citet{Elias_80_HV11417},
based on photometric and spectroscopic observations. It was the first
extragalactic red supergiant ever identified. This source is also an infrared variable \citep[][Riebel et al.~\emph{in prep.}]{Polsdofer_15_variables}.

\noindent \emph{HV 12122 (SMC IRS 116)} 
is an M-type AGB star (Spectral Type: M5) with a period of 544\,days
and an $I$-band amplitude of 1.43\,mag
\citep{Groenewegen_09_luminosities}. They measure a low value for $\dot{M}$ of $2.7\times10^{-8}$\,\mdot. The source is also known to vary in the infrared \citep[][Riebel et al.~\emph{in prep.}]{Polsdofer_15_variables}.
\citet{Smith_95_lithium} detect a Li overabundance in the optical
spectrum, indicating HBB. They also detect TiO \& ZrO, in the optical
spectrum.

\noindent \emph{PMMR 34 (SMC IRS 117)}, 
which is also known as  MSX SMC 096, is a red
supergiant
\citep[e.g.~][]{Prevot_83_latetype,Elias_85_supergiants,Massey_02_UBVR}.
\citet{Yang_12_PLrelation} found it to be a semi-regular pulsator
with a period of 381--388\,days using data from the ASAS project
\citep{Pojmanski_02_catalog}. Its IRS spectrum shows weak silicate
emission as well as PAH features. There is no apparent extended
emission in the region, so the PAHs must be local to the source.

\noindent \emph{SkKM 71 (SMC IRS 118)}, 
also known as MSX SMC 109,
is one of the many red supergiants discovered by Sanduleak in the
Magellanic Clouds
\citep{Humphreys_79_abundances,Elias_85_supergiants,Sanduleak_89_supergiants}. Using
ASAS data \citep{Pojmanski_02_catalog}, \citet{Yang_12_PLrelation}
found it to be a `long secondary period' variable with a period of
$\sim$565\,days. Its IRS spectrum shows 9 and 18-\um\ silicate
emission features as well as a PAH feature at 11.3\,\um. As with
PMMR 34, there is no obvious extended emission to contaminate the IRS
slit so we conclude that the PAHs are local to the RSG.
             
\noindent \emph{HV 2084 (SMC IRS 119)} 
is a known red supergiant
\citep[e.g.][]{Prevot_83_latetype,Elias_85_supergiants,Massey_02_UBVR}.

\noindent \emph{HV 1652 (SMC IRS 120)} 
 was observed during an objective-prism survey of the Small
Magellanic Cloud by \citet{Prevot_83_latetype}. It was identified as
a M0-1 supergiant.
 
\noindent \emph{2MASS J00463159$-$7328464 (SMC IRS 121)}, 
also known as MSX SMC 018,
is shown to be a Mira  with $P=897$\,days and
$\Delta I=2.47$\,mag based on its OGLE-III
light curve
\citep{Soszynski_11_OGLE-III}. It is also an infrared variable \citep{Polsdofer_15_variables}. \citet{Groenewegen_09_luminosities}
classify it as M-type (SpT $=$ M7) and measure $\dot{M}=8.1\times10^{-6}$\,\mdot.  \citet{vanLoon_08_dustproduction} also classify
the star as M-type via 3--4-\um\ spectra, consistent with our
classification as O-AGB.

\noindent \emph{2MASS J00470552$-$7321330 (SMC IRS 122)} 
is also known as MSX SMC 033.
Spectra from \citet{Groenewegen_09_luminosities},
\citet{vanLoon_08_dustproduction}, and \citet{Sloan_06_SMC} all
demonstrate the carbon-rich AGB nature of this object with the
presence of C$_2$H$_2$ features.  \citet{Sloan_06_SMC} also note a SiC
dust feature in the IRS spectrum.  OGLE-III
\citep{Soszynski_11_OGLE-III} measure a period of 535\,days, and an
$I$-band amplitude of 1.93\,mag. \citet{Groenewegen_09_luminosities}
find that $\dot{M}=3.7\times10^{-6}$\,\mdot. It is also variable in the infrared (Riebel et al.~\emph{in prep.}). We classify this
source as C-AGB, consistent with previous studies.

\noindent \emph{2MASS J00433957$-$7314576 (SMC IRS 123)}, 
also known as MSX SMC 044, is a C-rich AGB star. Spectra from
\citet{Groenewegen_09_luminosities},
\citet{vanLoon_08_dustproduction}, and \citet{Sloan_06_SMC} all
confirm the presence of C$_2$H$_2$.  \citet{Sloan_06_SMC} also note an
SiC dust feature in the IRS spectrum.  OGLE-III
\citep{Soszynski_11_OGLE-III} measure a period of 440\,days, and an
$I$-band amplitude of 1.39\,mag. This source is also known to vary in
the infrared \citep[][Riebel et al.~\emph{in
  prep.}]{Polsdofer_15_variables}. \citet{Groenewegen_09_luminosities}
find $\dot{M}=3.1\times10^{-6}$ \mdot. We classify this source as
C-AGB, consistent with previous studies.

\noindent \emph{2MASS J00424090$-$7257057 (SMC IRS 124)}, 
also known as MSX SMC 062,
is a C-rich AGB star. Spectra from \citet{Groenewegen_09_luminosities}
and \citet{Sloan_06_SMC} confirm the presence of C$_2$H$_2$.
\citet{Sloan_06_SMC} also notes an SiC dust feature in the IRS
spectrum. OGLE-III \citep{Soszynski_11_OGLE-III} measure a period of 550
days, and an $I$-band amplitude of 1.78\,mag
(Mira). \citet{Groenewegen_09_luminosities} measure that
$\dot{M}=1.8\times10^{-6}$\,\mdot. We classify this source as C-AGB, consistent
with these earlier studies.

\noindent \emph{2MASS J00365671$-$7225175 (SMC IRS 125)}, 
also known as MSX SMC 091,   is an AGB star with a thick
carbon-rich dust shell \citep{Sloan_06_SMC}.  They noted it has a 
particularly strong SiC emission feature for an SMC source but no 26--30-\um\ feature. It has a very limited presence in optical catalogues, with no 
detections blueward of \emph{R} band, where it has a magnitude of $\sim$18.2 or 18.5 
\citep[respectively]{Lasker_08_GuideStarCatalog,Monet_03_USNO-B}.

\noindent \emph{2MASS J00514047$-$7257289 (SMC IRS 126)}, 
also known as MSX SMC 142, was identified as a
carbon star candidate by \citet{Raimondo_05_Cstars} from its near-IR
photometry. The OGLE light curve shows it is a Mira
\citep{Soszynski_11_OGLE-III} with a period of $\sim$295\,days
\citep{Groenewegen_04_LPV,Soszynski_11_OGLE-III}. It is also variable in the infrared \citep[][Riebel et al.~\emph{in prep.}]{Polsdofer_15_variables}. It is the weakest
source in the \citet{Sloan_06_SMC} sample but clearly has a
carbon-rich dust spectrum in the IRS.

\noindent \emph{2MASS J00524017$-$7247276 (SMC IRS 127)}, 
also known as MSX SMC 162,
 is classified as a carbon-rich AGB star by
\citet{Sloan_06_SMC}, based on its IRS spectrum, its period
(520\,days) and its luminosity (11480\,\lsun). It is also known to vary in the infrared \citep[][Riebel et al.~\emph{in prep.}]{Polsdofer_15_variables}.

\noindent \emph{2MASS J00531013$-$7211547 (SMC IRS 128)}, 
also known as MSX SMC 202, is classified as a carbon-rich AGB star by
\citet{Sloan_06_SMC}, based on its IRS spectrum and its luminosity
(11460\,\lsun). This source is also an infrared variable
\citep{Polsdofer_15_variables}.

\noindent \emph{2MASS J00561639$-$7216413 (SMC IRS 129)}, 
also known as MSX SMC 209,
is classified as a carbon-rich AGB star by \citet{Sloan_06_SMC}, based
on its IRS spectrum, its period (520\,days) and its luminosity (16330\,\lsun). It is also in the \citet{Polsdofer_15_variables} and Riebel et al.~(\emph{in prep.}) lists of infrared variables.
         
\noindent \emph{2MASS J01080114$-$7253173 (SMC IRS 131)},             
 also NGC 419 LE 16, is a semi-regular
variable located near the cluster NGC 419. It was discovered by
\citet{Lloyd-Evans_78_NGC419} and classified as a carbon star based on colours.
The classification is confirmed from \spitzer\ spectroscopy
\citep{Lagadec_07_carbonstars}.  The cluster is approximately 2\,Gyr old
\citep{Kamath_10_pulsation}. The star shows several periods including a long
secondary one; the likely fundamental period is 416\,days
\citep{Soszynski_11_OGLE-III}. It is also variable in the infrared \citep{Polsdofer_15_variables}.

\noindent \emph{2MASS J01082495$-$7252569 (SMC IRS 134)},  
 also RAW 1553 and NGC 417 LE 18, is a
well-studied carbon star in the cluster NGC 419. It was first discovered by
\citet{Lloyd-Evans_78_NGC419} and spectroscopically observed by
\citet{Rebeirot_93_carbonstars}. The semiregular variable
\citep{Soszynski_11_OGLE-III} has a dominant period of 181\,days
\citep{Kamath_10_pulsation}.

\noindent \emph{2MASS J01082067$-$7252519 (SMC IRS 135)},  
 also NGC 419 LE 27, was discovered by
\citet{Lloyd-Evans_78_NGC419}.  The semiregular variable has a dominant period
of 311 day \citep{Soszynski_11_OGLE-III}.  It shows the characteristic carbon
star bands in the \spitzer\ SL spectrum \citep{Lagadec_07_carbonstars}.

\noindent \emph{2MASS J00353726$-$7309561 (SMC IRS 136)} 
was discovered as a variable by Glen Moore
\citep[see][]{Lagadec_07_carbonstars} using UK schmidt plates. It is
classified as a carbon star on the basis of both colours
\citep{Raimondo_05_Cstars} and \spitzer\ spectra
\citep{Lagadec_07_carbonstars}. A C/O ratio close to unity is indicated
by the peculiar 11-\um\ feature, lack of C$_2$H$_2$ features
\citep{Lagadec_07_carbonstars}, and high resolution spectra of
\citet{Abia_11_fluorine}.

\noindent \emph{PMMR 52 (SMC IRS 137)}, 
 is a well-studied SMC
RSG. \citet{Massey_03_RSGs} find \mbol $= -8.8$\,mag, \vrad $=159.1$\,\kms, and a spectral type of K5-7 I. The star shows low-amplitude
variability with a period of 483\,days \citep{Yang_12_PLrelation} and has
silicate dust \citep{Lagadec_07_carbonstars}. It is also an infrared variable \citep{Polsdofer_15_variables}. Note that the star should not be be
confused with the nearby carbon Mira J005304.7$-$730409
\citep{vanLoon_08_dustproduction} which was the original but missed target of the
\spitzer\ spectroscopy.
         
\noindent \emph{2MASS J00545410$-$7303181 (SMC IRS 139)} 
is a 541-day Mira variable \citep{Soszynski_11_OGLE-III}, classified as
a carbon star on the basis of optical and \spitzer\ spectroscopy
\citep{Rebeirot_93_carbonstars, Lagadec_07_carbonstars}. The 13.7-\um\
C$_2$H$_2$ absorption features are absent. It is also in the \citet{Polsdofer_15_variables} and Riebel et al.~(\emph{in prep.}) lists of infrared variables.

\noindent \emph{2MASS J00545075$-$7306073 (SMC IRS 140)} 
 is classified as a Mira with  a period of
430\,days \citep{Soszynski_11_OGLE-III}, and it is also an infrared variable \citep[][Riebel et al.~\emph{in prep.}]{Polsdofer_15_variables}. It is classified as a carbon star based
on photometry \citep{Raimondo_05_Cstars} and \spitzer\ spectroscopy
\citep{Lagadec_07_carbonstars}. 

\noindent \emph{2MASS J01015458$-$7258223 (SMC IRS 142)}   
is identified as a carbon star based on
the absorption bands in the \spitzer\ spectrum \citep{Lagadec_07_carbonstars}. 
\citet{Soszynski_11_OGLE-III} report a period of 339\,days. The source is also known to vary in the infrared \citep[][Riebel et al.~\emph{in prep.}]{Polsdofer_15_variables}.

\noindent \emph{LEGC 105 (SMC IRS 143)}    
is located near a small cluster, Bruck 80. The age of the
cluster is $5 \times 10^8\,$yr \citep{Chiosi_06_clusters}. The star was
discovered as a 310-day variable by \citet{Lloyd-Evans_88_LPV} (\#105 in
their paper). \citet{Polsdofer_15_variables} and Riebel et al. (\emph{in prep.}) also find it to be an infrared variable. The \spitzer\ spectrum classifies it as a carbon star but there is
little or no dust excess \citep{Lagadec_07_carbonstars}.

\noindent \emph{2MASS J00572054$-$7312460 (SMC IRS 144)} 
is a 350-day Mira variable \citep{Soszynski_11_OGLE-III}. It is also
known to vary in the infrared \citep[][Riebel et al.~\emph{in
  prep.}]{Polsdofer_15_variables}. \citet{Lagadec_07_carbonstars} find
the typical carbon-star bands in the \spitzer\ spectrum whilst
\citet{Raimondo_05_Cstars} classify it as a carbon star based on
infrared colours.

\noindent \emph{2MASS J00555464$-$7311362 (SMC IRS 145)}, 
also RAW 960, is a 315-day Mira
variable \citep{Soszynski_11_OGLE-III}, classified as a carbon star
based on optical photometry and spectroscopy \citep{Rebeirot_93_carbonstars},
 using the 516\,nm C$_2$
band. \citet{Lagadec_07_carbonstars} find little evidence for cool dust
but strong molecular absorption bands of C$_2$H$_2$. This source is also an infrared variable \citep[][Riebel et al.~\emph{in prep.}]{Polsdofer_15_variables}.

\noindent \emph{IRAS 00554$-$7351 (SMC IRS 146)} 
was identified by \citet{Whitelock_89_dustshell} as a very long period
variable near the tip of the AGB.  \citet{Soszynski_11_OGLE-III} report
a period of 610\,days. The \spitzer\ spectra \citep[this
work;][]{Lagadec_07_carbonstars} show the typical carbon-star features.

\noindent \emph{NGC 419 IR 2 (SMC IRS 147)} 
is located in a crowded region of NGC 419. \citet{Tanabe_97_superwind}
first discovered the self-obscured infrared source and the high
mass-loss rate was confirmed by \citet{vanLoon_05_clusters}. The
period is 381\,days \citep{Kamath_10_pulsation}. It is also in the \citet{Polsdofer_15_variables} of infrared variables. We classify it as C-AGB.

\noindent \emph{NGC 419 IR 1 (SMC IRS 148)}, 
also known as 2MASS J01081296$-$7252439, is a large-amplitude
\citep{Kamath_10_pulsation}, 461-day Mira variable
\citep{Soszynski_11_OGLE-III} first reported by
\citet{Tanabe_97_superwind}. It is classified as carbon-rich based on
the \spitzer\ spectra \citep[this work;][]{Lagadec_07_carbonstars},
and is also known to vary in the infrared
\citep{Polsdofer_15_variables}.

\noindent \emph{AzV 404 (SMC IRS 149)} 
is a B2.5 yellow supergiant \citep{Lennon_97_supergiants}. It has been
modelled to have a luminosity of 174,000\,\lsun\ and light
reddening of $E(B-V) = 0.05$\,mag \citep{Dufton_00_supergiants}. We have
classified it as STAR, due to the absence of emission features in the
IR.
            
\noindent \emph{LHA 115-N38 (SMC IRS 153)}, 
also known as SMP SMC 13, is a round PN with a radius of 0.19\,arcsec
\citep{Stanghellini_03_morphology}. The dust features point to a
carbon-rich chemistry, and the object shows intermediate excitation
lines. The temperature of the dust continuum is derived to be
\tcont\ = 190\,K \citep{Stanghellini_07_PNe}.  The abundance
of carbon is slightly elevated \citep{Stanghellini_09_carbon}, and LHA
115-N 38 is known to contain fullerenes
\citep{Garcia-Hernandez_11_fullerenes}.

\noindent \emph{LHA 115-N 40 (SMC IRS 154)}, 
also known as SMP SMC 14, is a round PN with a radius of 0.42\,arcsec,
showing some internal structure
\citep{Stanghellini_03_morphology}. The \spitzer\ spectrum reveals
carbon-rich dust features and high excitation lines, and a dust
temperature of \tcont\  = 150\,K can be
derived \citep{Stanghellini_07_PNe}.

\noindent \emph{LHA 115-N 43 (SMC IRS 155)}, 
which is also known as SMP SMC 15, is a round PN with a radius of
0.17\,arcsec \citep{Stanghellini_03_morphology}. The \spitzer\ spectrum
shows carbon-rich dust features, intermediate excitation lines, a dust
continuum with \tcont\  = 190\,K, and PAH-related emission
features at 15-21\,\um. The SiC feature is unusually broad, and the
spectrum is similar to that of SMP SMC 18 and 20
\citep{Stanghellini_07_PNe}. LHA 115-N 43 is known to contain
fullerenes \citep{Garcia-Hernandez_11_fullerenes}.
            
\noindent \emph{LHA 115-N 42 (SMC IRS 156)}, 
also known as SMP SMC 16, is an elliptical PN with a radius of
0.18\,arcsec \citep{Stanghellini_03_morphology}. The IRS spectrum shows
carbon-rich dust, and low excitation lines, while a dust temperature of
 \tcont\ = 180\,K can be derived
\citep{Stanghellini_07_PNe}. The object is known to contain fullerenes
\citep{Garcia-Hernandez_11_fullerenes}. All of this is consistent with
our classification as C-PN.

\noindent \emph{LHA 115-N 44 (SMC IRS 157)} 
is also known as SMP SMC 17. It is an elliptical PN with a radius of
0.25\,arcsec, with a faint detached halo
\citep{Stanghellini_03_morphology}. The carbon-rich nature is revealed
by its IRS spectrum, due to the presence of carbon-rich
dust. Intermediate excitation lines are present and the dust
temperature is \tcont\  = 160 K \citep{Stanghellini_07_PNe}.

\noindent \emph{LHA 115-N 47 (SMC IRS 158)}, 
also known as SMP SMC 18, is an unresolved PN
\citep{Stanghellini_03_morphology}. It is carbon-rich in nature, and
shows intermediate excitation lines. The dust temperature is derived
to be \tcont\  = 170\,K \citep{Stanghellini_07_PNe}.  The N
and O abundances are slightly low \citep{Shaw_10_abundances}, and it
contains fullerenes \citep{Garcia-Hernandez_11_fullerenes}.

\noindent \emph{Lin 239 (SMC IRS 159)}, 
also known as Jacoby SMC 20 and SMP SMC 19 is a round PN with a radius
of 0.30\,arcsec, showing some outer structure
\citep{Stanghellini_03_morphology}.  It is carbon-rich in nature and
shows very high excitation lines. The dust temperature is measured to
be \tcont\ = 150\,K \citep{Stanghellini_07_PNe}.

\noindent \emph{LHA 115-N 54 (SMC IRS 160)}, 
also known as SMP SMC 20, is an unresolved PN
\citep{Stanghellini_03_morphology}, with a carbon-rich nebula showing
intermediate excitation lines \citep{Stanghellini_07_PNe}. These last
authors have also derived the dust temperature to be
\tcont\ = 250\,K, which is very high, and suggest that it is
one of the least evolved PNe of their sample. LHA 115-N 54 shows
somewhat low N, O, Ne, S, and Ar abundances
\citep{Shaw_10_abundances}, and may also contain fullerenes
\citep{Garcia-Hernandez_11_fullerenes}, although \citet{Sloan_14_aromatics}
could not confirm this claim.

\noindent \emph{Lin 343 (SMC IRS 161)}, 
also known as SMP SMC 23 or Jacoby SMC 26, is a bipolar core PN with a
radius of 0.30\,arcsec \citep{Stanghellini_03_morphology}.  The IR
spectrum shows intermediate excitation lines, and a dust continuum
with a temperature of \tcont\ = 150\,K
\citep{Stanghellini_07_PNe}.

\noindent \emph{Lin 357 (SMC IRS 162)}, 
also known as SMP SMC 25, is an elliptical PN with a radius of
0.19\,arcsec \citep{Stanghellini_03_morphology}. The IRS spectrum reveals
an O-rich chemistry and shows very high excitation lines, on top of a
dust continuum of \tcont\  = 130\,K
\citep{Stanghellini_07_PNe}. The object shows a low carbon abundance
\citep{Stanghellini_09_carbon}.  The progenitor was probably a
high-mass star \citep{Villaver_04_centralstars}, causing HBB.

\noindent \emph{Lin 430 (SMC IRS 163)}, 
also known as SMP SMC 26, is a point-symmetric PN with a radius of
0.28\,arcsec \citep{Stanghellini_03_morphology}. Very high excitation
lines are present in the IR, on top of a dust continuum of
\tcont\  = 130\,K \citep{Stanghellini_07_PNe}. We classify it
as C-PN.

\noindent \emph{LHA 115-N 87 (SMC IRS 164)}, 
also known as SMP SMC 27, is a round PN with a radius of 0.23\,arcsec,
and an attached outer halo \citep{Stanghellini_03_morphology}.  The
IRS spectrum shows carbon-rich dust features and intermediate
excitation lines, on top of a dust continuum of \tcont\  =
180 K \citep{Stanghellini_07_PNe}. Fullerenes 
may also be present in this source \citep{Garcia-Hernandez_11_fullerenes},
although this is not confirmed by \citet{Sloan_14_aromatics}.

\noindent \emph{LHA 115-N2 (SMC IRS 165)}, 
also known as SMP SMC 2 or Lin 14, is a round PN with radius of
0.25\,arcsec \citep{Stanghellini_03_morphology}.  The IRS spectrum shows
carbon-rich dust and very high excitation lines on a dust continuum of
\tcont\  = 160\,K \citep{Stanghellini_07_PNe}.

\noindent \emph{LHA 115-N 5 (SMC IRS 166)}, 
or SMP SMC 5 or Line 32, is a round PN with a radius of 0.31\,arcsec \citep{Stanghellini_03_morphology}.  The IRS spectrum shows
carbon-rich dust and very high excitation lines, on a dust continuum
of \tcont\  = 180\,K \citep{Stanghellini_07_PNe}.

\noindent \emph{LHA 115-N 7 (SMC IRS 167)}, 
also SMP SMC 8 or Lin 43, is a round PN with a radius of 0.23\,arcsec
\citep{Stanghellini_03_morphology}.  In the infrared, intermediate
excitation lines sit on a featureless continuum of \tcont\ 
= 160\,K \citep{Stanghellini_07_PNe}.

\noindent \emph{BFM 1 (SMC IRS 170)}. 
The presence of the LaO band at 0.79\,\um\ indicates that BFM 1 is an S
star \citep{Blanco_81_Sstar}. It is known to show \ha\
emission \citep{Meyssonnier_93_Halpha}, and its MACHO lightcurve is used
to derive a period of 394--400\,days
\citep{Raimondo_05_Cstars,vanLoon_08_dustproduction,Sloan_08_zoo}.
OGLE variability data also exist \citep{Soszynski_11_OGLE-III}. It is also an infrared variable (Riebel et al.~\emph{in prep.}). BFM 1
is observed twice with \spitzer\ IRS \citep[2.ST,
2.NO;][]{Sloan_08_zoo} and the spectrum shows ZrO and LaO
bands, and a dust emission feature at 13--14\,\um, which has been
attributed to SiS \citep{Sloan_11_SiS}. The fitted blackbody
temperature is 1460 $\pm$ 110\,K or 1160 $\pm$ 10\,K
\citep{Sloan_08_zoo}.  We have classified this as OTHER - S
Star.

\noindent \emph{HV 1963 (SMC IRS 171)} 
 is a known LPV, with a spectral type consistent with an AGB
star and a period of 330 days
\citep{Catchpole_81_RSG,Wood_83_LPVs}, in agreement with our
O-EAGB classification. It is also known to vary in the infrared \citep{Polsdofer_15_variables}. HV 1963 appears to be enhanced in Lithium and s
process elements, while otherwise having low metallicity
\citep{Smith_89_s-process}.  The high Lithium abundance, indicative
of hot bottom burning in 4--8-\msun\ AGB stars, is confirmed by
\citet{Smith_90_lithium}, \citet{Plez_93_lithium} and
\citet{Smith_95_lithium}.  The red spectrum shows moderately strong
TiO and weak CN and ZrO bands, pointing towards carbon dredge up, and
possibly envelope burning \citep{Brett_91_envelopeburning}. These authors
also estimate that C/O ratio is around 0.4--0.8.  The spectral type of
HV 1963 is established to be M4.5s..., with \teff\ $= 3350$\,K, log $g = -0.27$ and [Fe/H] $=-0.43$ 
\citep{Soubiran_10_PASTEL}.  HV 1963 was observed twice with \spitzer\
IRS.  \citet{Sloan_08_zoo} report a spectral classification of
1.N:O(:) and fit a blackbody temperature of 2160 $\pm$ 120\,K or 1990
$\pm$ 50\,K.  A revised period of 249\,days is derived from the OGLE
data \citep{Sloan_08_zoo}.  The dust mass loss rate is found to
be $10^{-9.8}$\,\mdot \citep{Boyer_12_metallicity}.

\noindent \emph{HV 11329 (SMC IRS 172)} 
is reported to be a variable O-rich AGB star
\citep{Catchpole_81_RSG,Wood_83_LPVs}, with a 
period of 390\,days, in agreement with our classification of
O-EAGB. \citet{Lloyd-Evans_85_largeamplitude} revised the pulsational period to
380 days. It is also in the \citet{Polsdofer_15_variables} and Riebel et al.~(\emph{in prep.}) lists of infrared variables. Strong Li\,{\sc i} absorption at 6707 and 8126\,\AA\ indicates hot
bottom burning
\citep{Smith_90_lithium,Plez_93_lithium,Smith_95_lithium}.
The effective temperature of this member of NGC 292 is found to be
\teff\ = 3600\,K, with log $g = -0.05$  and
[Fe/H] = $-0.37$  \citep{Soubiran_10_PASTEL}.
\citet{Sloan_08_zoo} assign a spectral classification of 1.NO,
and fit a blackbody of 1600 $\pm$ 80\,K. The MACHO data indicate a
period of 377\,days \citep{Sloan_08_zoo,Yang_12_PLrelation} and
OGLE variability data also exist for this source
\citep{Soszynski_11_OGLE-III}.

\noindent \emph{2MASS J00445256$-$7318258 (SMC IRS 175)}. 
Variability data are available for this object in the OGLE and MACHO
surveys, and a period of 158\,days has been derived from the OGLE data
\citep{Groenewegen_04_LPV}, while \citet{Raimondo_05_Cstars} arrive
at a period of 129\,days using the MACHO data. We classify this object
as an O-AGB star. {The infrared
spectrum is dominated by crystalline silicate features at $\sim$19, 23,
28, and 33 \micron\ originating from the circumstellar dust shell, and
we classify this object
as an O-AGB star, despite the presence of the 7.5 and 13.7 \micron\ absorption
bands due to C$_2$H$_2$, pointing towards a carbon-rich central star. Like
RAW 631, this object shows a dual chemistry (see also Kraemer et al.~\emph{in prep.}).}

\noindent \emph{PMMR 24 (SMC IRS 176)}, 
also known as Massey SMC 11939, is known to be a RSG
\citep[e.g.~][]{Bonanos_10_SMC,Yang_12_PLrelation}.  Several
(inconsistent) radial velocity measurements exist
\citep{Maurice_87_CORAVEL,Maurice_89_latetypestars,Massey_03_RSGs},
and the period of 352\,days is determined using MACHO data
\citep{Cioni_03_ISO}. Atmospheric modeling of the optical
spectra yields \teff\  = 4025\,K, $A_V = 1.05$, log $g = 0.0$,
and $R = 750$\,\rsun. \citet{Polsdofer_15_variables} found it to be variable in the infrared.

\noindent \emph{BMB-B 75 (SMC IRS 177)} 
is a dust enshrouded O-rich AGB star of spectral type M6.0
\citep{Blanco_80_latetype}, showing long period variability in both the
MACHO \citep{Raimondo_05_Cstars} and OGLE surveys
\citep{Groenewegen_04_LPV,Soszynski_11_OGLE-III}.  It is also variable in the infrared \citep[][Riebel et al.~\emph{in prep.}]{Polsdofer_15_variables}. SiO absorption is
seen in the NIR spectroscopy \citep{vanLoon_08_dustproduction}.  This object
has also been observed with \spitzer 's MIPS-SED, and the data show a
rising continuum, indicative of cold dust (42 $\pm$ 2\,K), and no
emission lines \citep{vanLoon_10_MIPS-SMC}. 

\noindent \emph{IRAS F01066$-$7332 (SMC IRS 178)}  
 is a dust enshrouded oxygen-rich AGB star of spectral type M8
\citep{Groenewegen_98_IRASsources,vanLoon_08_dustproduction}, in agreement with our
classification of O-AGB.  SED fitting with a dusty circumstellar shell
indicates that its luminosity is 25,000\,\lsun, and the mass loss
rate is $5.0 \times 10^{-7}$\,\mdot \citep{Groenewegen_00_SMC}. A
search for SiO using NIR spectroscopy does not yield any results
\citep{vanLoon_08_dustproduction}. The source is also known as a long period
variable \citep{Soszynski_11_OGLE-III}, and is also known to vary in the infrared \citep{Polsdofer_15_variables}.

\noindent \emph{RMC 50 (SMC IRS 193)}, 
also known as LHA 115-S 65, SMC V2364, and IRAS 01432$-$7455, is a
B[e] supergiant, classified between B2 and B9, with \teff\ = 17,000\,K
and $L = 500,000$\,\lsun\
\citep{Ardeberg_77_supergiant,Zickgraf_86_Be,Zickgraf_00_B-bracket-e,Cidale_01_B-bracket-e,Kraus_10_Keplerian,Bonanos_10_SMC}. Although
the star is only slightly variable, the circumstellar environment
appears very dynamic, with 2.2-\um\ CO emission appearing during
2011 \citep{Oksala_12_CO}.

\noindent \emph{NGC 362 SAW V16 (SMC IRS 195)} 
is not located in the SMC, but rather a foreground object. It is a
long-period variable \citep[V16 designation from Hogg;
see][]{Lloyd-Evans_83_redgianttip} in the Galactic globular cluster
NGC\,362 ([Fe/H] = $-1.33$; \citealt{Shetrone_00_redgiants}), hence a
low-mass AGB star or a tip-RGB
star. \citet{Lloyd-Evans_83_redgianttip} detected significant
variations in the strength of the TiO bands during the pulsation cycle
equivalent to spectral type variations K4--M4. He also detected
hydrogen emission, presumably due to pulsation shocks;
\citet{McDonald_07_role} showed the \ha\ to be in self-absorption in an
\'Echelle spectrum, at a spectral type
K5.5. \citet{Lloyd-Evans_83_globularclusters} determined a period of
135 days \citep[138\,days in ][]{Sloan_10_globular} and an amplitude $\Delta
V=2.3$\,mag (similar to that of Mira variables), and confirmed cluster
membership on the basis of radial velocity measurements
\citep[cf.~][]{McDonald_07_role}. It is also known to be an infrared variable \citep{Polsdofer_15_variables}. We note that it should be an O-EAGB, but rather classify it is as
OTHER, since it is a foreground object.

\noindent \emph{HV 206 (SMC IRS 196)} 
is a foreground star, rather than an SMC object.  It is a long-period
variable \citep{Sawyer_31_lightcurves} in the Galactic globular
cluster NGC\,362 ([Fe/H] = $-1.33$; \citealt{Shetrone_00_redgiants}),
hence a low-mass AGB star or a tip-RGB star
\citep[cf.~][]{Frogel_83_globular}. \citet{Lloyd-Evans_83_redgianttip}
detected modest variations in the strength of the TiO bands during the
pulsation cycle. He also detected hydrogen emission, presumably due to
pulsation shocks \citep[cf.~][]{Smith_99_Lithium}. The pulsations are
semi-regular with periods quoted of 90\,days
\citep{Lloyd-Evans_83_globularclusters}; 89\,days
\citep{Sloan_10_globular}; 105\,days \citep{Lebzelter_11_LPV}. It has been
found to be lithium rich by \citet{Smith_99_Lithium}. Although it is
an O-EAGB, we classify this source as OTHER, since it is also a foreground
object.

\noindent \emph{Massey SMC 5822 (SMC IRS 197)}, 
is known as B4 in the sample of \citet{Sheets_13_dustyOB} and \citet{Adams_13_dustyOB},
who are studying OB-type stars with an 24 or 70\,\um\ excess detected by MIPS indicative of
dust emission. This star is also known as source \#5822 by
\citet{Massey_02_UBVR}, and is assigned a group identification number
of 29 in the OB sample table (Table 4) of \citet{Oey_04_massive}. It
has spectral type O9 \citep{Sheets_13_dustyOB}, and \citet{Adams_13_dustyOB}
determine $M_\mathrm{V} = -4.10$\,mag. We classify it as OTHER: dusty OB star.

\noindent \emph{Massey SMC 7776 (SMC IRS 198)}, 
corresponds to B9 in the sample of infrared excess emission stars 
\citep{Sheets_13_dustyOB,Adams_13_dustyOB}.  \citet{Adams_13_dustyOB} determine
$M_\mathrm{V} = -3.87$\,mag and  \citet{Evans_04_2dF} assign a spectral type
of B0 V. We classify it as OTHER: dusty OB star.

\noindent \emph{2MASS J00465728-7318087 (SMC IRS 199)}  
is entry B11 in the sample of \citet{Sheets_13_dustyOB,Adams_13_dustyOB} of 
MIPS-24 or MIPS-70 excess emission stars.  
\citet{Adams_13_dustyOB} determine $M_\mathrm{V} = -2.95$\,mag.  We classify it as
OTHER: dusty OB star.

\noindent \emph{2MASS J00471901-7307110 (SMC IRS 200)} 
is B14 in the sample of infrared excess emission stars
\citep{Sheets_13_dustyOB,Adams_13_dustyOB}.  \citet{Adams_13_dustyOB} determine
$M_\mathrm{V} = -4.52$\,mag, and the spectral type is thought to be B0--B2
\citep{Sheets_13_dustyOB}. We classify it as OTHER: dusty OB star.

\noindent \emph{Massey SMC 9114 (SMC IRS 201)} 
is also known as source B21 in the infrared excess sample of
\citet{Sheets_13_dustyOB} and  \citet{Adams_13_dustyOB}, as source 2dFS5016
\citep{Bonanos_10_SMC}, as OGLE 04.130984 \citep{Evans_04_2dF},
and as source 01507 by \citet{Parker_98_UV}.  \citet{Adams_13_dustyOB}
determine $M_\mathrm{V} = -5.00$\,mag,  \citet{Evans_08_kinematics} report
a radial velocity of 150 $\pm$ 6\,\kms and \citet{Evans_04_2dF} assign a
spectral type of B1--2 II.  We classify it as OTHER: dusty OB star.

\noindent \emph{Massey SMC 9265 (SMC IRS 202)}, 
also known as OGLE SMC-SC4 175188, is source
B24 in the sample of OB stars with infrared excess \citep{Sheets_13_dustyOB,Adams_13_dustyOB}.  \citet{Adams_13_dustyOB}
determine $M_\mathrm{V} = -3.56$\,mag and  \citet{Evans_04_2dF} assign a
spectral type of B1--5 III.   \citet{Wyrzykowski_04_eclipsing} report
this object is part of an eclipsing binary system with a period of P=1.3243\,days.
We classify this object as OTHER: dusty OB star.

\noindent \emph{SSTISAGEMA J004752.26-732121.8  (SMC IRS 203)}  
is entry B26 in the list of OB stars with an infrared excess \citep{Sheets_13_dustyOB,Adams_13_dustyOB}.  
\citet{Adams_13_dustyOB} determine $M_\mathrm{V} = -3.93$\,mag, and the spectral type is thought
to be B0--B2 \citep{Sheets_13_dustyOB}. We classify this source as OTHER: dusty OB star.

\noindent \emph{Massey SMC 10129 (SMC IRS 204)} 
is entry B29 in the sample of OB stars with an infrared excess
\citep{Sheets_13_dustyOB,Adams_13_dustyOB}, and is assigned a group identification
number of 50 in the OB sample table (Table 4) of
\citet{Oey_04_massive}.  \citet{Adams_13_dustyOB} determine $M_\mathrm{V} =
-5.04$\,mag, and the spectral type is thought to be B0 \citep{Sheets_13_dustyOB}.
We classify this source as OTHER: dusty OB star.

\noindent \emph{2MASS J00483000-7318096 (SMC IRS 205)} 
is entry B34 in the sample of OB stars with an infrared excess
\citep{Sheets_13_dustyOB,Adams_13_dustyOB}. The spectral type is thought to be 
B0 \citep{Sheets_13_dustyOB}. We classify this source as OTHER: dusty OB star.

\noindent \emph{Massey SMC 22613 (SMC IRS 206)} 
corresponds to B83 in the sample of OB stars with an infrared excess
\citep{Sheets_13_dustyOB,Adams_13_dustyOB}.  \citet{Adams_13_dustyOB} did not analyze this
source since it is spatially resolved by \spitzer-IRS.
\citet{Evans_04_2dF} assign a spectral type of B1-5 III, while
\citet{Evans_08_kinematics} report a radial velocity of 165 $\pm$ 6\,\kms. We classify this source as OTHER: dusty OB star.

\noindent \emph{Massey SMC 28845 (SMC IRS 207)}, 
is entry B96 in the sample of OB stars with an infrared excess
\citep{Sheets_13_dustyOB,Adams_13_dustyOB}. The spectral type is thought to be 
B0, and the star is known to be part of
an eclipsing binary system \citep{Sheets_13_dustyOB}. 
We classify this source as OTHER: dusty OB star.

\noindent \emph{NGC 330 ELS 57 (SMC IRS 208)}, 
also known as object B100 in the sample of \citet{Sheets_13_dustyOB} and
\citet{Adams_13_dustyOB}; SMC5\_009833 \citep{Bonanos_10_SMC}; and
Massey SMC 31632, although the \spitzer\ spectrum is extracted $\sim 1$ arcsec
away from the position of NGC 330 ELS 57.  According to
\citet{Evans_06_FLAMES}, ELS 57, located 7.70\,arcmin away from the center
of NGC 330, is of spectral type B0.5V and has a radial velocity of
124 $\pm$ 4\,\kms.  \citet{Hunter_08_FLAMES} assign the following stellar
parameters to this star: \teff\ = 29,000\,K, log $g =
4.15$, $L  = 10^{4.34}$ \lsun, projected rotational velocity
$v \mathrm{sin} i = 104$ \kms, and $M = 13$ \msun.  \citet{Hunter_09_FLAMES}
determine an atmospheric microturbulent velocity for the star of 5\,\kms, and abundances of 8.01 $\pm$ 0.18\,dex for Oxygen, 6.96 $\pm$ 0.26 for
Magnesium, and $< 7.48 \pm 0.29$ for Nitrogen.  In the study by
\citet{Koehler_12_Nitrogen} this star is known as S27, and they assign
a main sequence lifetime of 13.3\,Myr, an inclination angle of
sin $i_{\mathrm{N}}$ = 0.33, and a 10.2\,Myr isochrone age.
\citet{Adams_13_dustyOB} determine an absolute $V$ magnitude, $M_\mathrm{V}$,
of $-3.36$\,mag. We classify this object as OTHER: dusty OB star, based on 
these studies.

\noindent \emph{Massey SMC 32159 (SMC IRS 209)}, 
is entry B102 in the sample of dusty OB stars from \citet{Sheets_13_dustyOB}
and \citet{Adams_13_dustyOB}, and it is assigned a group identification
number of 195 in the OB sample table (Table 4) of
\citet{Oey_04_massive}.  \citet{Adams_13_dustyOB} determine $M_\mathrm{V} =
-4.43$\,mag, while \citet{Sheets_13_dustyOB} assing a spectral type of O9.

\noindent \emph{2MASS J00561161$-$7218244 (SMC IRS 210)} 
is entry B112 in the sample of OB stars with a far-infrared excess
\citep{Sheets_13_dustyOB,Adams_13_dustyOB}. The star is known to be part of
an eclipsing binary system with a period of 4.48\,days \citep{Bayne_02_MOA,Faccioli_07_MACHO}. 
We classify this source as OTHER: dusty OB star.

\noindent \emph{AzV 216 (SMC IRS 211)} 
also known as Massey SMC 44984, is entry B137 in the sample of OB
stars with an infrared excess \citep{Sheets_13_dustyOB,Adams_13_dustyOB}. The
spectral type is thought to be B1--3II \citep{Evans_04_2dF}
or B1 \citep{Sheets_13_dustyOB}, and the
star is known to be part of an eclipsing binary system with a period
of 1.84\,days \citep{Faccioli_07_MACHO}.  We classify this source as
OTHER: dusty OB star.

\noindent \emph{Massey SMC 50031 (SMC IRS 212)} 
is entry B148 in the sample of OB stars with an infrared excess
\citep{Sheets_13_dustyOB,Adams_13_dustyOB}. The spectral type is thought to be B0
\citep{Sheets_13_dustyOB}.  We classify this source as OTHER: dusty OB star.

\noindent \emph{Massey SMC 54281 (SMC IRS 213)} 
is entry B154 in the sample of OB stars with an infrared excess
\citep{Sheets_13_dustyOB,Adams_13_dustyOB}. The spectral type is thought to be B1
\citep{Sheets_13_dustyOB}.  We classify this source as OTHER: dusty OB star.

\noindent \emph{Massey SMC 55094 (SMC IRS 214)} 
is entry B159 in the sample of OB stars with an infrared excess
\citep{Sheets_13_dustyOB,Adams_13_dustyOB}. The spectral type is thought to be B0
\citep{Sheets_13_dustyOB}.  We classify this source as OTHER: dusty OB star.

\noindent \emph{Massey SMC 55634 (SMC IRS 215)} 
is also known as B161 in the sample of \citet{Sheets_13_dustyOB} and
\citet{Adams_13_dustyOB}.  The spectral type of this object is B0
\citep{Sheets_13_dustyOB}.  \citet{Adams_13_dustyOB} did not analyze this source
since it is spatially resolved by \spitzer-IRS, and there is a bright
source nearby in its IRS data.  We classify this source as OTHER:
dusty OB star.

\noindent \emph{Massey SMC 60439 (SMC IRS 216)} 
is entry B182 in the sample of \citet{Sheets_13_dustyOB} and
\citet{Adams_13_dustyOB}, and it is assigned a group identification number of
391 in the OB sample table (Table 4) of \citet{Oey_04_massive}.
\citet{Adams_13_dustyOB} determine $M_\mathrm{V} = -4.98$\,mag, while
\citet{Sheets_13_dustyOB} determine a spectral type of B0. We classify this
as OTHER: dusty OB star.

\noindent \emph{Massey SMC 67470 (SMC IRS 217)} 
is entry B188 in the sample of \citet{Sheets_13_dustyOB} and
\citet{Adams_13_dustyOB}. It is assigned a group identification number of 445
in the OB sample table (Table 4) of \citet{Oey_04_massive}.
\citet{Adams_13_dustyOB} determine $M_\mathrm{V} = -3.97$\,mag, while
\citet{Sheets_13_dustyOB} believe that this is a Be star with weak
\ha\  absorption. We group it in the OTHER: dusty OB star category.

\noindent \emph{Massey SMC 77248 (SMC IRS 218)} 
is entry B193 in the sample of OB stars with an infrared excess
\citep{Sheets_13_dustyOB,Adams_13_dustyOB}. The spectral type is thought to be B0
\citep{Sheets_13_dustyOB}.  We classify this source as OTHER: dusty OB star.

\noindent \emph{Massey SMC 25387 (SMC IRS 219)} 
is entry B87 in the sample of OB stars with a MIPS far-infrared excess
\citep{Sheets_13_dustyOB,Adams_13_dustyOB}. It is assigned a group identification
number of 149 in the OB sample table (Table 4) of
\citet{Oey_04_massive}.  \citet{Adams_13_dustyOB} determine $M_\mathrm{V} =
-4.63$\,mag, while \citet{Sheets_13_dustyOB} its spectral type to be B0. We
classify this source as OTHER: dusty OB star.

\noindent \emph{Massey SMC 55681 (SMC IRS 225)} 
is listed as an M3 star by \citet{Elias_85_supergiants} and M0--M1 in
the catalogue of \citet{Massey_03_RSGs}. The latter authors also
calculate \mbol  = $-10.53$\,mag, which supports our
classification of RSG. \citet{Polsdofer_15_variables} also found it to vary in the infrared.

\noindent \emph{Massey SMC 10889 (SMC IRS 226)} 
was confirmed to be a RSG in the SMC through high-accuracy radial
velocity measurements by \citet{Massey_03_RSGs}. A spectral type
of M0 Ia was assigned by \citet{Elias_85_supergiants} and it is listed
as having K7 I spectral type by \citet{Massey_03_RSGs}. This
all
supports our classification of RSG.

\noindent \emph{Massey SMC 11709 (SMC IRS 227)}. 
We classify this object as a RSG, in agreement with \citet{Massey_03_RSGs}.

\noindent \emph{Massey SMC 46662 (SMC IRS 228)} 
was identified as a late-type RSG by
\citet{Levesque_07_redsupergiants} due to the significant differences
in the spectral types (M2 I to K2--3 I) and effective temperatures
over short timescales. Its location in the Hayashi forbidden zone of
the H-R diagram indicates the star is no longer in hydrostatic
equilibrium and exhibits considerable variability in $V$
magnitudes. This supports our classification of RSG. It is also known
to vary in the infrared \citep{Polsdofer_15_variables}, and an
X-SHOOTER spectrum is available \citep{Chen_14_XSHOOTER}.

\noindent \emph{Massey SMC 52334 (SMC IRS 229)}. 
The preliminary spectral type assigned by \citet{Elias_85_supergiants}
is M0 Iab, and this object is also included in the RSG catalogue of
\citet{Massey_03_RSGs} who list the spectral class as K7 I and give
\mbol = -8.15\,mag. This is all in agreement with our classification
of RSG. An X-SHOOTER spectrum of this object has been obtained
\citep{Chen_14_XSHOOTER}.

\noindent \emph{HV 2232 (SMC IRS 230)} 
is assigned a spectral type of M2 in the catalogue of supergiants by
\citet{Elias_85_supergiants} and \citet{Levesque_07_redsupergiants}. However,
we classify this object as O-AGB, based on the bolometric luminosity.
An X-SHOOTER spectrum of this object has been obtained \citep{Chen_14_XSHOOTER}.

\noindent \emph{HV 11423 (SMC IRS 231)} 
is an unstable cool supergiant that was found by
\citep{Massey_07_HV11423} to have varied its spectral type between
K0--1 I and M4.5--5 I on a timescale of a few years. It was
originally classified as an M0 supergiant by
\citet{Humphreys_79_abundances} and \citet{Elias_85_supergiants} and listed
by \citet{Massey_03_RSGs} as a RSG in the SMC. We also classify it as
RSG.

\noindent \emph{Massey SMC 55188 (SMC IRS 232)}. 
\citet{Massey_03_RSGs} classify this source as a RSGs in the SMC, and
it is also identified as an unstable cool RSG with a spectral type (M2
I--4.5 I) by \citet{Levesque_07_redsupergiants}.  We classify this
object as RSG. It is in the \citet{Polsdofer_15_variables} list of
infrared variables, and an X-SHOOTER spectrum has been obtained
\citep{Chen_14_XSHOOTER}.

\noindent \emph{AzV 456 (SMC IRS 234)}, 
also known as Sk 143 \citep{Sanduleak_68_findinglist}, is a O9.5 or
B0--1 yellow supergiant with \teff\  = 23,000 -- 30,000\,K and
$L = 950,000$\,\lsun. Its spectral type was previously thought to
be even earlier \citep[O8 II:][]{Smith-Neubig_97_OB}.  The star is
significantly reddened by interstellar dust to $E(B-V)\approx0.37$\,mag, of which $\sim0.18$\,mag is attributed to foreground Galactic
dust, and hence has been used extensively to study the intervening
interstellar medium
\citep{Lequeux_82_SK143,Prinja_87_stellarwinds,Thompson_88_supergiants}. The
object shows a wind with a terminal velocity of 1450 \kms  and 
$\dot{M} = 7 \times 10^{-7}$\,\mdot 
\citep{Prinja_87_stellarwinds,Evans_04_supergiants}.  Membership of
the SMC is confirmed on the basis of its heliocentric radial velocity
$v_{\mathrm{hel}}=166\pm7$\,\kms  \citep{Evans_08_kinematics}.
Its proper motion has been measured at $\approx2\sigma$ significance
\citep[$\mu_\alpha=6.7\pm3.2$\,mas yr$^{-1}$, $\mu_\delta=-5.6\pm3.1$\,mas yr$^{-1}$;][]{Zacharias_04_UCAC2}.

\noindent \emph{NGC 346 MPG 293 (SMC IRS 235)} 
is a B3\,Ia \citep{Azzopardi_82_SMC}, B1\,Ia
\citep{Bouchet_85_SMC} or B2\,Ia \citep{Smith-Neubig_97_OB}
supergiant in the young massive cluster NGC\,346. It has a relatively
slow line-driven wind \citep[$v_\infty\approx900$\,\kms;][]{Prinja_87_stellarwinds}, attributed to the low metal content.

\noindent \emph{AzV 23 (SMC IRS 236)} 
is a B3\,Ia \citep{Azzopardi_82_SMC,Lennon_97_supergiants}
yellow supergiant, first catalogued as Sk\,17
\citep{Sanduleak_68_findinglist}. Membership of the SMC is confirmed
on the basis of its heliocentric radial velocity $v_{\rm hel}=178$\,\kms  \citep[who derived a spectral type of
B2\,I]{Neugent_10_yellowsupergiants}. It has been modelled to have a
luminosity of 230,000\,\lsun\ and moderate reddening of $E(B-V) =
0.21$\,mag \citep{Dufton_00_supergiants}.

\noindent \emph{IRAS 00350-7436 (SMC IRS 238)}, 
also known as LI-LMC 5, is a carbon-rich, high luminosity object
(\mbol = $-6.564$ -- $-6.82$\,mag;
\citealt{Whitelock_89_dustshell,vanLoon_98_OHIR,Tsalmantza_06_virtual}). \citet{Zijlstra_96_obscured}
designate it as a candidate AGB star, with a dust mass loss rate
derived from IR colours of $10^{-8.17}$-- $10^{-7.25}$\,\mdot, depending on
the specific colour. It is among the most luminous AGB stars in the
Magellanic Clouds, with only one object in the LMC having similar
luminosity, IRAS 04496$-$6958 \citep{vanLoon_98_OHIR}.
\citet{Matsuura_05_3umspectra} suggest that this is a post-AGB star,
based on the presence of the 3.3-\um\ PAH emission feature.  Because
of the presence of weak C$_2$H$_2$ absorption in the IRS spectrum, we
classify it as C-AGB.
 
\noindent \emph{NGC 330 SW 515 (SMC IRS 239)}  
is an irregular optical variable \citep{Sebo_94_NGC330}, with a
possible 45-day orbital period but also variability on both shorter and
longer timescales \citep{Schmidtke_08_NGC330}, and a possible member of
the cluster NGC\,330. It features broad \ha\  emission
\citep{Meyssonnier_93_Halpha,Oliveira_13_SMC}, which is double peaked
\citep{Hummel_99_Be}, as well as near-IR hydrogen emission
\citep{Tanabe_13_Paschen}. \citet{Martayan_07_FLAMES} classified it as
a HeB[e] object. Mid-IR excess emission was detected with the Infrared 
Space Observatory (\iso ) by
\citet{Kucinskas_00_NGC330} and with \spitzer\ by
\citet{Bolatto_07_S3MC}. The \spitzer\ IRS spectrum was first
published by \citet{Oliveira_13_SMC}, who recognised it as a young
star but with no trace of H$_2$O or CO$_2$ ices. 

\noindent \emph{Lin 517 (SMC IRS 240)} 
is also known as [BSS2007 282] and LHA 115-N 86.  It appears in
earlier lists of \ha\ sources and possible PNe,
e.g.~\citet{Henize_56_Halpha},
\citet{Lindsay_61_catalogue}. \citet{Kamath_14_post-AGBcandidates}
find that it may be a hot post-AGB star.  We classify this object as
YSO-4.

\noindent \emph{[MA93] 1771 (SMC IRS 241)} 
is also 2MASS J01134116$-$7250499 and SSTISAGEMA
J011341.20$-$725049.8.  It is a known \ha\ source
\citep{Meyssonnier_93_Halpha}, who believe it to be the red component
of a 2\,arcsec pair, and possibly related to Lin 497.  Indeed, the
MCPS survey \citep{Zaritsky_02_MCPS-SMC} shows three potential
counterparts within 3\,arcsec of the SAGE-SMC position
\citep{Gordon_11_SAGE-SMC}. This very red source is also detected
within 3\,arcsec by \herschel\ in the Herschel Inventory of the Agents
of Galaxy Evolution survey \citep[HERITAGE;][]{Meixner_13_HERITAGE} at
100, 160 and 250\,\um. \citet{Kamath_14_post-AGBcandidates} find that
this object may be a hot post-AGB star.

\noindent \emph{2MASS J00540342$-$7319384 (SMC IRS 242)} 
 is object 18 in the spectral catalogue by
\citet{Oliveira_13_SMC}. It is identified as a YSO candidate based on
MIPS SED data \citep{vanLoon_08_dustproduction} and the IRS spectrum
\citep[this work;][]{Oliveira_11_metallicity}.

\noindent \emph{2MASS J01054645$-$7147053 (SMC IRS 243)}. 
Initially proposed as a YSO \citep{Bolatto_07_S3MC}, this object has
since been classified as a carbon-rich proto-planetary nebula by
\citet{Volk_11_21micron}, who point out the strong mid-IR emission
features usually attributed to PAHs, and a hint of a weak 21-\um\
emission feature of which the carrier remains unknown. They derived a
luminosity of $4660\pm510$\,\lsun\ and birth mass of only
$1.00^{+0.12}_{-0.04}$\,\msun\ -- which seems low for the
carbon enrichment. \citet{Kamath_14_post-AGBcandidates} derive
$L = 4106$\,\lsun\ and, based on the presence of atomic
emission lines and a UV continuum, conclude that this is a hot post-AGB
star.

\noindent \emph{2MASS J00540236$-$7321182 (SMC IRS 244)} 
 is object 17 in the spectral catalogue by
\citet{Oliveira_13_SMC}. It is identified as a YSO candidate based on
MIPS SED data \citep{vanLoon_08_dustproduction} and the IRS spectrum
\citep[this work;][]{Oliveira_11_metallicity}.

\noindent \emph{OGLE SMC-SC10 107856 (SMC IRS 245)} 
is a (strong) candidate R CrB type star
\citep{Tisserand_09_EROS-2}, experiencing erratic, sudden obscuration
events by circumstellar carbonaceous dust clouds. \citet{Polsdofer_15_variables} and Riebel et al.~(\emph{in prep.}) list it as an infrared variable. It has a relatively
cool atmosphere, $\sim5000$\,K compared to most R CrB stars.

\noindent \emph{IRAS 00516$-$7259 (SMC IRS 246)} 
is entry 16 in the spectral catalogue of YSO candidates by \citet{Oliveira_13_SMC}.

\noindent \emph{IRAS 00509$-$7342 (SMC IRS 247)} 
is entry 15 in the spectral catalogue of YSO candidates by
\citet{Oliveira_13_SMC}, who also report extended radio continuum
emission at 1.42, 2.37, 4.80 and 8.64\,GHz at the same position.
\citet{Kamath_14_post-AGBcandidates} confirm the YSO nature of this source.

\noindent \emph{2MASS J00505814$-$7307567 (SMC IRS 248)} 
is entry 14 in the catalogue by \citet{Oliveira_13_SMC}. It is also
known as an \ha\ emitter \citep{Meyssonnier_93_Halpha}.
\citet{Kamath_14_post-AGBcandidates} confirm the YSO nature of this
source.

\noindent \emph{2MASS J00504326$-$7246558 (SMC IRS 249)} 
is entry 13 in the spectral catalogue of YSO candidates by
\citet{Oliveira_13_SMC}, while it is also know as an \ha\ emitter
\citep{Meyssonnier_93_Halpha}. It appears to be part of a star cluster
\citep{Bica_95_SMC,Chiosi_06_clusters}.
\citet{Kamath_14_post-AGBcandidates} confirm the YSO nature of this
source.

\noindent \emph{2MASS J00504042$-$7320369 (SMC IRS 250)} 
is entry 12 in the spectral atlas by \citet{Oliveira_13_SMC}.

\noindent \emph{2MASS J00494469$-$7324331 (SMC IRS 251)} 
is entry 11 in the spectral atlas by \citet{Oliveira_13_SMC}.

\noindent \emph{SMP SMC 21 (SMC IRS 252)} 
is a well known SMC planetary nebula first identified by
\citep{Lindsay_61_catalogue}.  Our classification as an O-rich PN is
consistent with the abundance analyses available in the literature
\citep[i.e.~][]{Leisy_96_PNe}.

\noindent \emph{Massey SMC 60447 (SMC IRS 253)} 
 was first catalogued by \citet{Basinski_67_SMC} where it is object 353 in their Table V.  
 The object was then classified as a red supergiant star in
 \citet{Sanduleak_89_supergiants} where it is object 276; the positions are
 slightly different in these two papers but it is clearly the same
 object.  The spectral type is given as K2I in \citet{Massey_02_UBVR}.
 This is all consistent with our classification of RSG. The source is also known to vary in the infrared \citep{Polsdofer_15_variables}.

 \noindent \emph{PMMR 145 (SMC IRS 254)} 
 was first noted by \citet{Prevot_83_latetype}, as a red supergiant
 of spectral type K5 to K8.  The object is also discussed by
 \citet{Elias_85_supergiants}.  An abundance analysis was carried out
 by \citet{Hill_97_Ksupergiants}, who gives parameters \teff\ = 4300\,K,
 log $g = +0.3$, and [M/H] = $-$0.6 along with a radial velocity of 160.6\,\kms.  Analysis of the CNO abundances by \citet{Hill_97_abundances}
 yields a C/O ratio of 0.3.  All of these observations are consistent
 with our classification of the object as RSG.

 \noindent \emph{PMMR 141 (SMC IRS 255)} 
 \citet{Prevot_83_latetype} catalogue this star as a late-type
 supergiant.  The spectral type is given as K7--M0I by
 \citet{Massey_03_RSGs}.  All the references to this object in
 the literature are consistent with our classification of the object
 as RSG.

 \noindent \emph{PMMR 132 (SMC IRS 256)} 
 was catalogued as a late-type supergiant in the SMC by
 \citet{Prevot_83_latetype} based upon objective prism observations.
 Photometry is given in \citet{Elias_85_supergiants}. We classify it as
 RSG.

 \noindent \emph{LHA 115-S 38 (SMC IRS 257)} 
 was identified as an emission line object by
 \citet{Henize_56_Halpha}.  It was then also catalogued by
 \citet{Lindsay_61_catalogue} as Lin 418.  The spectral type is given
 as between A3 and F0 in \citet[][object 1804]{Evans_04_2dF}. The star
 is also discussed by \citet{Raimondo_05_Cstars} as a likely
 carbon-rich object based on the $J - K$ colour.
 \citet{Kamath_14_post-AGBcandidates} identify this object as a
 post-AGB candidate (\# 38 in their Q1 list), showing a slow
 brightening in the optical combined with a long period of
 $\sim$900\,days, possibly due to changes in the dust obscuration or
 the accretion rate.  The IRS spectrum, reveals its O-rich nature and
 we are able classify this object as O-PAGB.

\noindent \emph{RAW 594 (SMC IRS 258)} 
was identified as a probable carbon-rich variable star by
\citet{Rebeirot_93_carbonstars} based on its near-infrared colours.  There
are 4 other papers in the literature concerning the variability or SED
properties of the object, which is a semi-regular pulsator from the
OGLE survey.  Our IRS spectrum classification confirms the carbon-rich
nature and we classify it as C-AGB.

\noindent \emph{2MASS J00444463$-$7314076 (SMC IRS 259)}. 
Nothing specific is known about this object, except that it is an infrared variable \citep{Polsdofer_15_variables}.

\noindent \emph{SSTISAGEMA J005419.21$-$722909.7 (SMC IRS 260)} 
was identified based on its IRAC colours as a YSO candidate by
\citet{Bolatto_07_S3MC}, as YSO candidate 139 in their list ([BSS2007]
139), and as such included in the YSO sample to be observed with
\spitzer-IRS by \citet[][entry \#19]{Oliveira_13_SMC}.  However, these
authors find that this object is not a YSO, but rather a D-type
symbiotic star, consisting of an AGB star and white dwarf, with mass
transfer between them.  There is also a nearby blue star (about 2.5\,arcsec
away from the symbiotic binary) contributing a continuum to the
spectrum that cannot be separated out \citep{Oliveira_13_SMC}.

\noindent \emph{2MASS J00432649$-$7326433 (SMC IRS 261)} 
is a semi-regular variable carbon star
\citep{Raimondo_05_Cstars,Soszynski_11_OGLE-III}. It is also in the \citet{Polsdofer_15_variables} list of infrared variables.

\noindent \emph{Lin 250 (SMC IRS 262)}, 
also known as LHA 115-S 18, is a well-known B[e] supergiant (see
e.g.~\citealt{Clark_13_LHA115-S18} and references therein for a
comprehensive view of this system). It is an infrared variable
\citep[][Riebel et al.~\emph{in prep.}]{Polsdofer_15_variables}.

\noindent \emph{IRAS 00471$-$7352 (SMC IRS 263)}, 
is known to be a long period variable
\citep{Groenewegen_04_LPV,Raimondo_05_Cstars,Soszynski_11_OGLE-III},
carbon-rich in nature \citep{Raimondo_05_Cstars}. It is
also an infrared variable \citep{Polsdofer_15_variables}.

\noindent \emph{SSTISAGEMA J004901.61$-$731109.5 (SMC IRS 264)} 
is entry 10 in the \spitzer-IRS spectral atlas of YSO candidates by \citet{Oliveira_13_SMC}.

\noindent \emph{LHA 115-N 31 (SMC IRS 265)}, 
also known as Lin 120, is an emission line star
\citep{Henize_56_Halpha,Meyssonnier_93_Halpha}.  It is classified as a
candidate YSO by \citet{Bolatto_07_S3MC}, while
\citet{Charmandaris_08_HII} included it in their sample of candidate
compact H{\sc ii} regions, as object \#5, and remark that the IRAC
colours set it apart from class I and class II YSOs.  LHA 115-N 31 is
also entry 9 in the \spitzer-IRS spectral atlas of YSO candidates by
\citet{Oliveira_13_SMC}.

\noindent \emph{Lin 60 (SMC IRS 266)} 
is also known as LHA 115-N 10 \citep[see e.g.][]{Meyssonnier_93_Halpha}.
It was one of the sample with MIPS spectra obtained by
\citet{vanLoon_10_MIPS-SMC}, and \citet{Oliveira_13_SMC} examined it
as a bright YSO as well.  Its spectrum has clear CO$_2$ ice absorption
at 15\,\um.  It appears in older lists, such as \citet{Henize_56_Halpha}
and \citet{Lindsay_61_catalogue}.

\noindent \emph{S3MC J004825.83$-$730557.29 (SMC IRS 267)} 
is found in a complex region, and is therefore not identified as a
point source by the SAGE-SMC team \citep{Gordon_11_SAGE-SMC}, but its
flux levels were extracted by the S3MC team \citep{Bolatto_07_S3MC},
who also classify it as a candidate YSO.  This is entry \#8 in the IRS
spectral catalogue of YSO candidates by \citet{Oliveira_13_SMC}, who
report detections of CO$_2$ ice and PAHs. \citet{Kamath_14_post-AGBcandidates}
offer an alternative view by classifying this as a candidate PN.
We classify this object as YSO-1.

\noindent \emph{2MASS J00444111$-$7321361 (SMC IRS 268)} 
is a small-amplitude semi-regular variable star, with $P\approx98$
days \citep{Soszynski_11_OGLE-III}, later refined to 96.338 days
\citep{Kamath_14_post-AGBcandidates}.  While it was selected as a
candidate YSO on the basis of mid-IR photometry
\citep{Bolatto_07_S3MC}, it has subsequently been shown to be a
post-AGB object on the basis of its 21-\um\ emission feature
\citep{Volk_11_21micron}. The post-AGB nature was confirmed by the
measurement of extreme $s$-process elemental enhancements, its carbon
enrichment (with C/O $> 1$), and iron depletion, but not the otherwise
expected lead overabundance; a birth mass of $\sim1.3$\,\msun\ was
inferred \citep{DeSmedt_12_21um,DeSmedt_14_lead}.

\noindent \emph{2MASS J00465185$-$7315248 (SMC IRS 269)} 
appears to be slightly extended at IRAC wavelengths, and is therefore
not identified as a point source by the SAGE-SMC team
\citep{Gordon_11_SAGE-SMC}. The S3MC team \citep{Bolatto_07_S3MC} was
able to fit a point source and extract the flux levesl.  They also
classify it as a candidate YSO.  This is entry \#7 in the IRS spectral
catalogue of YSO candidates by \citet{Oliveira_13_SMC}, who report the
presence of PAH emission.

\noindent \emph{S3MC J004624.46$-$732207.30 (SMC IRS 270)}  
is a YSO candidate \citep[][\#43]{Bolatto_07_S3MC}, observed
with MIPS SED \citep[][\#3]{vanLoon_10_MIPS-SMC} and IRS \citep[][\#6]{Oliveira_13_SMC}.

\noindent \emph{SSTISAGEMA J004547.53$-$732142.1 (SMC IRS 271)} 
is entry 5 in the IRS spectral catalogue by \citet{Oliveira_13_SMC}. It is an infrared variable \citep{Polsdofer_15_variables}.

\noindent \emph{2MASS J00452129$-$7312185 (SMC IRS 272)} 
is entry 4 in the IRS spectral catalogue by \citet{Oliveira_13_SMC}.
It also appears in the emission-line star catalogue by \citet{Meyssonnier_93_Halpha}.

\noindent \emph{IRAS 00429$-$7313 (SMC IRS 273)}, 
also known as LI-SMC 25, was originally thought of as an AGB star
candidate, based on its \iras\ detection
\citep{Loup_97_MagellanicClouds}. More recently, it was recognized as
an early-type YSO candidate, and it is included in the spectral
catalogues by \citet[\#1; who also provide a short literature review
on this source]{vanLoon_10_MIPS-SMC} and
\citet[\#2]{Oliveira_13_SMC}. \citet{Kamath_14_post-AGBcandidates}
also list it as YSO candidate, showing optical variability in the form
of Cepheid-like small oscillations, with a period of 22.5\,days.

\noindent \emph{LHA 115-N 8 (SMC IRS 274)}, 
also known as Lin 41, is entry \#1 in the IRS spectral catalogue
of YSO candidates by \citet{Oliveira_13_SMC}. It is also known 
as an emission-line star \citep{Meyssonnier_93_Halpha}.

\noindent \emph{2MASS J01061966$-$7155592 (SMC IRS 275)} 
is catalogued as a YSO candidate by \citet[entry \#257]{Bolatto_07_S3MC}.
Based on its UV continuum and emission-line characteristics,
\citet{Kamath_14_post-AGBcandidates} also list this as a YSO candidate.

\noindent \emph{2MASS J01052863$-$7159426 (SMC IRS 276)} 
is catalogued as a YSO candidate by \citet[entry \#251]{Bolatto_07_S3MC}. 
Based on its UV continuum and emission-line characteristics,
\citet{Kamath_14_post-AGBcandidates} also list this as a YSO candidate.

\noindent \emph{HV 12956 (SMC IRS 277)} 
is a luminous (\mbol$ \sim-6.5$ mag), Mira-type variable with a period
of 518\,days
\citep{Catchpole_81_RSG,Wood_83_LPVs,Soszynski_11_OGLE-III,Yang_12_PLrelation}. \citet{Polsdofer_15_variables}
also list it as an infrared variable. This M5e-type AGB star was
identified with the mid-IR source IRAS\,01074$-$7140
\citep{Whitelock_89_dustshell,Zijlstra_96_obscured,Groenewegen_98_IRASsources,vanLoon_98_OHIR},
and the mid-IR spectrum obtained with the \iso\ satellite was analysed
by \citet{Groenewegen_00_SMC}. Despite being one of the better
candidates in the SMC for circumstellar maser emission, none was
detected in deep searches \citep{vanLoon_01_masers}. The star was
found to be lithium-rich, probably as a result of nuclear processing
at the base of the convection zone \citep{Smith_95_lithium}. A 3--4
\um\ spectrum was presented in \citet{vanLoon_08_dustproduction}.

\noindent \emph{2MASS J01035898-7255327 (SMC IRS 278)} 
is entry \#238 in the YSO candidate list by \citet{Bolatto_07_S3MC}.

\noindent \emph{LHA 115-N 61 (SMC IRS 279)}, 
also known as Lin 321; IRAS 00557-7248 and LI-SMC 121; this is a
well-known emission line nebula in the SMC first discovered by
\citet{Henize_56_Halpha}.  It was classified as a planetary nebula by
\citet{Lindsay_61_catalogue} based on a lack of detected optical
continuum around H$\alpha$ and the presence of the [N\,{\sc ii}] 6548/6584
forbidden lines.  For this reason the object has sometimes been
considered as a PN in the literature
\citep[i.e.~][]{Jacoby_02_SMCPNe}, although it was classified as an
\hii\ region by \citet{Henize_63_SMC} and this was confirmed
spectroscopically by \citet{Dufour_77_PNe}.

\noindent \emph{Lin 238 (SMC IRS 280)} 
is a known emission-line star
\citep{Lindsay_61_catalogue,Meyssonnier_93_Halpha}.  More recently it was
included in the catalogue by \citet[][object 132]{Bolatto_07_S3MC} as
a YSO candidate. It is an infrared variable \citep{Polsdofer_15_variables}.

\noindent \emph{HD 5980 (SMC IRS 281)} 
is a famous 19-day eclipsing Wolf--Rayet (WR) binary
\citep{Breysacher_80_HD5980,Breysacher_82_HD5980} with estimated
masses of $M_{\rm A}=58$--79\,\msun\ and $M_{\rm B}=51$--67\,\msun,
respectively
\citep{Foellmi_08_HD5980}. \citet{Koenigsberger_14_multiple} report
that the system has in fact a third component, which is an O-star in a
eccentric orbit with a period of 96.56\,days.  HD 5980 had already
been recognised as peculiar well over a century ago
\citep{Pickering_01_peculiar}, and was identified as an emission-line
star half a century later \citep{Henize_56_Halpha}, with variable
emission-line width \citep{Feast_60_brightest}. It resides in the
massive cluster NGC\,346, but its proper motion of
\citep[$\mu_\alpha=-3.50\pm1.70$\,mas yr$^{-1}$,
$\mu_\delta=-2.40\pm1.60$\,mas yr$^{-1}$;][]{Hog_98_TYCHO} is
marginally significant. Component A, of WN type, underwent Luminous
Blue Variable (LBV)-type eruptions in 1993 and 1994
\citep{Barba_95_HD5980,Barba_96_HD5980,Cellone_96_HD5980,Eenens_96_HD5980,Heydari-Malayeri_97_HD5980,Sterken_97_HD5980,Koenigsberger_95_HD5980,Koenigsberger_96_HD5980,Koenigsberger_98_HD5980wind,Koenigsberger_98_HD5980Helium,Moffat_98_HD5980},
creating a circumstellar shell
\citep{Morris_96_transition,Koenigsberger_00_HD5980,Koenigsberger_01_HD5980,Gonzalez_14_LBVshell,Dopita_94_WRstars},
as well as showing longer term S\,Doradus-type variability
\citep{Koenigsberger_10_HD5980,Georgiev_11_HD5980} and spectral
changes preceding the outburst \citep{Koenigsberger_94_HD5980}.
HD\,5980 is a bright, variable X-ray source
\citep{Naze_04_XMM,Guerrero_08_Chandra} as a result of colliding winds
\citep{Breysacher_00_HD5980,Koenigsberger_04_HD5980,Koenigsberger_06_HD5980,Koenigsberger_08_HD5980};
polarimetric variability indicates complex structure and/or a neutron
star companion \citep{Villar-Sbaffi_03_HD5980}. Indeed, the supernova
remnant SNR\,0057$-$7226 lies in the direction towards HD\,5980
\citep{Hoopes_01_HD5980,Naze_02_HD5980,Velazquez_03_collision}. The
binary was detected at mid-IR wavelengths with \iso\ by
\citet{Contursi_00_N66}; the \spitzer\ mid-IR photometric properties
were analysed by \citet{Bonanos_10_SMC}, who point out that this is
the only WR star (but not the only LBV) in the SMC with a detection at
24\,\um. Being such bright, early-type, line-broadened star, HD\,5980
has been used extensively as a background probe for interstellar
medium studies \citep[despite its low $E(B-V)=0.05$\,mag
--][]{Schmutz_91_WRstars} -- e.g., polarization
\citep{Schmidt_70_polarization,Mathewson_70_polarization}; UV
\citep{deBoer_80_coronae,Savage_81_UV,Fitzpatrick_83_HD5980,Fitzpatrick_85_UV,Sembach_92_ionized}
including detection of H$_2$ \citep{Richter_98_H2,Shull_00_H2} -- see
also \citet{Cohen_84_SMC}; \citet{Songaila_86_gas};
\citet{Sembach_93_interstellar}; \citet{Lipman_95_titanium} and
\citet{Hoopes_02_interstellar}. \citet{Polsdofer_15_variables} list it
as an infrared variable.

\noindent \emph{HD 6884 (SMC IRS 282)} 
(or R\,40) is a luminous blue supergiant \citep{Feast_60_brightest} of
type B9\,Iae (\citealt{Garmany_87_OBstars}; B8\,Ia based on UV --
\citealt{Smith-Neubig_97_OB}) displaying \ha\ line emission
\citep[LHA\,115-S\,52;][]{Henize_56_Halpha}, with confirmed membership of
the SMC based on a radial velocity of 170\,\kms
\citep{Buscombe_62_supergiant}. It is the visually brightest B star
\citep{Stahl_85_peculiar} and the first recognised LBV in the SMC
\citep{Szeifert_93_R40}, explaining the later spectral type of A2
determined by \citep{Lennon_97_supergiants} and the S\,Dor-type
variability \citep{Sterken_98_R40}. Early optical photometry was
presented by \citet{Dachs_70_SMC}, \citet{Osmer_73_SMC},
\citet{Ardeberg_77_supergiant} and \citet{vanGenderen_82_VBLUW}; a
K-band spectrum was presented by \citet{Oksala_13_SINFONI}. Not
surprisingly, it has been used extensively for studies of the
interstellar medium
\citep[e.g.,][]{Cohen_84_SMC,Songaila_86_gas}; the 4430-\AA\
diffuse interstellar band was detected in its spectrum by
\citet{Hutchings_66_4430}, and the degree of polarization was measured
by \citet{Mathewson_70_polarization}. \citet{Bonanos_10_SMC}
detected mid-IR excess emission on the basis of \spitzer\
photometry. It is in the \citet{Polsdofer_15_variables} list of infrared variables.

\noindent \emph{2MASS J00531330$-$7312176 (SMC IRS 283)} 
is entry \#131 in the YSO candidate list by \citet{Bolatto_07_S3MC}.
The source is possibly also detected in the far-infrared \citep{Wilke_03_farinfrared},
as [WSH2003] b-63, although the coordinates are very coarse at these wavelengths. 
Within 10 arcsec, another far-infrared detection with \herschel\ has also been reported, again
with rather imprecise coordinates \citep{Meixner_13_HERITAGE}.

\noindent \emph{2MASS J00531330$-$7312176 (SMC IRS 284)} 
is entry \#129 in the YSO candidate list by \citet{Bolatto_07_S3MC}.

\noindent \emph{2MASS J00505425$-$7324170 (SMC IRS 286)}. 
 is YSO candidate 112 in the catalogue by \citet{Bolatto_07_S3MC}.  

 \noindent \emph{LHA 115-N 32 (SMC IRS 287)} 
 LHA 115-N 32 is a well known source, with more than 20 references,
 starting with \citet{Henize_56_Halpha} and \citet{Lindsay_61_catalogue}.
 It has been characterized more as an \hii\ region than a PN
 \citep[most recently by][]{Charmandaris_08_HII}. We classify the spectrum as 
YSO-3, but remark that it is due from a source located within an \hii\ region.

\noindent \emph{2MASS J00484901$-$7311226 (SMC IRS 288)} 
is entry \#83 in the list of YSO candidates by \citet{Bolatto_07_S3MC}.
In contrast, \citet{Kamath_14_post-AGBcandidates} catalogue this object
as a hot post-AGB candidate. We classify it as YSO-3.

\noindent \emph{2MASS J00465576$-$7331584 (SMC IRS 289)} 
is entry \#47 in the list of YSO candidates by \citet{Bolatto_07_S3MC}.
\citet{Kamath_14_post-AGBcandidates} also list this object as a candidate
YSO.

\noindent \emph{2MASS J00454296$-$7317263 (SMC IRS 290)} 
is entry \#27 in the list of YSO candidates by \citet{Bolatto_07_S3MC}.

\noindent \emph{IRAS 01042$-$7215 (SMC IRS 291)}. 
\citet{Wilke_03_farinfrared} determined the 25- and 60-\um\ flux
densities from high resolution \iras\ maps (source b-97), to be
$F_{25}=0.6\pm0.1$ and $F_{60}=8.4\pm0.3$\,Jy (cf.\
\citealt{Schwering_89_SMC}); they did not detect the source in their
\iso-PHOT maps at a wavelength of 170\,\um. A far-IR spectrum obtained
with the \spitzer\ SED mode was presented by
\citet{vanLoon_10_MIPS-SMC}. The near-IR counterpart was found by
\citet{Groenewegen_98_IRASsources}. While \citet{Groenewegen_00_SMC}
modelled the \iso\ mid-IR data as if it were a mass-losing cool
oxygen-rich AGB star, \citet{vanLoon_08_dustproduction} classified it
as a candidate YSO on the basis of a 3--4-\um\ spectrum displaying
water ice absorption and hydrogen recombination emission. Indeed,
\citet{Oliveira_11_metallicity} detected (weak) CO$_2$ ice absorption
in the \spitzer\ IRS spectrum confirming the YSO nature of this
object; CO ice was not detected in their groundbased M-band
spectrum. \citet{Oliveira_13_SMC} also presented an optical spectrum
showing just broad \ha\ emission, as well as new near-IR photometry
($J$\kband $L^\prime$). \citet{Polsdofer_15_variables} list it as an
infrared variable.

\noindent \emph{Lin 49 (SMC IRS 292)} 
has been known as an emission line source and/or a PN
\citep{Lindsay_61_catalogue,Henize_63_SMC,Dopita_85_PNe,Meyssonnier_93_Halpha,Morgan_95_UKST}.
While \citet{Bolatto_07_S3MC} included it in their YSO candidate list
(\#7), the IRS spectrum clearly shows that this is a C-PN, and a
fullerene source \citep[this work;][]{Sloan_14_aromatics}.

\noindent \emph{2MASS J00431490$-$7300426 (SMC IRS 294)} 
is entry \#4 in the list of YSO candidates by \citet{Bolatto_07_S3MC}.

\noindent \emph{NGC 330 BAL 555 (SMC IRS 296)} 
is another member of NGC 330, and is known to be a long period
semi-regular variable from OGLE observations
\citep{Soszynski_11_OGLE-III}.  The object appears to have first been
catalogued as a member of NGC 330 by \citet{Balona_92_NGC330} where it
is object 555 in Table 2. \citet{Kamath_14_post-AGBcandidates} list it
as a carbon star, consistent with our classification of this object as
C-AGB.

\noindent \emph{2MASS J01065966$-$7250430 (SMC IRS 298)} 
is an emission-line star \citep{Meyssonnier_93_Halpha}, listed
as a candidate PN by \citet{Kamath_14_post-AGBcandidates}.
\citet{vanLoon_10_MIPS-SMC} list it as a YSO (entry 11), and the IRS
spectrum confirms the YSO nature of this object
\citep[][\#30]{Oliveira_13_SMC}. We classify it as YSO-3.

\noindent \emph{2MASS J01053088$-$7155209 (SMC IRS 299)} 
is entry 29 in the IRS spectral catalogue of YSO candidates by \citet{Oliveira_13_SMC}.

\noindent \emph{2MASS J01050732$-$7159427 (SMC IRS 300)} 
is entry 8 in the study by \citet{vanLoon_10_MIPS-SMC}, who list it
as a YSO, and provide some further references. Its YSO nature
is confirmed by its IRS spectrum \citep[][\#28]{Oliveira_13_SMC}.

\noindent \emph{S3MC J010306.13$-$720343.95 (SMC IRS 301)} 
is entry 228 in the YSO candidate list by \citet{Bolatto_07_S3MC},
and its YSO nature is confirmed by its IRS spectrum
\citep[][\#27]{Oliveira_13_SMC}.

\noindent \emph{S3MC J010248.54$-$715317.98 (SMC IRS 302)} 
is an mid-infrared point source (also detected by the SAGE team as
SSTISAGEMA J010248.56$-$715318.0) part of star forming region A8
discovered by \citep{Livanou_07_SFregions}, which is about 240 pc in
size. Region A8 contains several \hii\ regions, and S3MC
J010248.54$-$715317.98 is located inside one of these \hii\
regions, namely DEM S 117b, also known as LHA 115-N 77A
\citep{Bica_95_SMC}, which has a size of 0.55\,arcmin. The IRS
spectrum therefore shows many emission lines due to the \hii\
region. We conclude that the point source is an YSO-3 object
located within an \hii\ region \citep[see
also][\#26]{Oliveira_13_SMC}.

\noindent \emph{S3MC J010131.69$-$715040.30 (SMC IRS 303)} 
is entry 25 in the IRS spectral catalogue of YSO candidates by
\citet{Oliveira_13_SMC}, and entry 210 in the list of YSO candidates
of \citet{Bolatto_07_S3MC}.

\noindent \emph{SSTISAGEMA J010022.34-720957.8 (SMC IRS 304)} 
is entry 24 in the IRS spectral catalogue of YSO candidates by
\citet{Oliveira_13_SMC}, and it is also entry 60 in the list of YSOs
in N66 compiled by \citet{Simon_07_NGC346}, who derive $L = 2910$\,\lsun\ and $M = 8$\,\msun\ for this source. It is also known to vary in the infrared \citep[][Riebel et al.~\emph{in prep.}]{Polsdofer_15_variables}.

\noindent \emph{IRAS 00563$-$7220 (SMC IRS 305)} 
is entry 23 in the IRS spectral catalogue of YSO candidates by
\citet{Oliveira_13_SMC}.  It is also \#174 in the YSO candidate list
published by \citet{Bolatto_07_S3MC}.

\noindent \emph{IRAS 00562$−$7255 (SMC IRS 306)} 
is entry 22 in the IRS spectral catalogue of YSO candidates by
\citet{Oliveira_13_SMC}.  It is also \#171 in the YSO candidate list
published by \citet{Bolatto_07_S3MC}.

\noindent \emph{2MASS J00560662$-$7247225 (SMC IRS 307)} 
is reported to be an emission-line star \citep{Meyssonnier_93_Halpha},
perhaps consisting of multiple components.  It is also entry 21 in the
IRS spectral catalogue of YSO candidates by
\citet{Oliveira_13_SMC}. \citet{Bolatto_07_S3MC} list it as \#146 in
their list of YSO candidates, and it is also included as a YSO
candidate in the work by \citet{Kamath_14_post-AGBcandidates}. We
classify it as YSO-1.

\noindent \emph{HV 11464 (SMC IRS 309)} 
is first classified as M0\,I by
\citet{Prevot_83_latetype} and \citet{Elias_85_supergiants}; SMC
membership was confirmed based on the heliocentric radial velocity of
$v_{\rm hel}=189$\,\kms\ \citep{Maurice_87_CORAVEL}; however the
spectral type was revised to K0\,I and the radial velocity to 182\,\kms\
by \citet{Massey_03_RSGs}, who also determined a bolometric
luminosity of \mbol\ $=-8.0$\,mag. \citet{Yang_12_PLrelation}
determined a long secondary period of 1500--1600\,days. The \spitzer\
mid-IR photometry was analysed by \citet{Bonanos_10_SMC}. 
It straddles the boundary between the RSG and AGB classification, and 
because of its \mbol\ value we classify it as O-AGB. 

\noindent \emph{IRAS 00436$-$7321 (SMC IRS 310)} 
is in a complex field.  The source is also an emission-line star and
listed as entry 108 in the \ha\  survey by
\citet{Meyssonnier_93_Halpha}. The \iras\ name is cross-identified with
this \spitzer\ source, as it is the only source with a MIPS-[24] flux
level comparable to the \iras-[25] flux within the error box of the
\iras-[25] position (no detection at \iras-[12]).  Keller et
al.~(\emph{in prep.}) believe this source to be a PN, however we
classify it as HII.

\noindent \emph{NGC 330 ARP 41 (SMC IRS 311)}, 
 a member of the NGC 330 cluster, has been known to be a 
late-type supergiant since the 1970's.  The star was first observed 
by \citet{Arp_59_NGC330} as object 41 in his list.  The spectral 
type was determined as G6Ib by \citet{Feast_79_RSGs}.   Our classification of this object as a red supergiant 
is (marginally) consistent with this spectral type.

\section*{Supporting Information}

Additional Supporting Information may be found in the online version of this article: \\

\noindent Online Table described by Table~\ref{tab:metatable}. Classification of point sources targeted in IRS staring mode. \\

\noindent Please note: Oxford University Press (OUP) are not responsible for the content or functionality of any supporting materials supplied by the authors. Any queries (other than missing material) should be directed to the corresponding author for the article.

\bsp  

\end{document}